\documentclass[fleqn,usenatbib]{mnras}

\usepackage{hyperref}
\hypersetup{
  colorlinks,
  citecolor=blue,
  linkcolor=red,
  urlcolor=blue}

\usepackage{newtxtext,newtxmath}
\usepackage[compatibility=false]{caption}
\usepackage{subcaption}
\usepackage{lscape}
\usepackage[LGRgreek]{mathastext}
\usepackage{enumitem}
\usepackage{amssymb}
\usepackage{pifont}
\usepackage{enumitem}

\usepackage[T1]{fontenc}
\usepackage{ae,aecompl}

\usepackage{cleveref}
\crefname{section}{$\S$}{$\S\S$}
\Crefname{section}{$\S$}{$\S\S$}

\usepackage{graphicx}   
\graphicspath{{Users/}{owenturner/}{Documents/}{PhD/}{KMOS/}{paper_1/}{Figures/}}
\usepackage{amsmath}    
\usepackage{amssymb}    
\usepackage[flushleft]{threeparttable}

\newcommand{\Sers}{S\'{e}rsic }
\newcommand{\Lagr}{\mathcal{L}}

\newcommand{\cmark}{\ding{51}}%
\newcommand{\xmark}{\ding{55}}%

\title[KDS I: dynamical properties of 77 $z\simeq3.5$ galaxies]{The KMOS Deep Survey (KDS) I: dynamical measurements of typical star-forming galaxies at $z\simeq3.5$ \thanks{Based on observations obtained at the Very Large Telescope of the European Southern Observatory. Programme IDs: 092.A-0399(A), 093.A-0122(A,B), 094.A-0214(A,B),095.A-0680(A,B),096.A-0315(A,B,C)}}

\author[O.J. Turner et al.]{
O. J. Turner,$^{1,2,\thanks{E-mail: ojamesturner@icloud.com (OJT)}}$
M. Cirasuolo,$^{2,1}$
C. M. Harrison,$^{2,3}$
R. J. McLure,$^{1}$
J. S. Dunlop,$^{1}$\newauthor
A. M. Swinbank,$^{3,4}$
H. L. Johnson,$^{3,4}$
D. Sobral,$^{5,6}$
J. Matthee,$^{6}$
R. M. Sharples$^{3,4}$
\\
$^{1}SUPA\thanks{Scottish Universities Physics Alliance}$, Institute for Astronomy, University of Edinburgh, Royal Observatory, Edinburgh EH9 3HJ\\
$^{2}$European Southern Observatory, Karl-Schwarzschild-Str. 2, 85748 Garching b. M{\"u}nchen, Germany\\
$^{3}$Centre for Extragalactic Astronomy, Durham University, South Road, Durham, DH1 3LE, U.K.\\
$^{4}$Institute for Computational Cosmology, Durham University, South Road, Durham, DH1 3LE, U.K.\\
$^{5}$Department of Physics, Lancaster University, Lancaster, LA1 4BY, U.K.\\
$^{6}$Leiden Observatory, Leiden University, PO Box 9513, NL-2300 RA Leiden, the Netherlands}

\date{Accepted XXX. Received YYY; in original form ZZZ}

\pubyear{2017}

\begin{document}
\label{firstpage}
\pagerange{\pageref{firstpage}--\pageref{lastpage}}
\maketitle

\begin{abstract}
We present dynamical measurements from the KMOS ({\it K}-band Multi-Object Spectrograph) Deep Survey (KDS), which is comprised of 77 typical star-forming galaxies at $z\simeq3.5$ in the mass range $9.0 < log(M_{\star}/M_{\odot}) < 10.5$.
These measurements constrain the internal dynamics, the intrinsic velocity dispersions ($\sigma_{int}$) and rotation velocities ($V_{C}$) of galaxies in the high-redshift Universe.
The mean velocity dispersion of the galaxies in our sample is $\sigma_{int} = 70.8^{+3.3}_{-3.1} km\,s^{-1}$, revealing that the increasing average $\sigma_{int}$ with increasing redshift, reported for $z\lesssim2$, continues out to $z\simeq3.5$.
Only $34 \pm 8\%$ of our galaxies are rotation-dominated ($V_{C}/\sigma_{int}$ > 1), with the sample average $V_{C}/\sigma_{int}$ value much smaller than at lower redshift.
After carefully selecting comparable star-forming samples at multiple epochs, we find that the rotation-dominated fraction evolves with redshift with a $z^{-0.2}$ dependence. 
The rotation-dominated KDS galaxies show no clear offset from the local rotation velocity-stellar mass (i.e. $V_{C}-M_{\star}$) relation, although a smaller fraction of the galaxies are on the relation due to the increase in the dispersion-dominated fraction.
These observations are consistent with a simple equilibrium model picture, in which random motions are boosted in high-redshift galaxies by a combination of the increasing gas fractions, accretion efficiency, specific star-formation rate and stellar feedback and which may provide significant pressure support against gravity on the galactic disk scale.
\end{abstract}




\section{INTRODUCTION}

The galaxy population at all redshifts appears to be bimodal in many physical properties \citep[e.g. as described in][]{Dekel2006}, with a preference for the most massive galaxies to lie on the red sequence, characterised by red optical colours, low star-formation rates (SFRs) and spherical morphologies, and less massive galaxies in the blue sequence, characterised by blue colours, high star-formation rates and disky morphologies. 
For these blue, star-forming galaxies (SFGs) there is a roughly linear correlation between SFR and stellar mass ($M_{\star}$) \citep[e.g.][]{Daddi2007,Noeske2007,Elbaz2007}, in the sense that galaxies which have already accumulated a larger stellar population tend to have higher SFRs.
This correlation, or `main-sequence', underpins the `equilibrium model', in which the SFR of galaxies is regulated by the availability of gas, with outflows and accretion events sustaining the galaxy gas reservoirs in a rough equilibrium as the galaxy evolves \citep[e.g.][]{Dave2012,Lilly2013,Saintonge2013}.
The main-sequence has been studied comprehensively, using multi-wavelength SFR tracers, between $0 < z < 3$ \citep[e.g.][]{Rodighiero2011,Karim2011,Whitaker2012,Behroozi2013b,Whitaker2014,Rodighiero2014,Speagle2014,Pannella2014,Sobral2014,Sparre2015,Lee2015,Schreiber2015,Renzini2015,Nelson2016}, showing evolution of the relation towards higher SFRs at fixed $M_{\star}$ with increasing redshift, reflecting the increase of the cosmic Star-Formation Rate Density (SFRD) in this redshift range \citep[e.g.][]{Madau_2014,Khostovan2015}.
At each redshift slice it has been suggested that galaxies on the main-sequence evolve secularly, regulated by their gas reservoirs, meaning that selecting such populations offers the chance to explore the evolution of the physical properties of typical SFGs across cosmic time.
This assumes that high-redshift, main-sequence galaxies are the progenitors of their lower redshift counterparts, which may not be the case \citep[e.g.][]{Gladders2013,Kelson2014,Abramson2016b} and also assumes that we can learn about galaxy evolution (i.e. how individual galaxies develop in physical properties over time) by studying the mean properties of populations at different epochs. 

The picture is complicated by the addition of major and minor galaxy mergers which can rapidly change the physical properties of galaxies \citep[e.g.][]{Toomre1977,Lotz2008,Conselice2011,Conselice2014} and the relative importance of in-situ, secular stellar mass growth vs. stellar mass aggregation via mergers is the subject of much work involving both observations and simulations \citep[e.g.][]{Robaina2009,Kaviraj2012,Stott2013,Lofthouse2017,Qu2017}. 
To account for the growing number density of quiescent galaxies from $z\simeq2.5$ to the present day \citep[e.g.][]{Bell2004,Faber2007,Brown2007,Ilbert2010,Brammer2011,Muzzin2013,Buitrago2013} there must also be processes which shut-off star-formation within main-sequence galaxies (i.e. quenching) which, to explain observations, must be a function of both mass and environment \citep{Peng2010,Darvish2016}.

Recent cosmological volume simulations provide subgrid recipes for the complex interplay of baryonic processes which are at work as galaxies evolve, and can track the development of individual galaxies from early stages, to maturity and through quenching \citep{Dubois2014,Vogelsberger2014b,Schaye2015}.
Observations can aid the predictive power of such simulations by providing constraints on the evolving physical properties of galaxy populations.
The observed dynamical properties of galaxies contain information about the transfer of angular momentum between their dark matter halos and baryons, and the subsequent dissipation of this angular momentum (through gravitational collapse, mergers and outflows e.g. \citealt{Fall1983,Romanowsky2012,Fall2013}), constituting an important set of quantities for simulations to reproduce.
Developments in both Integral-Field Spectroscopy (IFS) instrumentation and data analysis tools over the last decade have led to the observation of two-dimensional velocity and velocity dispersion fields for large samples of galaxies of different morphological types, spanning a wide redshift range \citep[e.g.][]{Sarzi2005,Flores2006,Epinat2008,ForsterSchreiber2009,Cappellari2011,Gnerucci2011,Epinat2012,Croom2012,Swinbank2012,Swinbank2012a,Bundy2015,Wisnioski2015,Stott2016,Harrison2017,Swinbank2017}.
When interpreted in tandem with high-resolution imaging data from the {\it Hubble Space Telescope (HST)}, these data provide information about the range of physical processes which are driving galaxy evolution.
In particular, in recent years the multiplexing capabilities of KMOS \citep{Sharples2013} have allowed for IFS kinematic observations for large galaxy samples to be assembled rapidly \citep{Sobral2013,Wisnioski2015,Stott2016,Mason2017,Harrison2017} providing an order-of-magnitude boost in statistical power over previous high-redshift campaigns.    

Random motions within the interstellar medium of SFGs appear to increase with increasing redshift between $0 < z < 3$, as traced by their observed velocity dispersions, $\sigma_{obs}$ \citep{Genzel2008,ForsterSchreiber2009,Law2009,Cresci2009,Gnerucci2011,Epinat2012,Kassin2012,Green2014,Wisnioski2015,Stott2016}.
This has been explained in terms of increased `activity' in galaxies during and before the global peak in cosmic SFRD \citep{Madau_2014}, in the form of higher specific star-formation rates (sSFRs) \citep{Wisnioski2015}, larger gas reservoirs \citep{Law2009,ForsterSchreiber2009,Wisnioski2015,Stott2016}, more efficient accretion \citep{Law2009}, increased stellar feedback from supernovae \citep{Kassin2012} and turbulent disk instabilities \citep{Law2009,Bournaud2007,Bournaud2016}, all of which combine to increase $\sigma_{obs}$ and complicate its interpretation.

There is also an increasing body of work measuring the relationship between the observed maximum rotation velocity of a galaxy, a tracer for the total dynamical mass, and its stellar mass, known as the stellar mass Tully-Fisher Relation (smTFR) \citep{Tully1977}, with surveys reporting disparate results for the evolution of this relation with redshift \citep[e.g.][]{Puech2008,Miller2011,Gnerucci2011,Swinbank2012,Simons2016,Tiley2016,Harrison2017,Straatman2017,Ubler2017}.
Systematic differences in measurement and modelling techniques at high-redshift, especially with regards to beam-smearing corrections, combine with our poor understanding of progenitors and descendants to blur the evolutionary picture which these surveys paint.
Additionally, there has been increasing focus in recent years on whether the measured velocity dispersions track random motions which provide partial gravitational support for high-redshift galaxy disks \citep[e.g.][]{Burkert2010,Wuyts2016b,Ubler2017,Genzel2017,Lang2017}.
These random motions may become an increasingly significant component of the dynamical mass budget with increasing redshift \citep{Wuyts2016b} and pressure gradients across the disk could result in a decrease in the observed rotation velocities \citep{Burkert2010}. 
Different interpretations of the gaseous velocity dispersions and their role in providing pressure support against gravity also complicate the evolutionary picture. \\

\noindent
In this paper we present new results from the KMOS Deep Survey (KDS), which is a guaranteed time programme focusing on the spatially-resolved properties of main-sequence SFGs at $z\simeq3.5$, a time when the universe was building to peak activity.
With this survey, we aim to complement existent studies by providing deep IFS data for the largest number of galaxies at this redshift.
By making use of KMOS (with integration times of 7.5-9 hours), we have been able to study [O~{\sc III}]$\lambda$5007 emission in 77 galaxies spanning the mass range $9.0 < log(M_{\star}/M_{\odot}) < 10.5$, roughly tripling the number of galaxies observed via IFS at $z > 3$.
In order to interpret the evolution of the physical properties of typical star-forming galaxies we have carefully constructed a set of comparison samples spanning $0 < z < 3$.
These samples use integral-field spectroscopy to track the ionised gas emission in star-forming galaxies and follow consistent kinematic parameter extraction methods.  
By doing this we seek to minimise the impact of systematic differences introduced by differing approaches to defining and extracting kinematic parameters.

There are still many open questions which we can begin to answer by studying the emission from regions of ionised gas within individual galaxies at these redshifts: 
\begin{enumerate}[label=(\roman*),align=left]
\item What are the dynamical properties of main-sequence galaxies at this early stage in their lifetimes?
\item  What are the radial gradients in metal enrichment within these galaxies and what can this tell us about the physical mechanisms responsible for redistributing metals?
\item What is the connection between the gas-phase metallicity and  kinematics, particularly in terms of inflows and outflows of material?
\end{enumerate}

This paper focuses on (i) by deriving and interpreting the spatially-resolved kinematics of the KDS galaxies, particularly the rotation velocities and velocity dispersions, using the [O~{\sc III}]$\lambda$5007 emission line, discussing what we can learn about the nature of galaxy formation at $z\simeq3.5$ and forming evolutionary links with lower redshift work. \\

\noindent
The structure of the paper is as follows.
In \cref{sec:Survey_and_data} we present the survey description, sample selection, observation strategy and data reduction, leading to stacked datacubes for each of the KDS galaxies.
In \cref{sec:analysis} we describe the derivation of morphological and kinematic properties for our galaxies, explaining the kinematic modelling approach and the beam-smearing corrections which lead to intrinsic measurements of the rotation velocities, $V_{C}$, and velocity dispersions $\sigma_{int}$ for each of the galaxies classified as morphologically isolated and spatially-resolved in the [O~{\sc III}]$\lambda$5007 emission line.
\cref{sec:results} presents an analysis of these derived kinematic parameters, comparing with lower redshift work where possible and drawing conclusions about the evolutionary trends and possible underlying physical mechanisms.
We discuss these results in \cref{sec:discussion} and present our conclusions in \cref{sec:conclusion}.
Throughout this work we assume a flat $\Lambda$CDM cosmology with (h, $\Omega_{m}$, $\Omega_{\Lambda}$) = (0.7, 0.3, 0.7). 

\section{SURVEY DESCRIPTION, SAMPLE SELECTION AND OBSERVATIONS}\label{sec:Survey_and_data}
\subsection{The KDS survey description and sample selection}\label{subsec:survey_intro}
The KDS is a KMOS study of the gas kinematics and metallicity in 77 SFGs with a median redshift of $z\simeq3.5$, probing a representative section of the galaxy main-sequence.
The addition of these data approximately triples the number of galaxies observed via IFS at this redshift \citep{Cresci2010,Lemoine-Busserolle2010,Gnerucci2011}, and will allow for a statistically-significant investigation of the dynamics and metal content of SFGs during a crucial period of galaxy evolution. 
The key science goals of the KDS are to investigate the resolved kinematic properties of high-redshift galaxies in the peak epoch of galaxy formation (particularly the fraction of rotating disks and the degree of disk turbulence) and also to study the spatial distribution of metals within these galaxies in the context of their observed dynamics.
We seek to probe both a `field' environment in which the density of galaxies is typical for this redshift and a `cluster' environment containing a known galaxy over-density, in order to gauge the role of environment in determining the kinematics and metallicities of SFGs during this early stage in their formation history.
To achieve this we require very deep exposure times in excess of 7 hours on source to reach the signal-to-noise required to detect line emission in the outskirts of the galaxies where the rotation curves begin to flatten, and to achieve adequate signal-to-noise across several ionised emission lines within individual spatial pixels (spaxels).
Consequently, the KDS is one of the deepest spectroscopic datasets available at this redshift. 

\subsubsection{Sample selection}\label{subsubsec:sample_selection}
Target selection for the KDS sample is designed to pick out SFGs at $z\simeq3.5$, supported by deep, multi-wavelength ancillary data.
Within this redshift range the [O~{\sc III}]$\lambda$4959,$\lambda$5007 doublet and the $H\,\beta$ emission lines are visible in the {\it K}-band and the [O~{\sc II}]$\lambda$3727,$\lambda$3729 doublet is visible in the {\it H}-band, both of which are observable with KMOS.
From these lines, [O~{\sc III}]$\lambda$5007 generally has the highest signal-to-noise and so is well suited to dynamical studies, whereas [O~{\sc III}]$\lambda$4959, $H\,\beta$ and the [O~{\sc II}]$\lambda$3727,$\lambda$3729 doublet complement [O~{\sc III}]$\lambda$5007 as tracers of the galaxy metallicities. 
To ensure a high detection rate of the ionised gas emission lines in the KDS we select galaxies in well-studied fields that have a wealth of imaging and spectroscopic data.
Most of the galaxies for the KMOS observations had a confirmed spectroscopic redshift (see below).
A subset of the selected cluster galaxies in the SSA22 field were blindly-detected in Ly$\alpha$ emission during a narrow-band imaging study of a known overdensity of Lyman Break Galaxies (LBGs) at $z\simeq3.09$ \citep{Steidel2000}.
In each pointing, few sources had no spectroscopic redshift and were selected on the basis of their photometric redshift.
We make no further cuts to the sample on the basis of mass and SFR, in order to probe a more representative region of the star-forming main-sequence at this redshift (see Figure \ref{fig:main_sequence}).

\subsubsection{GOODS-S}\label{subsubsec:sample_selection_goods}

\begin{table*}
\centering
\begin{threeparttable}
\caption{This table summarises the KDS pointing statistics for the full observed sample of 77 galaxies.
The columns list the pointing name and galaxy environment probed, the central pointing coordinates, the number of observed, detected, resolved and merging objects as described in \protect\cref{subsubsec:datareduction}, the waveband observed with KMOS, the exposure time and the PSF measured in the {\it K}-band.}
\label{tab:pointings}
\begin{tabular}{l l l l l l l l l l l l l}

 \hline
Pointing & RA & DEC & N$_{obs}$ & N$_{Det}$ & $\%$ Det. & N$_{Res}$ & $\%$ Res & N$_{Merg}$ & $\%$ Merg$^{*}$ & Band$^{a}$ & Exp (ks) & PSF ($^{\prime\prime}$)$^{b}$  \\
 \hline
 \scriptsize{\bf GOODS-S-P1} & 03:32:25.9 & $-$27:51:58.7 & 20 & 16 & 80 & 13 & 65 & 2 & 17 & {\it K} & 32.4 & 0.50 \\
\scriptsize{\bf GOODS-S-P2} & 03:32:32.2 & $-$27:43:08.0 & 17 & 14 & 82 & 13 & 76 & 2 & 18 & {\it K} & 31.8 & 0.52 \\
\scriptsize{\bf SSA22-P1} & 22:17:11.9 & $+$00:15:44.7 & 21 & 15 & 71 & 9 & 46 & 8 & 89 & {\it HK} & 38.1 & 0.62 \\
\scriptsize{\bf SSA22-P2} & 22:17:35.1 & $+$00:09:30.5 & 19 & 17 & 89 & 12 & 63 & 2 & 18 & {\it HK} & 27.8 & 0.57 \\

 \hline
\end{tabular}
\begin{tablenotes}
      \small
      \item $^{a}$ We also observed the two GOODS-S pointings in the {\it H}-band to cover the [O~{\sc II}]$\lambda$3727,3729 emission lines and a description of these observations will be given in a future work.
      \item $^{b}$ The PSF values correspond to measurements in the {\it K}-band.
      \item $^{*}$ Note that the Merger percentage is computed with respect to the number of resolved galaxies; the other percentages are computed with respect to the total number of galaxies observed in that pointing.
    \end{tablenotes}
  \end{threeparttable}
  \end{table*}

Two of the three field environment pointings are selected within the GOODS-S region \citep{Guo2013}; accessible from the VLT and with excellent multi-wavelength coverage, including deep {\em HST WFC3 F160W} imaging with a $0.06^{\prime\prime}$ pixel scale and $\simeq0.2^{\prime\prime}$ PSF, which is well suited for constraining galaxy morphology \citep{Grogin2011,Koekemoer2011}.
We selected targets from the various spectroscopic campaigns which have targeted GOODS-S, including measurements from VIMOS \citep{Balestra2010,Cassata2014}, FORS2 \citep{Vanzella2005,Vanzella2006,Vanzella2008} and both LRIS and FORS2 as outlined in \cite{Wuyts2009}.
These targets must be within the redshift range $3 < z < 3.8$, have high spectral quality (as quantified by the VIMOS redshift flag equal `3' or `4', and the FORS2 quality flag equal `A') and we carefully excluded those targets for which the [O~{\sc III}]$\lambda$5007 or $H\,\beta$ emission lines, observable in the {\it K}-band at these redshifts, would be shifted into a spectral region plagued by strong OH emission.
The galaxies which remain after imposing these criteria are distributed across the GOODS-S field, and we selected two regions where $\simeq20$ targets could be allocated to the KMOS IFUs (noting that the IFUs can patrol a $7.2^{\prime}$ diameter patch of sky during a single pointing).
We name these GOODS-S-P1 and GOODS-S-P2, which which we observe 20 and 17 galaxies respectively (see Table \ref{tab:pointings}).\footnote{We note that the number of observed galaxies quoted for GOODS-S-P2 does not include two observed targets which were later found to have $z < 0.5$.}

\subsubsection{SSA22}\label{subsubsec:sample_selection_ssa}
A single cluster environment pointing was selected from the SSA22 field, \citep{Steidel1998,Steidel2000,Steidel2003,Shapley2003}, which, as mentioned above, is an overdensity of LBG candidates at $z\simeq3.09$.
Hundreds of spectroscopic redshifts have been confirmed for these LBGs with follow-up observations using LRIS \citep{Shapley2003,Nestor2013}.
A combination of deep {\em B,V,R} band imaging with the Subaru Suprime-Cam \citep{Matsuda2004}, deep narrow-band imaging at 3640$\AA$ \citep{Matsuda2004} and at 4977$\AA$ \citep{Nestor2011,Yamada2012a} and archival {\em HST ACS} and {\it WFC3} imaging provides ancillary data in excellent support of integral field spectroscopy, albeit over a shorter wavelength baseline and with shallower exposures than in the GOODS-S field.
Fortunately at $z\simeq3.09$ the [O~{\sc III}]$\lambda$5007 line is shifted into a region of the {\it K}-band which is free from OH features and so for the cluster environment pointing we filled the KMOS IFUs with galaxies located towards the centre of the SSA22 protocluster (SSA22-P1).

We also added a further field environment pointing to the south of the main SSA22 spatial overdensity where the density of galaxies is typical of the field environment (SSA22-P2).
In SSA22-P1 and SSA22-P2 we observe 19 and 21 galaxies respectively.
In summary, we have chosen three field environment pointings and a single cluster environment pointing across GOODS-S and SSA22, comprising a total of 77 galaxies, as described in Table \ref{tab:pointings}. \footnote{Additional pointings in the COSMOS and UDS fields were originally scheduled as part of the GTO project, however 50$\%$ of the observing time was lost to bad weather during these visitor mode observations.}

\subsection{Observations and data reduction}\label{subsubsec:observations_and_dr}

Our data for the 77 KDS targets were observed using KMOS \citep{Sharples2013}, which is a second generation IFS mounted at the Nasmyth focus of UT1 at the VLT.
The instrument has 24 moveable pickoff arms, each with an integrated IFU, which patrol a region 7.2$^{\prime}$ in diameter on the sky, providing considerable flexibility when selecting sources for a single pointing.
The light from a set of 8 IFUs is dispersed by a single spectrograph and recorded on a 2k$\times$2k Hawaii-2RG HgCdTe near-IR detector, so that the instrument is comprised of three effectively independent modules.
Each IFU has 14$\times$14 spatial pixels which are 0.2$^{\prime\prime}$ in size, and the central wavelength of the {\it K}-band grating has a spectral resolution of $R\simeq4200$ ({\it H}-band $R\simeq4000$, {\it HK}-band $R\simeq2000$).
\subsubsection{Observations}\label{subsubsec:Obs}

To achieve the science goals of the KDS, the target galaxies at $3 < z < 3.8$ were observed in both the {\it K}-band, into which the [O~{\sc III}]$\lambda$5007 and $H\,\beta$ lines are redshifted, and the {\it H}-band, into which the [O~{\sc II}]$\lambda$3727,$\lambda$3729 doublet is redshifted, allowing both dynamical and chemical abundance measurements.
The GOODS-S pointings were observed in the {\it H} and {\it K}-bands separately, however, due to loss of observing time the SSA22 galaxies were observed with the KMOS {\it HK} filter, which has the disadvantage of effectively halving the spectral resolution, but allows for coverage of the {\it H}-band and {\it K}-band regions simultaneously. \\

\noindent
We prepared each pointing using the KARMA tool \citep{Wegner2008}, taking care to allocate at least one IFU to observations of a `control' star closeby on the sky to allow for precise monitoring of the evolution of seeing conditions and the shift of the telescope away from the prescribed dither pattern (see \cref{subsubsec:datareduction}).
For the four pointings described above and summarised in Table \ref{tab:pointings}, we adopted the standard object-sky-object (OSO) nod-to-sky observation pattern, with 300s exposures and alternating $0.2^{\prime\prime}$/$0.1^{\prime\prime}$ dither pattern for increased spatial sampling around each of the target galaxies.
This procedure allowed for datacube reconstruction with 0.1$^{\prime\prime}$ size spaxels as described in \cref{subsubsec:datareduction}. \\

\noindent
The observations were carried out during ESO observing periods P92-P96 using Guaranteed Time Observations (Programme IDs: 092.A-0399(A), 093.A-0122(A,B), 094.A-0214(A,B),095.A-0680(A,B),096.A-0315(A,B,C)) with excellent seeing conditions.
In GOODS-S-P1 and GOODS-S-P2 the median {\it K}-band seeing was $\simeq0.5^{\prime\prime}$ and for the SSA22-cluster and SSA22-field pointings the {\it K}-band seeing ranged between $\simeq0.55-0.65^{\prime\prime}$.
We observed 17-21 $z\simeq3.5$ targets in each field (see Table \ref{tab:pointings}), with these numbers less than the available 24 arms for each pointing due to the combination of three broken pickoff arms during the P92/93 observing semesters and our requirement to observe at least one control star throughout an Observing Block (OB). \\

\noindent
This paper is concerned with the spatially-resolved kinematics of the KDS galaxies.
Consequently, we now focus exclusively on the spatially-resolved [O~{\sc III}]$\lambda$5007 measurements in the {\it K}-band spectral window. 
The details of the {\it H}-band data reduction and corresponding metallicity analyses will be described in a future study.

\subsubsection{Data reduction}\label{subsubsec:datareduction}

\begin{figure*}
    \centering \hspace{-1.3cm}
    \begin{subfigure}[h!]{0.5\textwidth}
        \centering
        \includegraphics[height=3.5in]{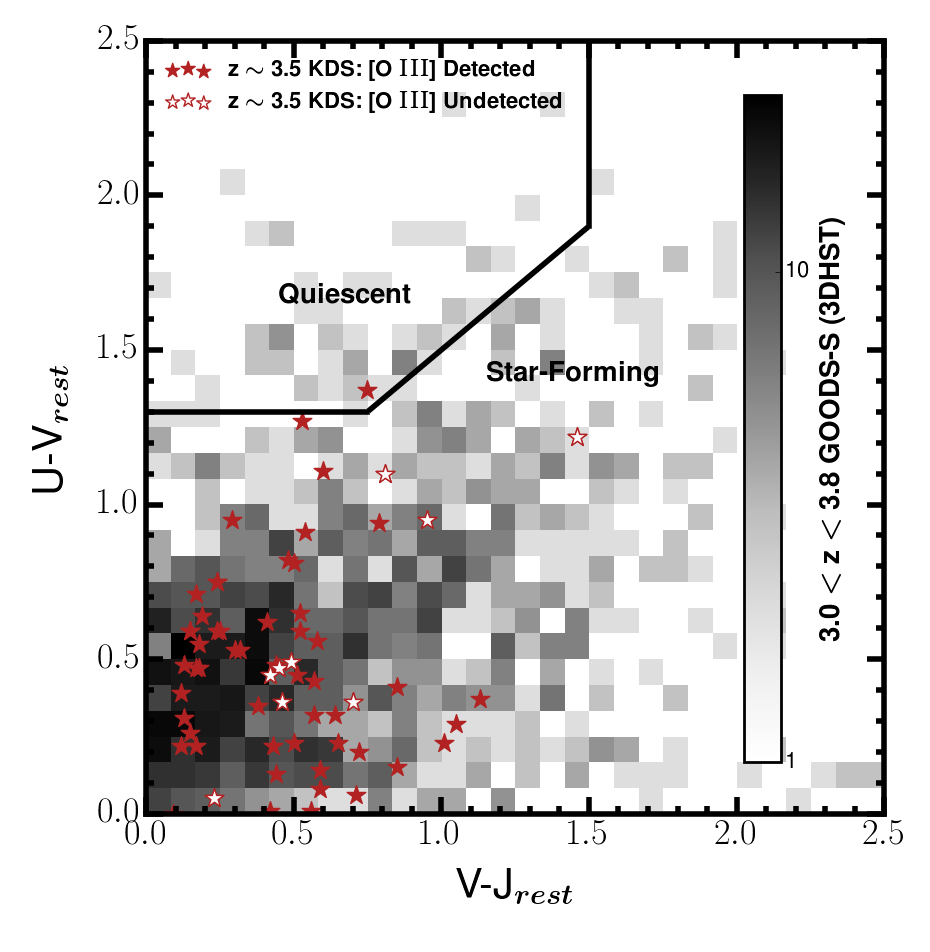}
    \end{subfigure} \hspace{+0.4cm}
    \begin{subfigure}[h!]{0.5\textwidth}
        \centering
        \includegraphics[height=3.5in]{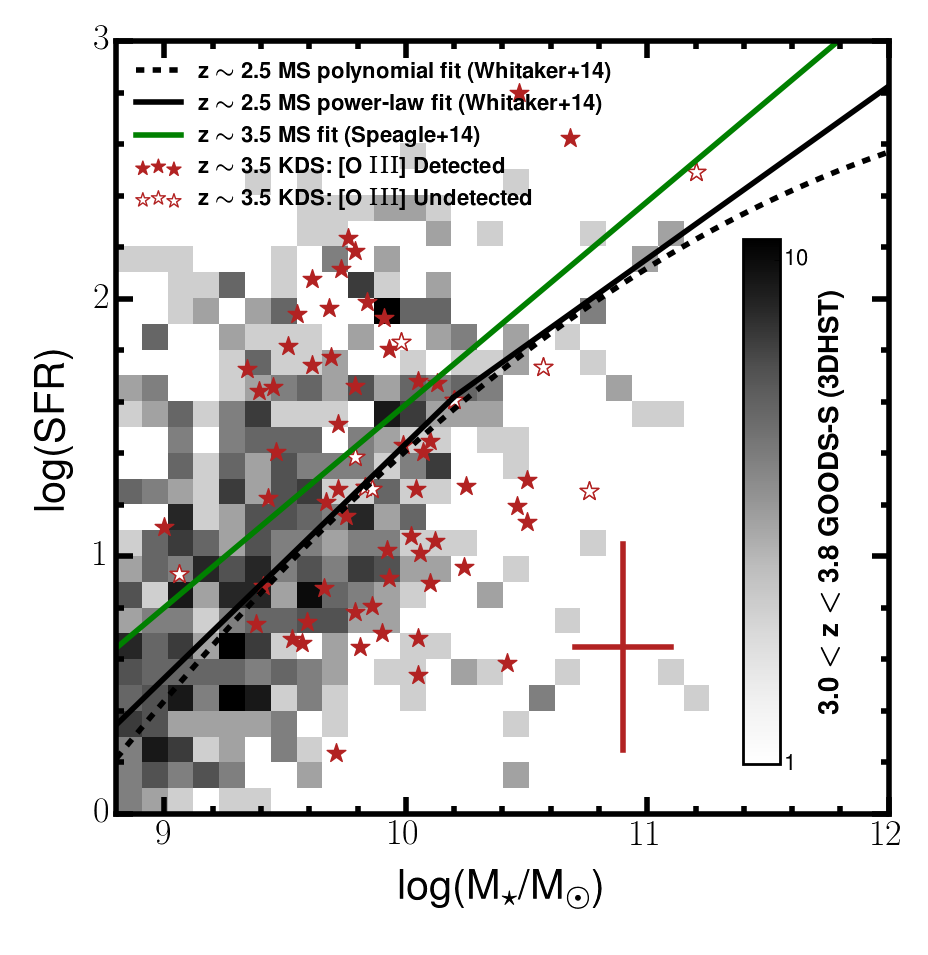}
    \end{subfigure}
    \caption{{\it Left:} The distribution of all 77 KDS galaxies in rest-frame UVJ colour space is plotted, with filled symbols showing galaxies detected in [O~{\sc III}]$\lambda$5007 and open symbols showing those which were not detected.
    Also plotted in this plane are $\simeq4000$ galaxies in GOODS-S with $3.0 < z < 3.8$ (mirroring the KDS redshift range) from the 3D-{\em HST} survey  \protect\citep{Brammer2012,Momcheva2016}, with the filled squares denoting the density of galaxies in that region.
    We use the galaxy selection criteria defined in \protect\cite{Whitaker2012a} to highlight star-forming and quiescent regions (motivated by the age sequence of quiescent galaxies), finding that all but one of the KDS galaxies are clearly in the star-forming region and overlap with the highest density of 3D-{\em HST} targets.
    {\it Right:} We plot the location of the KDS galaxies in the SFR versus M$_{\star}$ plane, using the same symbol convention. 
    The same GOODS-S galaxies from the 3D-{\em HST} survey as in the left panel are plotted with the filled squares, as a reference for the typical relationship between SFR and $M_{\star}$.
    The black solid line and the dashed line show the $z\simeq2.5$ broken power-law and quadratic fit to the main-sequence respectively, described in \protect\cite{Whitaker2014}.
    We include the MS relation evaluated at $z=3.5$ (the median redshift of the KDS sample) described in \protect\cite{Speagle2014} and given in Equation \protect\ref{eq:speagle_ms} as the green line, with the discrepancy between the two relations representating the expected main-sequence evolution between these redshifts. 
    Within the typical uncertainties (see error bars) the KDS sample is representative of $z\simeq3.5$ SFGs.}
    \label{fig:main_sequence}
\end{figure*}

The data reduction process primarly made use of the Software Package for Astronomical Reduction with KMOS, (SPARK; \citealt{Davies2013}), implemented using the ESO Recipe Execution Tool (ESOREX) \citep{Freudling2013}.
In addition to the SPARK recipes, custom Python scripts were run at different stages of the pipeline and are described throughout this section.

The SPARK recipes were used to create dark frames and to flatfield, illumination correct and wavelength calibrate the raw data.
An additional step, which is not part of the standard reduction process, was carried out at this stage, which is to address readout channel bias.
Differences in the readout process across each 64-pixel wide channel on the detector image lead to varying flux baselines in these channels.
We corrected back to a uniform flux baseline across the detector image for each object exposure by identifying pixels which are not illuminated in every readout channel and subtracting their median value from the rest of the pixels in the channel.

\noindent
Standard star observations were carried out on the same night as the science observations and were processed in an identical manner to the science data.
Following this pre-processing, each of the object exposures was reconstructed independently, using the closest sky exposure for subtraction, to give more control over the construction of the final stacks for each target galaxy.
Each 300s exposure was reconstructed into a datacube with interpolated $0.1\times0.1^{\prime\prime}$ spaxel size, facilitated by the subpixel dither pattern discussed in \cref{subsubsec:Obs} which boosts the effective pixel scale of the observations. \\

\noindent
Sky subtraction was enhanced using the SKYTWEAK option within SPARK \citep{Davies2007}, which counters the varying amplitude of OH lines between exposures by scaling `families' of OH lines independently to match the data.
Wavelength miscalibration between exposures due to spectral flexure of the instrument is also accounted for by applying spectral shifts to the OH families during the procedure, and in general the use of the SKYTWEAK option in the {\it K}-band greatly reduces the sky-line residuals. 
We monitored the evolution of the atmospheric PSF and the position of the control stars over the OBs, to allow us to reject raw frames where the averaged {\it K}-band seeing rose above $0.8^{\prime\prime}$ and to measure the spatial shifts required for the final stack more precisely.
The PSF was determined by fitting the collapsed {\it K}-band image of the stacked control stars in each pointing with an elliptical gaussian, with the values reported in Table \ref{tab:pointings}.
The telescope tends to drift from its acquired position over the course of an OB and the difference between the dither pattern shifts and the measured position of the control stars provides the value by which each exposure must be shifted to create the stack. \\

\noindent
We stacked all 300s exposures for each galaxy which pass the seeing criteria using 3-sigma clipping, leaving us with a flux and wavelength-calibrated datacube for every object in the KDS sample.
We have found that the thermal background is often under-subtracted across the spatial extent of the cube following a first pass through the pipeline, leading to excess flux towards the long wavelength end of the {\it K}-band.
To account for this, a polynomial function is fit, using the python package {\scriptsize LMFIT} \citep{Newville2014}, which makes use of the Levenberg-Marquardt algorithm for non-linear curve fitting, to the median stacked spectrum from spaxels in the datacube which contain no object flux and then subtracted from each spaxel in turn. \\

\noindent
The central coordinates of each pointing, the number of target galaxies observed, N$_{obs}$, the number of galaxies with [O~{\sc III}]$\lambda$5007 detected as measured by attempting to fit the redshifted line in the integrated galaxy spectrum using the known redshift value, N$_{Det} = 62/77$ $(81\%)$, the number with spatially-resolved [O~{\sc III}]$\lambda$5007 emission, N$_{Res}$ (see \cref{subsubsection:spaxel_fitting}), the on source exposure time and the averaged seeing conditions are listed in Table \ref{tab:pointings}.

\subsection{Stellar masses and SFRs}\label{subsec:stellar_masses_and_sfrs}

The wealth of ancillary data in both fields allows for a consistent treatment of the SED modelling, providing physical properties which are directly comparable between the cluster and field environments.
These derived properties are considered in the context of the galaxy main-sequence, to verify that the KDS sample contains typical SFGs at $z\simeq3.5$.

\subsubsection{SED fitting and main-sequence}\label{subsubsec:sed_fitting}
In order to constrain their SFRs and stellar masses, the available photometry for the KDS targets was analysed using the SED fitting software described in \cite{McLure2011} and \cite{McLeod2015}.
The photometry for each target was fit with the same set of solar metallicity BC03 \citep{Bruzual2003} templates adopted by the 3D-HST team \cite{Momcheva2016}, and derived stellar masses and SFRs were based on a Chabrier IMF.
In addition, the SED fitting software accounts for the presence of strong nebular emission lines according to the line ratios determined by \cite{Cullen2014}.
During the SED fitting process, dust attenuation was accounted for using the \cite{Calzetti2000} reddening law, with dust attenuation allowed to vary freely within the range $0.0<A_{V}<4.0$.
Based on the adopted template set, the median stellar mass for the full observed sample is $\log(M_{\star}/{\rm M}_{\odot}) = 9.8$. 
Fitting the photometry of the KDS targets with 0.2 $Z\odot$ templates, rather than solar metallicity templates, typically reduces the derived stellar masses by $\simeq0.1$ dex, but this change does not affect the conclusions of this work.
In GOODS-S, we have compared our derived stellar masses to those in \cite{Santini2015} (which presents an average result from 10 different sets of analyses), finding a median difference between the two sets of values of $\Delta \log(M_{\star}/{\rm M}_{\odot}) = 0.009$.
We also note that using the star-forming galaxy templates described in \cite{Wuyts2011} typically leads to stellar masses which are 0.2 dex higher. 

In the left panel of Figure \ref{fig:main_sequence} we plot the KDS galaxies in the rest-frame {\it U-V} vs. {\it V-J} colour space.
This is a commonly used diagnostic plane for selecting star-forming and quiescent galaxies \citep[e.g.][]{Williams2009,Brammer2011,Whitaker2012a} with the age gradients of the stars within quiescent galaxies placing them in a different region of the plane to those which are actively forming stars.
The selection criteria defined in \cite{Whitaker2012a} (which evolve only gently with redshift), separate quiescent and star-forming regions, which are indicated by the black wedge.
We also make use of the rest-frame colours of $\simeq4000$ primarily star-forming galaxies located in GOODS-S between $3.0 < z < 3.8$ (based upon the `z\_best classification flag') from the 3D-{\em HST} survey \citep{Brammer2012,Momcheva2016}.
The filled squares indicate the density of 3D-{\em HST} targets in colour space and we observe that the peak density location is consistent with the location of the KDS targets, all but one of which are in the star-forming region. 

In the right panel of Figure \ref{fig:main_sequence} we plot the M$_{\star}$ and SFR `main-sequence' for the KDS galaxies with SFR measurements, in combination with the derived physical properties of the same GOODS-S galaxies as in the left panel.
We also plot both the linear-break and quadratic $z\simeq2.5$ main-sequence fits to the 3D-{\em HST} data described in \cite{Whitaker2014} with the solid and dashed black lines as well as the main-sequence relation described in \cite{Speagle2014} given in Equation \ref{eq:speagle_ms}, evaluated at $z\simeq3.5$ (where the age of the universe is 1.77 Gyr) with the green line.

\begin{equation}\label{eq:speagle_ms}
logSFR(M_{\star}, t) = (0.84 - 0.026 \times t)log(M_{\star}/{\rm M}_{\odot}) - (6.51 - 0.11 \times t)
\end{equation}

\noindent
The difference in position of these relations highlights the main-sequence evolution towards higher SFRs at fixed $M_{\star}$ between $z\simeq2.5-3.5$.
The KDS galaxies scatter, within the errors, consistently above and below the $z\simeq3$ main-sequence.

When taken together, both panels indicate that the KDS sample is representatitve of typical star-forming galaxies at $z>3$.

\subsubsection{Sample summary}\label{subsubsec:sample_summary}
We have observed 77 SFGs spanning $3.0 < z < 3.8$ with the KMOS IFUs as part of the KMOS Deep Survey.
By processing the observations, we have constructed a datacube for each of these galaxies, finding integrated [O~{\sc III}]$\lambda$5007 in 62/77 targets.
We have also derived $M_{\star}$ values and SFRs from SED modelling, finding that these values place the KDS galaxies into regions of parameter space spanned by typical star-forming galaxies at this redshift.
Three KMOS pointings cover the field environment and one pointing covers a cluster environment at the heart of the SSA22 protocluster, and with these data we seek gauge the role of environment in shaping the early evolutionary history of SFGs.
Throughout the following analysis sections we describe the use of high-resolution imaging to make morphological measurements for the KDS galaxies and the extraction of kinematic properties for the galaxies using the [O~{\sc III}]$\lambda$5007 emission line.
Chemical abundance measurements for the KDS galaxies in both the field and cluster environments will be presented in a future work.

\section{ANALYSIS}\label{sec:analysis}

\subsection{Morphological Measurements}\label{subsec:morphological_measurements}
For a robust interpretation of the observed velocity fields, it was necessary to separately determine the morphological properties of the galaxies from high-resolution images.
This imaging was used primarily to determine morphological parameters which characterise the size (quantified here through the half-light radius, $R_{1/2}$), morphological position angle, $PA_{morph}$, and axis ratio, $b/a$, of the galaxies. 
In the following sections we describe the approach chosen to recover these parameters, also describing comparisons with matched galaxies in the morphological parameter catalogue of \cite{VanderWel2012}.
At $3 < z < 4$ and $0.1 < z < 1$ we made use of secure spectroscopic redshifts obtained for SFGs during the ESO public surveys zCOSMOS \citep{Lilly2007}, VUDS \citep{Tasca2016}, GOODS\_FORS2 \citep{Vanzella2005,Vanzella2006,Vanzella2008} and GOODS\_VIMOS \citep{Balestra2010} to cross-match with \cite{VanderWel2012}.
This allowed us to investigate the morphological properties of typical star-forming galaxy populations at two redshift slices in comparison with those determined for the KDS sample.
The imaging also helped to distinguish multiple `merging' components with small angular separations from objects which are morphologically isolated, which we discuss in \cref{subsubsection:spaxel_fitting} and \cref{subsec:morpho-kin-class} where we refine our sample for dynamical analysis.

\subsubsection{Applying {\scriptsize GALFIT} to the imaging data}\label{subsubsec:galfitting}
We used {\scriptsize GALFIT} \citep{Peng2010_galfit} to fit 2D analytic functions, convolved with the PSF, to the observed {\em HST} images of the KDS field galaxies across GOODS-S and SSA22 in a consistent way.
The GOODS-S imaging data used is the latest release of the total field in {\it WFC3 F160W} band, which traces the rest-frame near-UV at $z\simeq3.5$, available via the CANDELS \citep{Grogin2011,Koekemoer2011} data access portal\footnote{\url{http://candels.ucolick.org/data_access/Latest_Release.html}}.
For SSA22 we made use of archival {\em HST} imaging\footnote{\url{https://archive.stsci.edu/hst/search.php}} data in the {\it WFC3 F160W} band (P.I. Lehmer: PID 13844; P.I. Mannucci: PID 11735) and the {\it ACS F814W} band, tracing $\simeq$ 2500$\AA$ light at $z\simeq3.1$ (P.I. Chapman: PID 10405; P.I. Abraham: PID 9760; P.I. Siana: PID 12527).
The {\em HST} coverage is shallower in SSA22 (exposure times of $\simeq5$ ks) and the {\it F160W} coverage is concentrated on the SSA22-cluster and so we resorted to the bluer {\it ACS F814W} data to derive morphological parameters in SSA22-P2 (also in SSA22-P2, 4 galaxies do not have any {\em HST} coverage as indicated in Table \ref{tab:phys-props}). \\

\noindent
We first ran SExtractor \citep{Bertin1996} on the relevant images to recover initial input parameters and segmentation maps for running {\scriptsize GALFIT}, and then extracted postage stamp regions around the galaxies in the KDS sample.
At this redshift, the galaxies are more compact and generally we could not resolve more complicated morphological features such as spiral arms and bars and so we followed the simple method of fitting \Sers profiles, with the \Sers index fixed to the exponential disk value of $n=1$.
In the {\it F160W} band, the adopted PSF was a hybrid between the Tiny Tim {\it H}-band model \citep{Krist2011} in the PSF centre and an empirical stack of stars observed in the {\it H}-band for the wings \citep{VanderWel2012} and in the {\it F814W} band we used the pure Tiny Tim {\it ACS} high-resolution PSF model.
During the fitting process all other morphological parameters, including $R_{1/2}$, the central x and y coordinates, $PA_{morph}$ and inclination, were free to vary.

This method was justified by the recovery of similar mean $\chi^{2}$ values when fitting floating \Sers index models and bulge/disk models with both an $n=1$ and $n=4$ component (following the procedure described in \citealt{Bruce2012}) to those recovered from the fixed exponential disk fit.
Additionally, 22 GOODS-S objects were also detected in the \cite{VanderWel2012} catalogue, for which the median \Sers index value is $n=1.2$.
The use of exponential profiles also facilitates comparison with recent large surveys such as KMOS$^{3D}$ \citep{Wisnioski2015} and KROSS \citep{Harrison2017} in which beam-smearing correction factors were applied to the derived kinematic parameters as a function of exponential disk scale length, $R_{D}$, defined as $R_{1/2}\simeq1.68 R_{D}$ (see Appendix \ref{subsec:KROSS}). \\

\noindent
This analysis provided us with three crucial morphological parameters required to support the kinematic analysis of \cref{subsec:3d_modelling}, namely the axis ratio $b/a$, half-light radius $R_{1/2}$ and the morphological position angle $PA_{morph}$.
Example {\em HST} images and {\scriptsize GALFIT} outputs are given in Figure \ref{fig:main_body_kinematic_grids} for selected galaxies, and maps for the full isolated field sample (see \cref{subsubsection:spaxel_fitting}) are plotted throughout Appendix \ref{app:kinematics_plots}. 

The errorbars produced by {\scriptsize GALFIT} are purely statistical and are determined from the covariance matrix used in the fitting, resulting in unrealistically small uncertainties on the derived galaxy parameters \citep{Hausler2007,Bruce2012}.
Throughout the following subsections we discuss the more reasonable adopted errors on each of the relevant morphological parameters in turn.
In each of the following sections we also discuss the approach followed to recover the morphological parameters for the galaxies without {\em HST} coverage.
The interpretation of the sample as a whole is not affected by the small number of galaxies without {\em HST} imaging.

\begin{table*}
\centering
\begin{threeparttable}
\caption{Physical properties of the resolved and morphologically isolated KDS field galaxies as measured from SED fitting and from applying {\scriptsize GALFIT} \protect\citep{Peng2010_galfit}.}
\label{tab:phys-props}
\begin{tabular}{llllllllll}

 \hline
ID              & RA       & Dec       & z     & K$_{AB}$     & $log(M_{\star}/M_{\odot})^{a}$ & $b/a$ & i$^{\circ}$$^{b}$ & PA$_{morph}^{\circ}$ & R$_{1/2}$(kpc)$^{c}$ \\
 \hline
b012141\_012208 & 03:32:23.290 & $-$27:51:57.348 & 3.471        & 24.12  & 9.8 & 0.36        & 72        & 9     & 1.57      \\
b15573          & 03:32:27.638 & $-$27:50:59.676 & 3.583        & 23.6   & 9.8  & 0.28        & 78        & 146   & 0.52      \\
bs006516        & 03:32:14.791 & $-$27:50:46.500 & 3.215        & 23.94  & 9.8 & 0.50         & 61        & 146   & 1.91      \\
bs006541        & 03:32:14.820 & $-$27:52:04.620 & 3.475       & 23.44  & 10.1  & 0.44        & 66        & 168   & 1.83      \\
bs008543        & 03:32:17.890 & $-$27:50:50.136 & 3.474        & 22.73  & 10.5 & 0.50         & 61        & 67    & 1.59      \\
bs009818        & 03:32:19.810 & $-$27:53:00.852 & 3.706        & 24.18  & 9.7  & 0.80         & 37        & 148   & 1.24      \\
bs014828        & 03:32:26.760 & $-$27:52:25.896 & 3.562        & 23.58  & 9.7  & 0.31        & 76        & 63    & 1.61      \\
bs016759        & 03:32:29.141 & $-$27:48:52.596 & 3.602       & 23.85  & 9.9   & 0.65        & 50        & 49    & 0.87      \\
lbg\_20         & 03:32:41.244 & $-$27:52:20.676 & 3.225        & 24.97  & 9.5  & 0.64        & 52        & 1     & 1.28      \\
lbg\_24         & 03:32:39.754 & $-$27:39:56.628 & 3.279       & 24.67  & 9.6   & 0.53        & 60        & 34    & 1.27      \\
lbg\_25         & 03:32:29.189 & $-$27:40:22.476 & 3.322        & 24.95  & 9.4  & 0.30         & 76        & 78    & 1.18      \\
lbg\_30         & 03:32:42.854 & $-$27:42:06.300 & 3.419        & 23.85  & 10.0  & 0.79         & 38        & 66    & 0.95      \\
lbg\_32         & 03:32:34.399 & $-$27:41:24.324 & 3.417      & 23.84  & 9.9    & 0.60         & 54        & 40    & 1.88      \\
lbg\_38         & 03:32:22.474 & $-$27:44:38.436 & 3.488       & 24.58  & 9.9   & 0.58        & 56        & 137   & 0.92      \\
lbg\_91         & 03:32:27.202 & $-$27:41:51.756 & 3.170       & 24.65  & 9.8   & 0.58        & 56        & 79    & 0.89 \\
lbg\_94         & 03:32:28.949 & $-$27:44:11.688 & 3.367       & 24.54  & 9.9   & 0.22        & 84        & 81    & 1.16      \\
lbg\_105        & 03:32:24.005 & $-$27:52:16.140 & 3.092      & 23.79  & 9.3    & 0.56        & 57        & 128   & 1.72      \\
lbg\_109        & 03:32:20.935 & $-$27:43:46.344 & 3.600       & 24.63  & 9.7   & 0.60         & 54        & 119   & 1.98      \\
lbg\_111        & 03:32:42.497 & $-$27:45:51.696 & 3.609       & 24.01  & 9.7   & 0.74        & 42        & 80    & 0.64      \\
lbg\_112         & 03:32:17.134 & $-$27:42:17.784 & 3.617        & 25.16  & 9.6 & 0.79         & 38        & 43    & 0.46      \\
lbg\_113        & 03:32:35.957 & $-$27:41:49.956 & 3.622      & 24.01  & 9.6    & 0.52        & 60        & 15    & 0.87      \\
lbg\_124         & 03:32:33.324 & $-$27:50:07.332 & 3.794        & 24.96  & 9.0  & 0.56         & 57        & 50    & 0.78      \\
$^{*}$n3\_006       & 22:17:24.859 & +00:11:17.620 & 3.069     & 22.98 & 10.5   & 0.57         & 57        & 156    & 2.52      \\
n3\_009       & 22:17:28.330 & $+$00:12:11.600 & 3.069 & -    & 8.7                       & 0.75        & 42        & 84    & 1.06      \\
n\_c3         & 22:17:32.585 & $+$00:10:57.180 & 3.096       & 24.96 & 9.8                & 0.71        & 46        & 94    & 0.56      \\
lab18         & 22:17:28.850 & $+$00:07:51.800 & 3.101           & -    & 8.2             & 0.85        & 32        & 27    & 0.46      \\
$^{*}$lab25         & 22:17:22.603 & $+$00:15:51.330 & 3.067      & -    & 8.4            & 0.57         & 57        & 87    & 2.18      \\
s\_sa22a-d3   & 22:17:32.453 & $+$00:11:32.920 & 3.069 & 23.46 & 9.7                       & 0.39        & 70        & 125   & 1.78      \\
s\_sa22b-c20  & 22:17:48.845 & $+$00:10:13.840 & 3.196     & 23.91 & 9.5                   & 0.57        & 57        & 76    & 1.59      \\
$^{*}$s\_sa22b-d5  & 22:17:35.808 & $+$00:06:10.340 & 3.175 & 23.72 & 10.2                & 0.57         & 57        & 60    & 3.30      \\
s\_sa22b-d9   & 22:17:22.303 & $+$00:08:04.130 & 3.084    & 24.25 & 10.1                  & 0.65        & 50        & 60   & 0.50       \\
$^{*}$s\_sa22b-md25 & 22:17:41.690 & $+$00:06:20.460 & 3.304     & 24.62 & 8.6             & 0.57         & 57        & 9    & 1.39      

\end{tabular}
\begin{tablenotes}
      \small
      \item $^{a}$ Representative error of 0.2 dex from SED modelling used throughout this study
      \item $^{b}$ Fixed 10 percent inclination error assumed (i.e. $\delta i = i / 10$, see \protect\cref{subsubsection:inclination_angle})
      \item $^{c}$ Fixed 10 percent $R_{1/2}$ error assumed, see \protect\cref{subsubsection:half-light_radii}
      \item $^{*}$ No {\em HST} coverage: $PA_{morph}$ set to $PA_{kin}$ value; $b/a$ and $R_{1/2}$ estimated as explained throughout the text.
    \end{tablenotes}
  \end{threeparttable}
  \end{table*}

\subsubsection{Inclination angles}\label{subsubsection:inclination_angle}
We used the derived $b/a$ values to determine galaxy inclination angles.
As suggested in \cite{Holmberg1958}, by modelling the disk galaxies as an oblate spheroid, the inclination angle can be recovered from the observed axis ratio as shown in Equation \ref{eq:holmberg_i}, where $\frac{b}{a}$ is the ratio of minor to major axis of an ellipse fit to the galaxy profile on the sky, $i$ is the inclination angle and $q_{0}$ is the axis ratio of an edge-on system.

\begin{equation}\label{eq:holmberg_i}
   cos^{2}i = \frac{\left(\frac{b}{a}\right)^{2} - q_{0}^{2}}{1 - q_{0}^{2}}
\end{equation}

\noindent
To derive the inclination, we selected a value for $q_{0}$, and following the discussion in \citep{Law2012a} we chose a value appropriate for thick disks, $q_{0} = 0.2$ \citep[e.g.][]{Epinat2012,Harrison2017}.
However, as discussed in \cite{Harrison2017}, varying the $q_{0}$ value by a factor of 2 makes only a small change to the final inclination corrected velocity values, with the difference $<10\%$ in the case of the KDS galaxies in the isolated field sample (see \cref{subsec:morpho-kin-class}).

\begin{figure}
\centering \hspace{-1.13cm}
\includegraphics[width=0.49\textwidth]{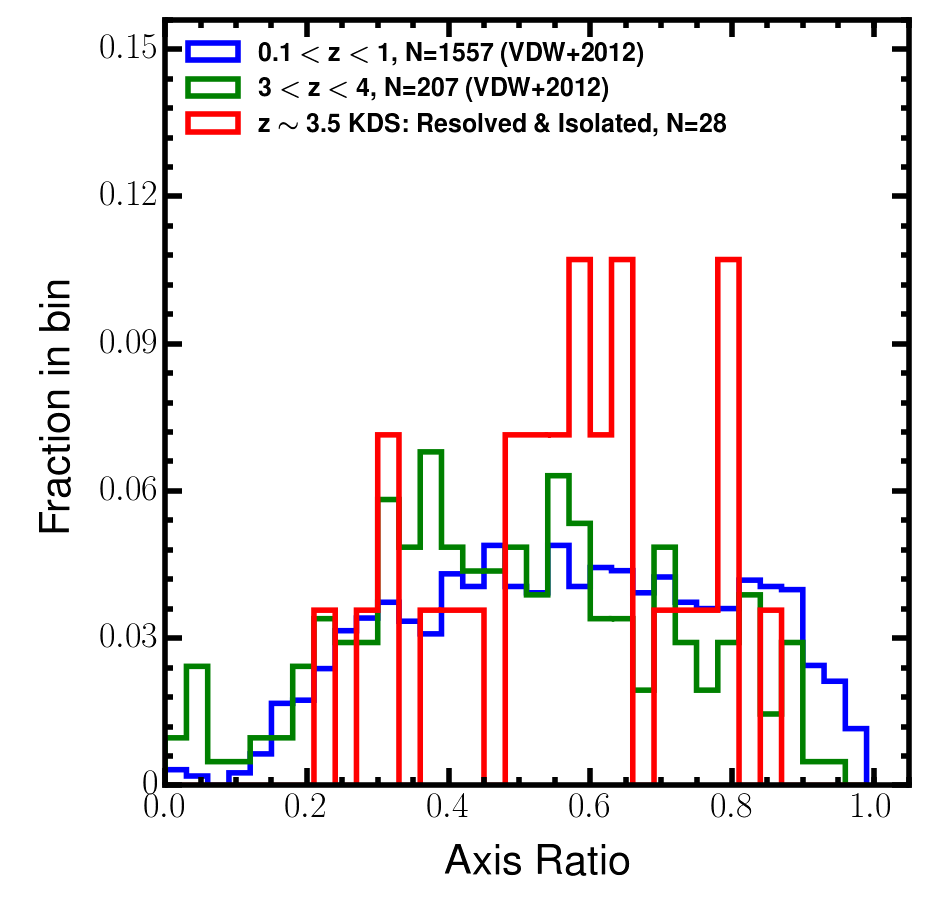}
\caption{The normalised counts of KDS galaxies in bins of axis ratio is plotted, along with the normalised counts from the morphological catalogue presented in \protect\cite{VanderWel2012} in two redshift ranges.
The axis ratio distribution appears to be constant with time as traced by the reference samples, and the KDS values are in good agreement with a relatively uniform distribution spanning $0.3 < b/a < 0.9$.
This suggests that we are not biased towards deriving a particular value for the axis ratio.}
\label{fig:morpho-distributions}
\end{figure}

The inclination angle calculated for each galaxy is used to correct the observed velocity field, which is the line of sight component of the intrinsic velocity field, with the correction factor increasing with increasing $b/a$.
The median difference between the derived $b/a$ values and those presented in \cite{VanderWel2012} for the 22 matched GOODS-S galaxies is $\Delta b/a = 0.0002$.
The $b/a$ distribution shown in Figure \ref{fig:morpho-distributions} (plotted for the 28 galaxies in the isolated field sample with {\em HST} imaging, \cref{subsec:morpho-kin-class}) is consistent with being uniform between $0.3 < b/a < 0.9$, with a median value of 0.58, corresponding to i = 57.0$^{\circ}$ using the conversion given in Equation \ref{eq:holmberg_i}, with $q_{0}=0.2$.
This compares well with the theoretical mean value of 57.3$^{\circ}$ computed for a population of galaxies with randomly drawn viewing angles (see e.g. the appendix in \citealt{Law2009}), which reassured us that we were not biased towards deriving either particularly low or particularly high inclination angles.

As mentioned in \cref{subsubsec:galfitting}, the error reported by {\scriptsize GALFIT} on the model $b/a$ value is unrealistically small.
Guided by the results of the Monte Carlo approach described in \cite{Epinat2012} we adopted a conservative nominal inclination angle uncertainty of 10$\%$, i.e. $\delta i = i / 10$.
For the galaxies without {\em HST} coverage, the inclination angle was set to the KDS sample median of 57.0$^{\circ}$ and we used the standard deviation of the KDS inclinations, 13.2$^{\circ}$, for the inclination uncertainty.
This added an additional factor of $\simeq1.3$ to the uncertainty in the derived velocities for these galaxies.

\subsubsection{Position angles}\label{subsubsection:position_angle}
The second {\scriptsize GALFIT} parameter was $PA_{morph}$, which is the direction of the photometric major axis of the galaxy on the sky.
We visually inspected each of the {\scriptsize GALFIT} maps to check that the derived $PA_{morph}$ follows the galaxy light distribution (with $PA_{morph}$ indicated by the orange dashed line in Figure \ref{fig:main_body_kinematic_grids}) and when this was not the case it was usually an indication of multiple distinct components or tidal streams (see \cref{subsec:morpho-kin-class} where we remove multiple-component objects from the final sample).
We note that discrepancy between $PA_{morph}$ and the kinematic position angle, $PA_{kin}$, is an indicator of sub-structure in the morphology \citep[e.g.][]{Queyrel2012,Wisnioski2015,Rodrigues2017}, and deviations can indicate clumps or mergers which may influence the underlying kinematics or bias the derived $PA_{morph}$ towards a particular direction.
We discuss this topic further in \cref{subsubsection:kin_and_phot}.
For the galaxies without {\em HST} coverage we fixed $PA_{morph}$ equal to $PA_{kin}$, for the analysis described in \cref{subsec:3d_modelling}.

\subsubsection{Half-light radii}\label{subsubsection:half-light_radii}
The half-light radius, $R_{1/2}$, provides an indication of the disk sizes, and hence gave us a common fiducial distance from the centre of the galaxy at which to extract rotation velocities.
As discussed throughout the introduction and in Appendix \ref{app:comparison_samples}, methods used to derive the intrinsic rotation velocity, $V_{C}$, vary widely between studies, and so to compare with previous work it is necessary to extract kinematic parameters consistently from the derived rotation curves.
Knowledge of $R_{1/2}$ gives a well defined and consistent point across the extent of the galaxy at which to extract velocities, which is important when making comparisons to previous IFU studies (see Appendix \ref{app:comparison_samples} and e.g. \citealt{ForsterSchreiber2009,Epinat2012,Wisnioski2015,Stott2016,Harrison2017,Swinbank2017}).
Throughout our dynamical modelling, and to derive a beam-smearing correction to the rotation velocities and velocity dispersions (see \cref{subsec:3d_modelling}), we also required the $R_{1/2}$ value for each galaxy.
The median difference between the derived $R_{1/2}$ values and those presented in \cite{VanderWel2012} for the 22 matched GOODS-S galaxies is $\Delta R_{1/2} = 0.0054^{\prime\prime}$ (corresponding to 0.04kpc at $z\simeq3.5$). \\

\noindent
\cite{Bruce2012} present a detailed error analysis of the morphological parameters of $1 < z < 3$ galaxies in the CANDELS fields (which includes {\em HST} data covering the GOODS-S field which we exploited during this work), reporting that the magnitude of the error on $R_{1/2}$ is typically at the 10$\%$ level.
We adopted a nominal error of 10$\%$ for our recovered $R_{1/2}$ values and note that the errors on $V_{C}$ and $\sigma_{int}$ are dominated by measurement errors and uncertainties connected with assumptions about the inclination angle correction and the velocity dispersion distribution in high-redshift galaxies (see Appendix \ref{app:kin_error_estimates}).

\noindent
To measure $R_{1/2}$ for the 4 galaxies without {\em HST} coverage, we fit the [O~{\sc III}]$\lambda$5007 narrow-band images with an exponential profile, and subtracted the appropriate PSF from the recovered $R_{1/2}$ value.
We note that the extracted rotation velocities are not sensitive to the precise $R_{1/2}$ values, since we measure these at $2R_{1/2}$ (see \cref{subsubsec:beam_smearing_corrected_velocities}) where the intrinsic model rotation curves have already reached flattening.

\subsection{Kinematic Measurements}\label{subsection:kinematic_measurements}

We proceeded to extract 2D kinematic measurements from the 62 KDS datacubes detected in integrated [O~{\sc III}]$\lambda$5007 emission by fitting the [O~{\sc III}]$\lambda$5007 emission line profiles within individual spaxels.
The sample was then refined to galaxies which are spatially-resolved in the [O~{\sc III}]$\lambda$5007 emission line in order to make dynamical measurements.
We also describe our dynamical modelling and the approach we have followed to extract $V_{C}$ and $\sigma_{int}$, using the models to derive beam-smearing corrections for each galaxy.
We made use of the imaging data described in the previous sections to reduce the number of free parameters in the model, an approach which is necessary when working in this signal-to-noise regime.  

\subsubsection{Spaxel Fitting}\label{subsubsection:spaxel_fitting}
We aimed to extract two-dimensional maps of the [O~{\sc III}]$\lambda$5007 flux, velocity and velocity dispersion from each of the 62 stacked and calibrated KDS datacubes with an [O~{\sc III}]$\lambda$5007 detection.
These properties are extracted via modelling emission line profiles within each spaxel, with a set of acceptance criteria imposed to determine whether the fit quality is high enough to allow the inferred properties to pass into the final 2D maps of the flux and kinematics.
We concentrated solely on the [O~{\sc III}]$\lambda$5007 emission line which always had signal-to-noise higher than both [O~{\sc III}]$\lambda$4959 and $H\,\beta$ and was detected over a larger area of each IFU. \\

\noindent
Each 0.1$^{\prime\prime}$ spaxel across a datacube was considered in turn.
The redshift value determined from fitting the [O~{\sc III}]$\lambda$5007 line in the integrated galaxy spectrum, $z_{sys}$, was used to determine the central wavelength of the redshifted [O~{\sc III}]$\lambda$5007 line, $\lambda_{obs_{5007}}$. 
We searched a narrow region around $\lambda_{obs_{5007}}$ for the peak flux, assumed to correspond to the ionised gas emission, and extracted a 5$\AA$ region (corresponding to 20 spectral elements) centred on this peak.
This truncated spectrum was then used in the fitting procedure for each spaxel.
The width of the extraction region was large enough to encompass unphysically large velocity shifts, but not so large as to compromise the fitting by potentially encompassing regions of the spectrum plagued by strong sky-line residuals.
A single gaussian model was fit to the extraction region, again using {\scriptsize LMFIT} \citep{Newville2014}, returning the values of the best-fitting flux, $F_{g}$, central wavelength, $\lambda_{g}$, and dispersion, $\sigma_{g}$.
These parameters, along with the estimate of the noise in the datacube, were used to assess whether the fitting of the [O~{\sc III}]$\lambda$5007 line is acceptable or not. 

The noise was estimated by masking all datacube spaxels containing object flux and computing the standard deviation of the flux values in the spectral extraction region for all remaining spaxels inwards of 0.3$^{\prime\prime}$ from the cube boundary (over the same spectral region as considered to find the peak object flux).
This boundary constraint was chosen to mitigate edge effects, where the noise increased sharply due to fewer individual exposures constituting the final stack in these regions.
The final noise value was then taken as the standard deviation of the results from each unmasked spaxel, as shown in Equation \ref{eq:noise}, where $F_{i,background}$ are the flux values across the spatial regions containing no object flux in the spectral extraction region, and $N_{m}$ denotes the noise estimated from masking the datacube.

\begin{equation}\label{eq:noise}
    N_{m} = STD\left(\sum_{i}F_{i,background}^{2}\right)
\end{equation}

\noindent
The signal in each spaxel was taken as the sum of the flux values across the spectral extraction region, $F_{i,object}$, as shown in Equation \ref{eq:signal}, where S denotes the signal.

\begin{equation}\label{eq:signal}
    S = \sum_{i}F_{i,object}
\end{equation}

\noindent
The criteria for accepting the fit within a given spaxel are as follows:

\begin{enumerate}[label=(\roman*),align=left]
\item The uncertainties on any of the parameters of the gaussian-fit as returned by {\scriptsize LMFIT} must not exceed 30\%
\item $\frac{S}{N_{m}} > 3$
\end{enumerate}

\noindent
We imposed a further criterion to help remove fits to spectral features unrelated to the galaxy (e.g. fits to skyline residuals), which was important for `cleaning up' accepted spaxels in the 2D maps which were clearly unrelated to the galaxy. For this test we subtracted the galaxy continuum and imposed that $0.7 < F_{g}/{S} < 1.3$, where $F_{g}$ is the flux from the gaussian-fit to the [O~{\sc III}]$\lambda$5007 line and S is the signal in the extraction region from Equation \ref{eq:signal}. If all three criteria were satisfied, the spaxel was accepted. \\

\noindent
From the gaussian-fit, we recovered the centre, $\lambda_{g}$, the width, $\sigma_{g}$, and the area, $F_{g}$.
The velocity was computed from $\lambda_{g}$ using Equation \ref{eq:velocity_comp} and the velocity dispersion was computed from $\sigma_{g}$ using Equation \ref{eq:dispersion_comp}, where $V_{obs}$ and $\sigma_{obs}$ are the observed rotation velocity and velocity dispersion respectively, c is the speed of light, $z_{sys}$ is the redshift determined from the integrated galaxy spectrum, $\lambda_{5007} = 0.500824\mu m$ is the rest wavelength of [O~{\sc III}]$\lambda$5007 in vacuum and $\sigma_{inst}$ is a measure of the KMOS instrumental resolution value averaged across gaussian-fits to several skylines (equal to 31.1 km\,s$^{-1}$ in the {\it K}-band and 55.4 km\,s$^{-1}$ in the {\it HK}-band).

\begin{equation}\label{eq:velocity_comp}
   v_{obs} = \frac{\lambda_{g} - (1 + z_{sys})\lambda_{5007}}{(1 + z_{sys})\lambda_{5007}} \times c
\end{equation}

\begin{equation}\label{eq:dispersion_comp}
   \sigma_{obs} = \sqrt{\left(\frac{\sigma_{g}}{(1 + z_{sys})\lambda_{5007}} \times c\right)^{2} - \sigma_{inst}^{2}}
\end{equation} \\

\noindent
To estimate the errors on the observed quantities, the gaussian-fit was repeated 1000 times, with each flux value in the extraction region perturbed by the noise value, $N_{m}$.
For each of these 1000 fits the gaussian parameters were used to compute the observed properties.
The resultant distributions of observed quantities were gaussian and we measured the median as the final value for $V_{obs}$, $\sigma_{obs}$ and $F_{g}$ within that spaxel, and the standard deviation of the distribution as the observational uncertainty.
A map of each of these observed parameters was constructed by applying this procedure to each spaxel. \\

\noindent
If a spaxel did not meet the criteria, we follow the procedure described in \cite{Stott2016} and binned with neighbouring spaxels to create a box of area $0.3\times0.3^{\prime\prime}$.
This was carried out by median stacking the extraction region spectra in each of the neighbouring spaxels.
The criteria were then re-examined, and if satisfied, the Monte-Carlo procedure was carried out. 
If the binned spaxels fail the criteria, one final iteration was carried out to bin to a box of area $0.5\times0.5^{\prime\prime}$, roughly equivalent to the size of the seeing disk (see Table \ref{tab:pointings}).
If after the widest binning the criteria weren't satisfied, no value was assigned to that spaxel in the extracted maps. \\

\noindent
The process was automated and applied to each of the 62/77 KDS galaxies with an [O~{\sc III}]$\lambda$5007 detection.
Following the fitting procedure, the galaxies were clasified as spatially-resolved in [O~{\sc III}]$\lambda$5007 emission or not if the spatial extent of the accepted spaxels in the [O~{\sc III}]$\lambda$5007 flux map was equivalent to or greater than the seeing disk.
For each pointing, the number of resolved galaxies, N$_{Res}$, was added to Table \ref{tab:pointings}.
We find that 47/77 (76\%) of the detected KDS galaxies across both the field (38/47) and cluster (9/47) environments were spatially-resolved and showed clear velocity structure.
The 15/77 galaxies which were not spatially-resolved were dropped at this stage due to the uncertainties associated with deriving kinematic properties and classifications from a single resolution element.
The physical properties of all spatially-resolved galaxies in both GOODS-S and SSA22 are listed in Table \ref{tab:phys-props}. \\ 

\noindent
The morphological information of \cref{subsec:morphological_measurements} was combined with this spatially-resolved condition to define two galaxy subsamples.
6/38 (16\%) of the resolved galaxies across the 3 field environment pointings and almost all (8/9) of the cluster environment galaxies in SSA22-P1 have multiple photometric components, which complicate the interpretation of the observed dynamics for these targets.
We dropped the cluster environment pointing from the full dynamical analysis, although we show examples of the {\em HST} imaging, flux, dispersion and velocity maps of the cluster target galaxies in Figure \ref{fig:merger_galaxies}.
The remaining 6/38 merging candidates in the field environment are referred to as the `merger field sample' throughout the remainder of the paper.
The 32/38 morphologically isolated field galaxies which are spatially resolved are referred to as the `isolated field sample' throughout the remainder of the paper.

\subsubsection{3D modelling and beam-smearing corrections}\label{subsec:3d_modelling}

As mentioned in the introduction, there is no universal standard for defining $V_{C}$ and $\sigma_{int}$, particularly when a dynamical model is used to provide beam-smearing corrections for $\sigma_{int}$ and to extrapolate beyond the data in order to measure $V_{C}$.
Physically motivated models take into account the potential-well of the galaxy through knowledge of the mass distribution \citep[e.g.][]{Genzel2008,ForsterSchreiber2009,Gnerucci2011,Wisnioski2015,Swinbank2017}.
Phenomenological models assume a fixed function known to reproduce observed galactic rotation curves which flatten at large radii \citep[e.g.][]{Epinat2010,Epinat2012,Swinbank2012,Stott2016,Harrison2017}.
Both of the above examples fit to 1D or 2D fields extracted from the datacubes using line-fitting software.
Recently, algorithms such as Galpak$^{3D}$ \citep{Bouche2015} and 3D-Barolo \citep{DiTeodoro2015} have emerged, fitting directly to the flux in the datacube in order to constrain the kinematic properties.

There are three reasons to construct model velocity fields for the isolated field sample galaxies; the first is to interpolate between data points; the second is to extrapolate the observations smoothly to some fiducial radius from which $V_{C}$ can be extracted; the third is to estimate the effect of beam-smearing on both the velocity and velocity dispersion fields so that we recover the set of intrinsic galaxy parameters which best describe the observations.
As a brief overview, we used a Markov chain Monte Carlo (MCMC) procedure to construct a series of intrinsic model datacubes for each galaxy, convolved these with the atmospheric PSF and then fit to the data.
The beam-smearing corrected kinematic parameters were then extracted from the intrinsic galaxy models as described in more detail below.
In this analysis, we minimised the number of free parameters by making use of information from {\em HST} imaging and by assuming a fixed function to describe the rotation curves as described in the following subsection.    
This is particularly important when dealing with galaxies at $z > 3$, since often there is not enough signal-to-noise away from the centre of the galaxy to observe the flattening of the rotation curve, consequently requiring extrapolation to determine $V_{C}$.  
The following section describes our 3D modelling procedure, and the validation of our results through comparison with the techniques used to determine the dynamical properties presented in \cite{Harrison2017}.

\begin{figure*}
    \centering

    \includegraphics[width=0.95\textwidth]{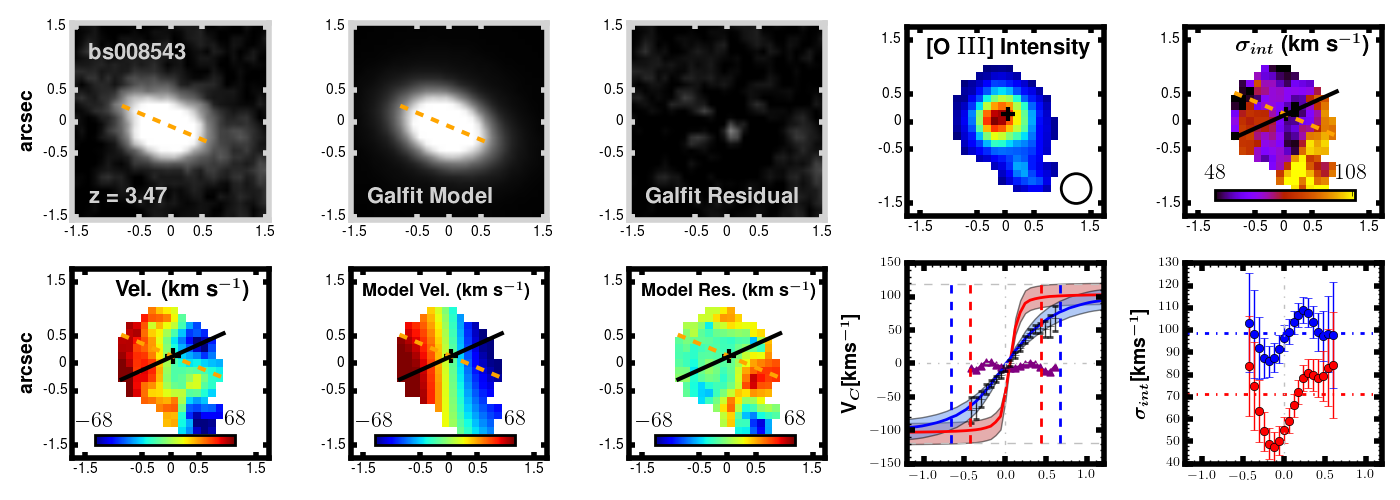}
    \includegraphics[width=0.95\textwidth]{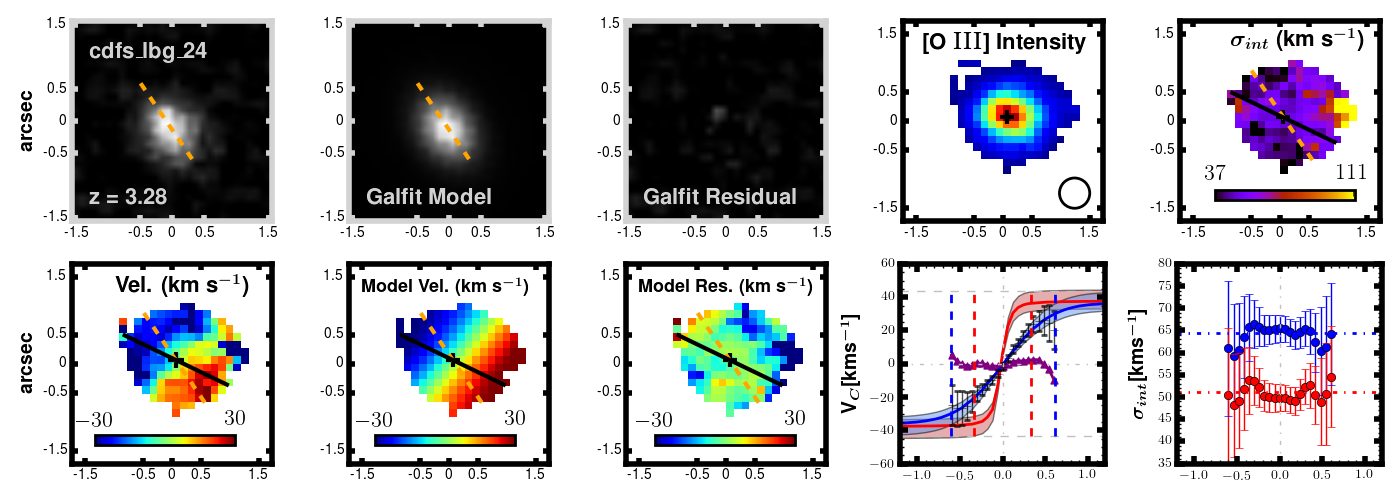}
    \includegraphics[width=0.95\textwidth]{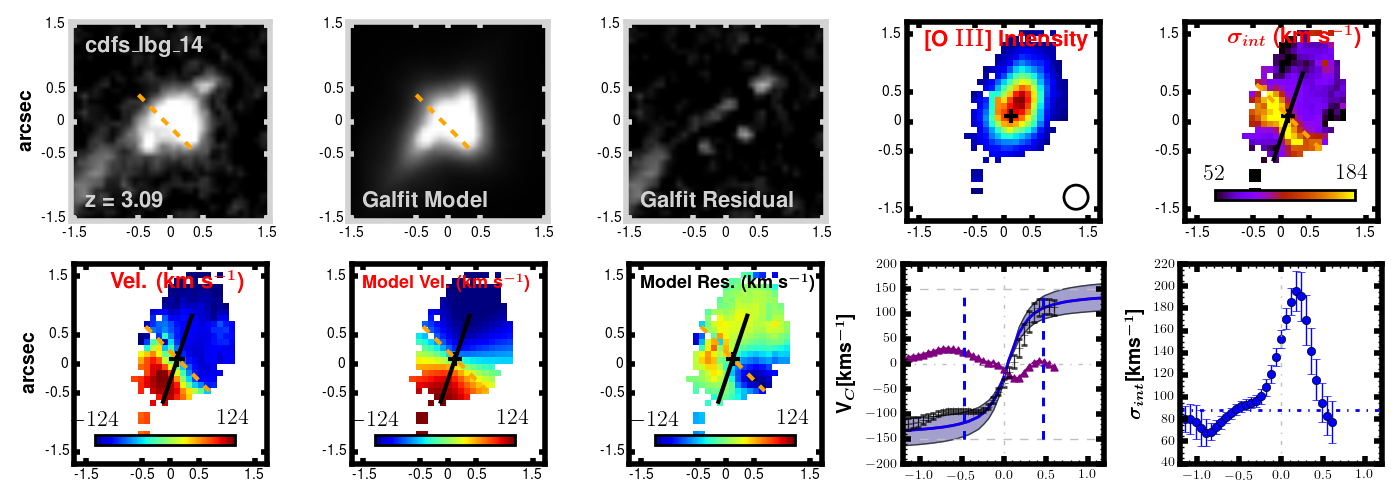}

    \caption{We plot the kinematic grids for one rotation-dominated galaxy (top grid), one dispersion-dominated galaxy (middle grid) and one merger candidate (bottom grid) (see \protect\cref{subsec:morpho-kin-class}) from the KDS.
    In Appendix \protect\ref{app:kinematics_plots} we include these plots for each of the spatially-resolved galaxies in the KDS, divided into rotation-dominated (13/38), dispersion-dominated (19/38) and merger categories (6/38).
    For each galaxy we plot on the top row the {\em HST} image, {\scriptsize GALFIT} model and {\scriptsize GALFIT} residuals, along with the observed [O~{\sc III}]$\lambda$5007 intensity map and velocity dispersion map. 
    On the bottom row we plot the observed velocity map, the beam-smeared model velocity and data-model velocity residuals with colour bar limits tuned to the velocity model.
    The solid black line and dashed orange line plotted in various panels show $PA_{kin}$ and $PA_{morph}$ respectively.
    Also on the bottom row, we plot extractions along $PA_{kin}$ for both the velocity and velocity dispersion.
    The grey points with errorbars on the velocity extraction plot are from the observed velocity map, the blue line and blue shaded regions represent the beam-smeared model fit to the data and errors respectively, whereas the red line and shaded regions are the intrinsic model from which $V_{C}$ is extracted.
    The purple symbols represent the extraction along $PA_{kin}$ from the residual map and the red and blue dashed lines denote the intrinsic and convolved $2R_{1/2}$ values respectively, with $V_{C}$ extracted at the intrinsic value and $V_{BS}$ extracted at the convolved value of the intrinsic and observed profiles respectively.
    The intrinsic models flatten at small radii, and so small changes in $R_{1/2}$ have negligible impact on the final, extracted $V_{C}$ values.
    The blue points on the velocity dispersion extraction plot show the values extracted from the observed dispersion map and the red are from the beam-smearing corrected map as per Equation \protect\ref{eq:dispersion_comp}.
    Also on the 1D dispersion plot we include with the blue and red dot-dash horizontal lines median values of the observed and intrinsic (i.e. $\sigma_{int}$) dispersion maps respectively.
    The merger candidate in the bottom grid requires several {\scriptsize GALFIT} components to describe the data, mimics rotation from a purely kinematic perspective and we plot the 1D extractions without beam-smearing corrections, which are not warranted for multiple component objects.}
    \label{fig:main_body_kinematic_grids}

\end{figure*}

\subsubsection{Constructing model datacubes}\label{subsubec:model_cube}
For each galaxy with spatially-resolved [O~{\sc III}]$\lambda$5007 emission we constructed a series of model datacubes with the same spatial dimensions as the observed datacube and populated each spaxel with an [O~{\sc III}]$\lambda$5007 emission line that has central wavelength determined using $z_{sys}$.
This central wavelength was then shifted using the velocity derived at each spaxel from the velocity field model described below.  
Following the procedure adopted by numerous authors \citep[e.g.][]{Epinat2010,Epinat2012,Swinbank2012,Stott2016,Mason2017} the velocity field of the gas was modelled as a thin disk with the discrete velocity points along $PA_{kin}$ determined by Equation \ref{eq:arctangent}, which has been found to reproduce the rotation curves of galaxies in the local universe \citep[e.g.][]{Courteau1997}:

\begin{equation}\label{eq:arctangent}
   v_{r} = \frac{2}{\pi}v_{asym}arctan\left(\frac{r}{r_{t}}\right)
\end{equation}

\noindent
with `r' measured as the distance from the centre of rotation.
This model was then projected onto the cube using the inclination angle determined in \cref{subsubsec:galfitting} and sampled with 4 times finer pixel scale than the KMOS raw data before binning back to the previous resolution.
The pixel scale refinement was necessary to capture the steep velocity gradients across the central regions.

\noindent
The velocity decreases with a factor of cos($\phi$) from the kinematic position axis, where $\phi$ is the angle measured clockwise from the axis, which points to the direction of the positive side of the velocity field.
The 6 free parameters of the model; $(xcen, ycen), i, PA_{kin}, r_{t}, v_{asym}$, were first reduced to 5 by using the inclination from the {\em HST} axis ratio.
We also fixed the (xcen,ycen) parameters to the location of the stellar continuum peak within the collapsed cubes, a proxy for the centre of the gravitational potential-well, to leave only 3 free parameters to vary in the MCMC sampling.
Due to the faintness of the continuum, there are several galaxies for which we could not reliably estimate the rotation centre using this method and instead set (xcen,ycen) to coincide with the peak [O~{\sc III}]$\lambda$5007 flux location (approach described in \citealt{Harrison2017}).
Using the continuum to fix the rotation centre was generally more successful for the SSA22 galaxies, since the use of the {\it HK} filter increases the signal-to-noise of the continuum by collapsing over twice the wavelength range. \\

\noindent
The intrinsic flux profile of these simulated [O~{\sc III}]$\lambda$5007 lines in the spatial direction was determined using the {\scriptsize GALFIT} model derived in \cref{subsubsec:galfitting} (which assumes that the [O~{\sc III}]$\lambda$5007 emission profile follows the stellar profile) and the intrinsic velocity dispersion of the line was set to follow a uniform distribution with a default width of 50 km\,s$^{-1}$.
This default line width was selected following examination of figure 10 of \cite{Wisnioski2015}, which places observational constraints on the velocity dispersions of typical star-forming galaxies at $z\simeq3.0$, and we discuss the impact of this assumption in Appendix \ref{appsubsec:model_errors} by using sensible alternative values.
The reported errors on $\sigma_{int}$ take this assumption into account. \\

\noindent
The simulated, intrinsic [O~{\sc III}]$\lambda$5007 cube was then convolved slice by slice using fast Fourier transform libraries with the {\it K}-band atmospheric seeing profile (Table \ref{tab:pointings}).
The velocity and velocity dispersion values were then re-measured in each spaxel to produce `beam-smeared' 2D maps of the kinematic parameters.
We fit the beam-smeared velocity field to the observed velocity field, using MCMC sampling with the python package `emcee' \citep{Foreman-Mackey2013} to vary the intrinsic model parameters, seeking the combination of parameters which maximises the log-Likelihood function given by Equation \ref{eq:likelihood}, which fully accounts for the errors on the observed velocity field:

\begin{equation}\label{eq:likelihood}
   ln\Lagr = \frac{-0.5\sum_{i=1}^{N}(d_{i} - M_{i})^{2}}{\delta_{v_{i}}^{2}} - ln\left(\frac{1}{\delta_{v_{i}}^{2}}\right)
\end{equation}

\noindent
where $\delta_{v_{i}}$ are the observed velocity errors, $d_{i}$ are the datapoints and $M_{i}$ are the convolved model velocity values.

The MCMC sampling provided a distribution of parameter values around those which give the maximum-likelihood.
We confirmed also by examining the individual parameter chains that the MCMC run has been properly `burned-in' and that the parameter estimates converged around the 50th percentile values.
The 16th and 84th percentiles of these distributions were used to assess the model uncertainties as described in Appendix \ref{appsubsec:model_errors}.

Following the construction of a convolved model cube and the measurement of the beam-smeared maps, we extract beam-smearing corrected values for $V_{C}$ and $\sigma_{int}$, as described throughout the following subsections.
There have been several different methods used to extract $V_{C}$ and $\sigma_{int}$ in previous surveys at different redshifts.
We select a set of comparison samples with similar galaxy selection criteria and kinematic parameter extraction methods to those described above in order to make comparisons with the KDS across different redshift slices (see \cref{sec:results}).
We have re-evaluated the means and medians of the kinematic properties in these samples in order to make fair and reliable comparisons.
An overview of the comparison samples is provided in Appendix \ref{app:comparison_samples}, which also describes the approach used to compute average parameter values.

\subsubsection{Beam-smearing corrected rotation velocities}\label{subsubsec:beam_smearing_corrected_velocities}
The intrinsic velocity, $V_{C}$, was found by reading off from the best-fitting intrinsic 2D model grid at the intrinsic $2R_{1/2}$ value ($\simeq3.4R_{D}$, where $R_{D}$ is the scale length for an exponential disk).
This extraction radius is shown with the vertical red dashed line in the panels of Figure \ref{fig:main_body_kinematic_grids} showing the 1D velocity extraction.
The $V_{C}$ value extracted at $2R_{1/2}$ is a commonly used measure of the `peak' of the rotation curve \citep[e.g.][]{Miller2011,Stott2016,Pelliccia2017,Harrison2017,Swinbank2017} and extracting the velocity at $\geqslant 2R_{1/2}$ can be crucial for measuring the majority of the total angular momentum \citep[e.g.][]{Obreschkow2016,Harrison2017,Swinbank2017}.
This is similar to the measurement of the velocity at the radius enclosing 80$\%$ of the total light, $V_{80}$ ($\simeq3R_{D}$), \citep{Tiley2016}.

The mean and median $V_{C}$ values for the isolated field sample are $76.7^{+4.9}_{-4.5}km\,s^{-1}$ and $57.0^{+6.6}_{-6.3}km\,s^{-1}$ respectively, which are less than observed for lower redshift typical star-forming galaxies (see \cref{subsec:results_rotation_vel} for a discussion).
In order to determine the beam-smearing correction factor, we also extracted the velocity from the beam-smeared model at the convolved $2R_{1/2}$ radius, represented by the vertical blue dashed line in Figure \ref{fig:main_body_kinematic_grids}, to recover the beam-smeared velocity, $V_{BS}$.
The magnitude of the beam-smearing correction, ($V_{C}/V_{BS}$), has a mean and median value of 1.29 and 1.21 for the isolated field sample and we discuss the observational and model uncertainties which combine to give the $V_{C}$ errors in Appendix \ref{appsubsec:model_errors}. \\

\noindent
In 13/38 of the isolated field sample galaxies, the intrinsic $2R_{1/2}$ fiducial radius used to extract $V_{C}$ from the model was greater than the last observed radius, however only 2 galaxies required extrapolation > 0.4$^{\prime\prime}$ (i.e. 2 KMOS pixels) with a mean extrapolation of 0.17$^{\prime\prime}$.
The extent of the extrapolation for each galaxy is made clear throughout the kinematics plots in Appendix \ref{app:kinematics_plots}, with the intrinsic and convolved $2R_{1/2}$ extraction radii marked on each velocity thumbnail.

\subsubsection{Beam-smearing corrected intrinsic dispersions}\label{subsubsec:beam_smearing_corrected_dispersions}
Convolution of the model velocity field with the seeing profile produces a peak in the beam-smeared velocity dispersion map at the centre of rotation.
The model therefore provides a beam-smearing correction for the velocity dispersion in every spaxel.
This is dependent primarily on the magnitude of the velocity gradient across the parameter maps, where larger velocity gradients result in larger central velocity dispersions.

By comparing the best-fitting simulated datacube with the assumption for the isotropic velocity dispersion, we recovered the beam-smearing correction for each galaxy.
The 2D beam-smearing correction map, $\sigma_{BS}$, was computed on a spaxel-by-spaxel basis as the difference between the beam-smeared dispersion produced by the model and the assumed intrinsic velocity dispersion of 50 km\,s$^{-1}$ i.e. $\sigma_{BS,i} = \sigma_{simulated,i} - \sigma_{assumed,i}$ (where $\sigma_{assumed,i}$ was fixed at 50 km\,s$^{-1}$ as described in \cref{subsubec:model_cube} and $\sigma_{simulated,i}$ is the simulated velocity dispersion recovered under this assumption).
The impact of the value of $\sigma_{assumed}$ on the magnitude of the $\sigma_{BS,i}$ correction is discussed further in Appendix \ref{appsubsec:model_errors} and the $\sigma_{int}$ errors take this into account.
We note also that not all of the observed $\sigma_{int}$ profiles peak in the centre, and since we did not attempt to fit the velocity dispersions, this is a feature the model could not reproduce. 

To measure the $\sigma_{int}$ value for each galaxy in our sample, we applied Equation \ref{eq:owen_sigma} in each spaxel to derive an intrinsic sigma map, where $\sigma_{g,i}$ is the value in the ith spaxel recovered from the gaussian-fit to the [O~{\sc III}]$\lambda$5007 emission line.

\begin{equation}\label{eq:owen_sigma}
\sigma_{cor,i} = \sqrt{\left(\sigma_{g,i} - \sigma_{BS,i}\right)^{2} - \sigma_{inst,i}^{2}}
\end{equation}

\noindent
The median of the $\sigma_{cor,i}$ map is taken as the $\sigma_{int}$ value for each galaxy and the median of the observed map is the $\sigma_{obs}$ value (see Table \ref{tab:kin_props}).
The mean and median $\sigma_{int}$ values in the isolated field sample are $70.8^{+3.3}_{-3.1} km\,s^{-1}$ and $71.0^{+5.0}_{-4.8} km\,s^{-1}$ respectively, higher than for typical star-forming galaxies at lower redshift as has been extensively reported \citep[e.g.][]{Genzel2006,Genzel2008,ForsterSchreiber2009,Law2009,Gnerucci2011,Epinat2012,Wisnioski2015}.
The magnitude of the beam-smearing correction, ($\sigma_{int}/\sigma_{obs}$), has a mean and median value of 0.81 and 0.79 respectively for the isolated field sample. \\

\noindent
To test the robustness of our beam-smearing correction approach, we also computed $V_{C}$ and $\sigma_{int}$ values using the independently derived correction factors, which are a function of the velocity gradient and the ratio $R_{1/2}/R_{psf}$, detailed in Johnson et al. (in prep) and used to evaluate the final kinematic parameters in \cite{Harrison2017}.
These correction factors were applied to $V_{BS}$ and $\sigma_{obs}$ in order to compute $V_{C}$ and $\sigma_{int}$.
When following this approach, for the isolated field sample, the derived $V_{C}$ values agree with our own corrections to within $\simeq10\%$, and the $\sigma_{int}$ values to within $\simeq5\%$, which is within the errors on these quantities. 
This confirmed that the choice of modelling and kinematic parameter extraction described above is consistent with the $z\simeq0.9$ KROSS methods \citep{Harrison2017}.
We also observed that the shape and width of the observed and intrinsic kinematic parameter distributions are equivalent, suggesting that the corrections do not over-extrapolate from the raw numbers.

We present the observed and model determined kinematic parameters in Table \ref{tab:kin_props} for the isolated field sample.

\subsubsection{Kinematic and photometric axes}\label{subsubsection:kin_and_phot}
The dynamical model parameter $PA_{Kin}$ was compared with $PA_{morph}$ computed by {\scriptsize GALFIT} in \cref{subsubsec:galfitting} to indicate the degree of misalignment between the gaseous and stellar components of the galaxies.
This follows the analysis described in e.g. \citealt{Epinat2008,Epinat2012,Barrera-Ballesteros2014,Barrera-Ballesteros2015,Wisnioski2015,Harrison2017,Swinbank2017}, where rotating galaxies generally have $\Delta PA \equiv |PA_{kin}-PA_{morph}| < 30^{\circ}$, with the discrepancy increasing as a function of axis ratio due to $PA_{morph}$ becoming more difficult to measure as the galaxy becomes more `face-on'.
Significant misalignment between the two axes can be a signature of galaxy interaction.

The resolved KDS galaxies at $z\simeq3.5$ generally have larger $\Delta PA$ values than those at lower redshift (median $\Delta PA = 37^{\circ}$) and this observation has several potential explanations.
For the majority of galaxies we are not in a regime where prominent morphological features are resolved in the {\em HST} imaging, so that $PA_{morph}$ is biased by single \Sers fits as described in \cite{Rodrigues2017}.
Another factor which could be partially driving increased misalignment in the KDS is the use of the full observed 2D velocity field to determine the kinematic position angle.
This generally does not give the same answer as simply connecting the velocity extrema, as we verified by rotating $PA_{Kin}$ in 1$^{\circ}$ increments and choosing the value which maximises the velocity gradient, calling this $PA_{rot}$.
If the direction of the `zero-velocity strip' across the centre of the velocity maps is not perpendicular to the line connecting the velocity extrema, the $PA_{Kin}$ determined from MCMC will be different to the $PA_{Kin}$ determined from rotation (e.g. GOODS-S galaxy bs006541 in Appendix \ref{fig:dispersion_dominated_galaxies}).
As a test, we have verified that extracting $V_{C}$ values along $PA_{rot}$ (which differs from $PA_{Kin}$ typically by only a few degrees) makes only a small difference to the final $V_{C}$ values, with the errors dominated by measurement uncertainties and model assumptions.

\subsection{Morpho-Kinematic classification}\label{subsec:morpho-kin-class}

\begin{table*}
\centering
\begin{threeparttable}
\caption{The dynamical properties for the KDS isolated field sample galaxies.
The observed properties are measured directly from the extracted two-dimensional maps (with $\sigma_{obs}$ corrected for the KMOS instrumental resolution), whereas $V_{C}$ and $\sigma_{int}$ have been corrected for the effects of beam-smearing as described throughout the text.
Each galaxy is classified as rotation-dominated or dispersion-dominated if the ratio $V_{C}/\sigma_{int}$ is greater than or less than 1 respectively, as highlighted in the `Class' column.}
\label{tab:kin_props}
\begin{tabular}{llllllll}

 \hline
ID & $V_{obs}$(km\,s$^{-1}$) & $V_{C}$(km\,s$^{-1}$) & $\sigma_{obs}$(km\,s$^{-1}$) & $\sigma_{int}$(km\,s$^{-1}$) & $V_{C}/\sigma_{int}$ & PA$_{kin}^{\circ}$ & Class$^{a}$ \\
 \hline
b012141\_012208 & 32$\pm$13                    & 93$^{+48}_{-40}$                     & 90$\pm$18                 & 63$^{+28}_{-30}$                 & 1.47$^{+1.04}_{-0.92}$             & 122$^{+9}_{-9}$ & RD   \\[1ex]
b15573          & 61$\pm21$                    & 81$^{+30}_{-33}$                     & 107$\pm23$                 & 84$^{+28}_{-26}$          & 0.97$^{+0.47}_{-0.52}$            & 118$^{+10}_{18}$ & DD   \\[1ex]
bs006516        & 49$\pm4$                    & 61$^{+10}_{-9}$                     & 58$\pm8$                  & 45$^{+10}_{-9}$              & 1.35$^{+0.37}_{-0.38}$         & 144$^{+4}_{-4}$ & RD   \\[1ex]
bs006541        & 33$\pm8$                      & 65$^{+24}_{-20}$                  & 92$\pm7$                  & 83$^{+9}_{-16}$             & 0.78$^{+0.34}_{-0.26}$          & 23$^{+3}_{-4}$ & DD    \\[1ex]
bs008543        & 68$\pm13$                    & 111$^{+20}_{-16}$                    & 94$\pm7$                  & 71$^{+13}_{-10}$         & 1.56$^{+0.37}_{-0.38}$                & 114$^{+1}_{-1}$  & RD  \\[1ex]
bs009818        & 39$\pm16$                    & 84$^{+39}_{-23}$                     & 92$\pm19$                  & 79$^{+21}_{-31}$       & 1.06$^{+0.65}_{-0.41}$                   & 72$^{+13}_{-15}$ & RD    \\[1ex]
bs014828        & 31$\pm15$                    & 46$^{+23}_{-23}$                     & 93$\pm15$                  & 88$^{+15}_{-17}$      & 0.53$^{+0.28}_{-0.28}$               & 44$^{+15}_{-11}$  & DD  \\[1ex]
bs016759        & 55$\pm9$                    & 110$^{+30}_{-21}$                    & 77$\pm11$                  & 57$^{+17}_{-22}$         & 1.93$^{+0.93}_{-0.7}$               & 75$^{+6}_{-6}$ & RD    \\[1ex]
lbg\_20         & 20$\pm6$                    & 56$^{+30}_{-19}$                     & 47$\pm6$                  & 40$^{+7}_{-21}$         & 1.39$^{+1.04}_{-0.53}$              & 41$^{+9}_{-10}$ & RD   \\[1ex]
lbg\_24         & 26$\pm6$                    & 42$^{+9}_{-7}$                     & 57$\pm5$                  & 51$^{+5}_{-6}$           & 0.82$^{+0.21}_{-0.17}$          & 64$^{+3}_{-4}$ & DD   \\[1ex]
lbg\_25         & 14$\pm5$                    & 33$^{+13}_{-13}$                     & 47$\pm6$                  & 41$^{+6}_{-8}$         & 0.8$^{+0.37}_{-0.34}$                  & 87$^{+7}_{-12}$  & DD   \\[1ex]
lbg\_30         & 20$\pm6$                    & 66$^{+19}_{-17}$                    & 89$\pm11$                 & 77$^{+14}_{-17}$       & 0.86$^{+0.32}_{-0.27}$               & 46$^{+6}_{-42}$ & DD \\[1ex]
lbg\_32         & 65$\pm5$                    & 121$^{+13}_{-12}$                    & 100$\pm13$                 & 76$^{+16}_{-15}$       & 1.58$^{+0.36}_{-0.37}$               & 11$^{+2}_{-1}$ & RD \\[1ex]
lbg\_38         & 26$\pm7$                    & 97$^{+72}_{-34}$                     & 71$\pm10$                  & 53$^{+16}_{-53}$       & 1.82$^{+2.3}_{-0.86}$               & 47$^{+2}_{-3}$  & RD   \\[1ex]
lbg\_91         & 48$\pm10$                    & 91$^{+15}_{-18}$                     & 94$\pm8$                  & 71$^{+16}_{-14}$        & 1.28$^{+0.33}_{-0.39}$              & 40$^{+5}_{-3}$  & RD  \\[1ex]
lbg\_94         & 18$\pm8$                    & 34$^{+19}_{-15}$                     & 60$\pm6$                  & 55$^{+7}_{-8}$         & 0.62$^{+0.36}_{-0.29}$                & 41$^{+18}_{-9}$  & DD  \\[1ex]
lbg\_105        & 11$\pm8$                    & 32$^{+17}_{-16}$                     & 61$\pm18$                  & 59$^{+18}_{-18}$        & 0.54$^{+0.34}_{-0.32}$                & 170$^{+4}_{-7}$ & DD  \\[1ex]
lbg\_109        & 26$\pm8$                    & 53$^{+21}_{-15}$                     & 96$\pm16$                 & 91$^{+16}_{-18}$      & 0.58$^{+0.26}_{-0.2}$                & 157$^{+11}_{-11}$ & DD  \\[1ex]
lbg\_111        & 36$\pm12$         & 58$^{+24}_{-17}$                     & 85$\pm14$                  & 81$^{+14}_{-16}$                   & 0.71$^{+0.33}_{-0.25}$                 & 168$^{+4}_{-5}$ & DD  \\[1ex]
lbg\_112         & 27$\pm3$                    & 63$^{+19}_{-11}$                    & 88$\pm5$                 & 78$^{+8}_{-11}$       & 0.81$^{+0.28}_{-0.17}$               & 4$^{+34}_{-12}$ & DD \\[1ex]
lbg\_113        & 50$\pm20$           & 41$^{+41}_{-29}$                     & 113$\pm22$                 & 108$^{+23}_{-25}$                  & 0.39$^{+0.39}_{-0.29}$                & 1$^{+34}_{-26}$  & DD   \\[1ex]
lbg\_124         & 9$\pm8$                    & 32$^{+27}_{-18}$                    & 52$\pm19$                 & 46$^{+19}_{-23}$       & 0.71$^{+0.69}_{-0.5}$               & 39$^{+20}_{-14}$ & DD \\[1ex]
$^{*}$n3\_006         & 36$\pm23$                    & 81$^{+47}_{-29}$                     & 133$\pm26$                 & 125$^{+27}_{-31}$          & 0.65$^{+0.41}_{-0.27}$                & 156$^{+13}_{-12}$ & DD   \\[1ex]
n3\_009         & 21$\pm9$                    & 49$^{+26}_{-16}$                     & 56$\pm13$                  & 48$^{+14}_{-45}$          & 1.01$^{+1.1}_{-0.45}$                & 20$^{+28}_{-25}$  & RD   \\[1ex]
n\_c3           & 26$\pm8$                    & 51$^{+17}_{-18}$                     & 84$\pm11$                  & 72$^{+12}_{-13}$          & 0.72$^{+0.28}_{-0.29}$                & 137$^{+4}_{-15}$  & DD  \\[1ex]
lab18           & 23$\pm11$                   & 41$^{+17}_{-14}$                     & 69$\pm6$                  & 62$^{+7}_{-15}$           & 0.67$^{+0.32}_{-0.24}$                 & 54$^{+18}_{-26}$ & DD    \\[1ex]
$^{*}$lab25           & 37$\pm11$                    & 48$^{+22}_{-16}$                     & 58$\pm8$                  & 49$^{+10}_{-19}$          & 0.98$^{+0.59}_{-0.4}$                & 72$^{+14}_{-13}$ & DD   \\[1ex]
s\_sa22a-d3     & 53$\pm17$                    & 109$^{+52}_{-32}$                    & 115$\pm13$                 & 93$^{+20}_{-36}$          & 1.16$^{+0.72}_{-0.43}$               & 30$^{+2}_{-2}$ & RD    \\[1ex]
s\_sa22b-c20    & 66$\pm18$                    & 139$^{+33}_{-30}$                    & 97$\pm17$                 & 48$^{+34}_{-51}$          & 2.87$^{+3.12}_{-2.14}$                & 37$^{+26}_{-11}$ & RD   \\[1ex]
$^{*}$s\_sa22b-d5     & 27$\pm10$                    & 57$^{+26}_{-23}$                     & 38$\pm20$                  & 23$^{+23}_{-18}$          & 2.45$^{+3.44}_{-2.41}$               & 60$^{+19}_{-19}$ & RD   \\[1ex]
s\_sa22b-d9     & 32$\pm14$                    & 46$^{+22}_{-15}$                     & 83$\pm8$                  & 73$^{+10}_{-17}$             & 0.63$^{+0.33}_{-0.23}$                 & 12$^{+23}_{-7}$ & DD   \\[1ex]
$^{*}$s\_sa22b-md25   & 44$\pm15$           & 57$^{+34}_{-24}$                     & 87$\pm19$   & 75$^{+21}_{-30}$                   & 0.76$^{+0.56}_{-0.39}$                & 9$^{+12}_{-7}$ & DD

\end{tabular}
\begin{tablenotes}
      \small
      \item $^{a}$ RD = Rotation-Dominated; DD = Dispersion-Dominated
      \item $^{*}$ No {\em HST} coverage; see Table \ref{tab:phys-props}
    \end{tablenotes}
  \end{threeparttable}
  \end{table*}

We proceeded to further divide the isolated field sample galaxies into two subsamples on the basis of the ratio $V_{C}/\sigma_{int}$.
This is a simple empirical diagnostic for whether a galaxy is `rotation-dominated' with $V_{C}/\sigma_{int} > 1$, or `dispersion-dominated' with $V_{C}/\sigma_{int} < 1$  \citep[i.e. a method to measure the prevalence of rotational and random motions;][]{Epinat2012,Wisnioski2015,Stott2016,Harrison2017}.
Previous studies have used various classification schemes of varying complexity to identify rotation-dominated galaxies which also appear to be `disk-like' and we briefly mention some of these in \cref{subsubsection:isolated_classification}.
We describe our simple classification criteria for the isolated field sample based on a joint consideration of the high-resolution photometry and derived kinematic parameters in \cref{subsubsection:isolated_classification} and mention the KDS merger candidates in more detail in the following section.

\subsubsection{KDS merger candidates}\label{subsubsection:merger_candidates}
The KDS galaxies were classified as mergers when more than one component was identified in the high-resolution imaging.  
One benefit of studying the {\em HST} photometry in tandem with the kinematics was to aid the interpretation of peculiarities in the velocity and velocity dispersion fields.
These peculiarities are usually unexpected discontinuities in the rotation fields, or extreme broadening of the dispersion field above that expected from beam-smearing or in regions spatially offset from the centre of the galaxy.
A second benefit was to uncover `rotation doppelg{\"a}ngers', which mimic disk rotation from a purely kinematic perspective (i.e. can be fit with the arctangent model (see \cref{subsec:3d_modelling}) with small residuals), but are clearly two or more potentially interacting components at slightly different redshifts in the imaging (e.g. the bottom grid in Figure \ref{fig:main_body_kinematic_grids}).

This has been discussed in \cite{Hung2015}, where redshifted local interacting galaxies are misclassified as single rotating disks on the basis of their kinematics. 
In short, the galaxies with multiple photometric components do not always have disordered kinematics and so the kinematics alone are not sufficient for classification.
It was difficult to distinguish in some cases between two {\em HST} components rotating together in a single disk (i.e. clumpy disk galaxies e.g. \citealt{Elmegreen2004,Bournaud2007}) with a receding and approaching side and two [O~{\sc III}]$\lambda$5007 emitting blobs which are offset along the line of sight.
Perhaps the most telling diagnostic was the presence of two flux peaks in the object spectrum at the kinematic centre viewed in combination with a double peak in the high-resolution imaging flux map. \\

\noindent
In 6/38 of the resolved KDS field galaxies (one count per merger) we clearly identified signs of multiple components or merging in the {\em HST} high-resolution imaging, which translates into a merger rate of 16\%.
In 8/9 resolved KDS cluster galaxies we identified multiple photometric components, translating into a merger rate of 89\%  (see Table \ref{tab:pointings}).
The higher merger rate observed for the cluster galaxies is perhaps unsurprising since they were drawn from a higher density environment, and these interactions will inevitably have major consequences for the consumption and stripping of gas, as well as the morphological and dynamical evolution within these galaxies.
The observation of frequent mergers, which cause major kinematic disturbances, is also in support of the evidence that galaxies living at the peak of the cosmic density field at lower redshift tend to be redder, less active and morphologically early-type, having presumably rapidly exhausted their gas supply throughout these violent interactions at high-redshifts \citep[e.g.][]{Steidel1998,White2007,Kodama2007,Zheng2009}.
As mentioned above, we omit galaxies in the cluster pointing (SSA22-P1) from the results presented in \cref{sec:results} due to the complexity added by galaxy interactions to the interpretation of cluster galaxy velocity fields.
We keep the KDS galaxies without {\em HST} coverage on the basis that we do not identify double components in the galaxy continuum maps. 
In the kinematic grids throughout \ref{fig:merger_galaxies} (as well as the bottom grid in \ref{fig:main_body_kinematic_grids}) we present the {\em HST} imaging, kinematic maps and one-dimensional kinematic extractions for the KDS merger candidates in both the field and cluster environments.

\subsubsection{Isolated field sample classification}\label{subsubsection:isolated_classification}
As mentioned above, various methods have been used throughout the literature to quantify the kinematic structure of star-forming galaxies at different redshifts. 
These include the kinemetry approach which quantifies aysmmetries in the 2D moment maps described in \cite{Shapiro2008} and applied throughout in \cite{ForsterSchreiber2009} and \cite{Cresci2009}.
In \cite{Wisnioski2015} a set of 5 criteria (see their section 4) identify rotation-dominated and disky galaxies by assessing the smoothness of the velocity gradient, the degree of rotational support through the ratio of the velocity to velocity dispersion, $V_{rot}/\sigma$, the $\Delta PA$ misalignment and the position of the kinematic centre in relation to both the peak in the velocity dispersion map and the centroid of the continuum centre (also see \citealt{Rodrigues2017} for a re-analysis of KMOS$^{3D}$ data at $z\simeq1$ using the same criteria).   
In \cite{Epinat2012}, a detailed description of a morpho-kinematic classification is outlined in their section 4, based on the proximity of counterparts in the imaging, $\Delta PA$ and $V_{C}/\sigma_{int}$.
In \cite{Gnerucci2011} an inclined plane is fit to the velocity map and the $\chi^{2}$ of this fit is evaluated in order to test for the presence of smooth velocity gradients, which they find for the 11/33 galaxies described in Appendix \ref{subsec:AMAZE}.
We used a simple approach (dictated by the signal-to-noise of the isolated field sample) and calculated rotation and dispersion-dominated percentages on the basis of $V_{C}/\sigma_{int}$ alone with repsect to the total number of resolved galaxies in the field environment (38, see Table \ref{tab:pointings}).
The morphologically isolated KDS field galaxies were classified as rotation-dominated if $V_{C}/\sigma_{int} > 1$, 13/38 or 34\% of the sample, and dispersion-dominated if $V_{C}/\sigma_{int} < 1$, 19/38 or 50\% of the sample, as indicated in the final column of Table \ref{tab:kin_props}.
We plot the best-fitting smeared and intrinsic models for the isolated field galaxies in Appendix \ref{app:kinematics_plots}, where the galaxies have been divided into rotation and dispersion-dominated categories in Figures \ref{fig:rotation_dominated_galaxies} and \ref{fig:dispersion_dominated_galaxies} respectively.

\subsubsection{Summary of the final sample for further analyses}\label{subsubsec:kin_sample_summary}
In summary, we analysed 77 typical $z\simeq3.5$ SFGs spanning both cluster and field environments, finding 62 with integrated [O~{\sc III}]$\lambda$5007 emission (\cref{subsubsec:datareduction}).
This sample was refined to 47 galaxies (61\%) with spatially-resolved [O~{\sc III}]$\lambda$5007 emission (\cref{subsubsection:spaxel_fitting}).
Due to the unusually high merger-candidate rate of 89\% in the cluster environment (identified by searching the high-resolution {\em HST} imaging for photometric counterparts), we omitted the 9 spatially-resolved galaxies in SSA22-P1 from further consideration.
In the remaining sample of 38 field environment galaxies we identified 6 merger candidates, giving a field galaxy merger rate of 16\%, which constitute the `merger field sample'.
The remaining 32 galaxies constitute the `isolated field sample', for which we measured morphological parameters using {\scriptsize GALFIT} to fit exponential disks to the {\em HST} imaging (see \cref{subsubsec:galfitting} and Table \ref{tab:phys-props}) and derived beam-smearing corrected measurements of the rotation velocities extracted at $2R_{1/2}$ and the intrinsic velocity dispersions (see \cref{subsec:3d_modelling} and Table \ref{tab:kin_props}). \\

\noindent
The isolated field sample was further divided into 13/38 rotation-dominated galaxies (34\% with respect to full field sample) and 19/38 dispersion-dominated galaxies (50\%) on the basis of the ratio $V_{C}/\sigma_{int}$ (see \cref{subsubsection:isolated_classification}), with the remaining 6/38 (16\%) classified as mergers.

Throughout the remainder of the paper we discuss the derived kinematic properties of the isolated field sample in further detail.

\section{RESULTS}\label{sec:results}

In the previous sections we presented our morphological and kinematic analyses for the 77 galaxies covering both cluster and field environments in the KDS survey.
These galaxies are representative of typical star-forming galaxies at $z\simeq3.5$, as verified by comparing with fits to the galaxy main-sequence at this redshift.
Through a combined morpho-kinematic classification described in \cref{subsec:morpho-kin-class}, we first reduced the sample to 38 spatially resolved field environment galaxies and then to 32 isolated field sample galaxies which have spatially-resolved [O~{\sc III}]$\lambda$5007 emission and are morphologically isolated (i.e. not merger candidates), which is the largest sample of galaxies observed with integral-field spectroscopy at this redshift.
For these galaxies we measured morphological properties as listed in Table \ref{tab:phys-props} and extracted intrinsic, beam-smearing corrected measurements of the rotation velocities, $V_{C}$, and velocity dispersions, $\sigma_{int}$, as listed in Table \ref{tab:kin_props}, which we used to classify the galaxies as either rotation or dispersion-dominated.
The morphological and kinematic grids for these 32/38 isolated field sample galaxies and the 6/38 merger candidates are plotted in Appendix \ref{app:kinematics_plots}.

Throughout the following sections we analyse the derived kinematic parameters of the isolated field sample, placing the galaxies in the context of galaxy evolution using the comparison samples spanning a wide redshift baseline described in Appendix \ref{app:comparison_samples}.
In Appendix \ref{app:comparison_samples} we describe our careful assessment of lower redshift samples from the literature, using tabulated data where possible to re-compute mean and median $V_{C}$ and $\sigma_{int}$ values.
These fair comparisons help to disentangle true evolution in the kinematic properties of typical star-forming galaxies from systematic effects originating in disparate measurement techniques or galaxy selection criteria.
During the following sections we will make statements about dynamical evolution by connecting the dots of these different surveys, assuming that they trace the average properties of typical star-forming galaxies at their respective mean redshifts.

We proceed first to investigate the relationship between rotation velocity and stellar mass for both the rotation and dispersion-dominated isolated field galaxies.

\subsection{Rotation velocity and stellar mass}\label{subsec:results_rotation_vel}

\begin{figure*}
    \centering \hspace{-1.3cm}
    \begin{subfigure}[h!]{0.5\textwidth}
        \centering
        \includegraphics[height=3.5in]{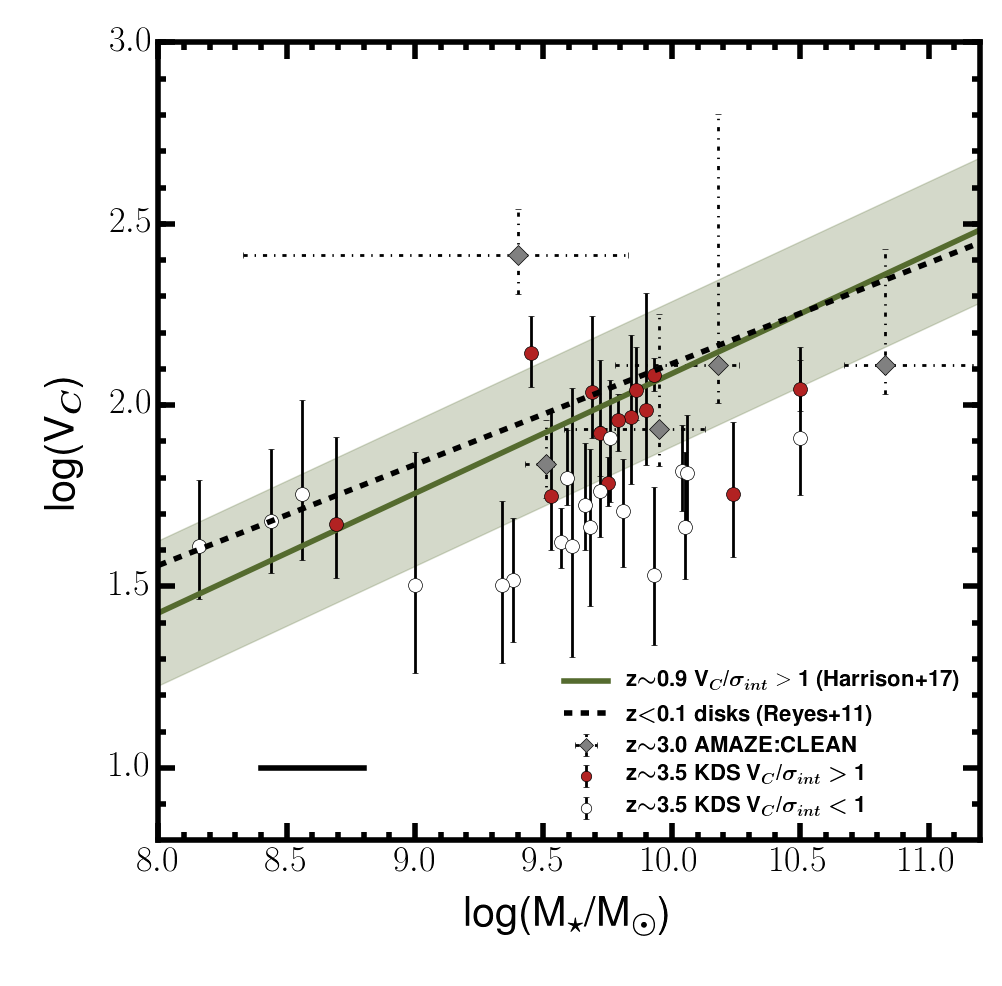}
    \end{subfigure} \hspace{0.4cm}
    \begin{subfigure}[h!]{0.5\textwidth}
        \centering
        \includegraphics[height=3.5in]{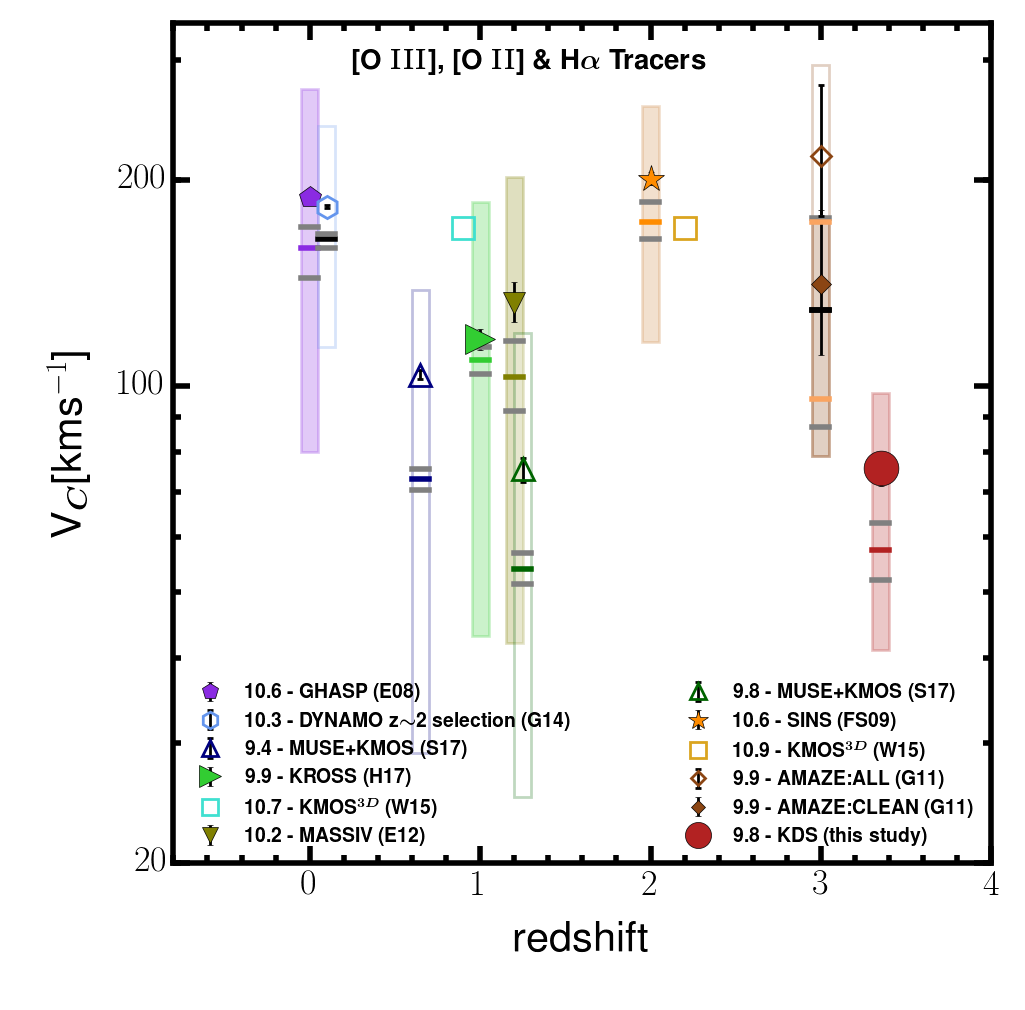}
    \end{subfigure}
    \caption{{\it Left:} We plot log($V_{C}$) vs. log($M_{\star}/M_{\odot}$) for the isolated field sample, with the rotation-dominated galaxies in red and dispersion-dominated galaxies with the hollow symbols.
    We also plot with the solid green line the $z\simeq0.9$ relation recovered from a fit to $\simeq400$ rotation-dominated galaxies from the KROSS survey \protect\cite{Harrison2017}, with the shaded region showing the associated $\simeq$0.2 dex scatter on the fit.
    This $z\simeq0.9$ relation shows no signifcant evolution from the fit to a local sample of spiral galaxies with $z < 0.1$, presented in \protect\cite{Reyes2011}.
    The rotation-dominated KDS galaxies, and galaxies classed as reliable from the AMAZE survey \protect\citep{Gnerucci2011}, also show no evolution from this relation within the scatter, with the dispersion-dominated galaxies scattering below the relation.
    This highlights the need for careful sample selection when constructing this relation.
    {\it Right:} The mean, median and distributions of the comparison sample $V_{C}$ measurements are plotted against redshift.
    The symbol positions with black errorbars (some not visible), show the mean values and the horizontal filled and grey bars show the median values and associated errors.
    Note that the median errors for the AMAZE Clean sample are shaded orange to avoid confusion with the full sample.
    The filled symbols and shaded regions denote surveys deemed directly comparable (as explained in Appendix \protect\ref{app:comparison_samples}), and the mean mass of each survey (in units of log($M_{\star}/M_{\odot}$)) is listed beside the survey name.
    The $V_{C}$ distributions in each of the individual studies are wide, reflecting the galaxy diversity within each sample.
    The low mean value of $V_{C}$ in the KDS isolated field sample is driven mainly by averaging over a sample which contains a higher fraction of dispersion-dominated galaxies with lower rotation velocities.}
    \label{fig:tf_relation}
\end{figure*}

In the $\Lambda CDM$ paradigm, galaxies form when baryons accumulate and cool in the centres of dark matter halos, subsequently forming a rotating disk of stars and gas.
The relationship between stellar mass and rotation velocity describes the connection between the luminous mass and the total dynamical mass in a system.
The connection has been studied extensively at low redshift \citep[e.g.][]{Bell2000a,Rhee2003,Reyes2011} with fits to late-type, rotating galaxies in the $V_{C}-M_{\star}$ plane having a well-defined slope and normalisation.
Observing this `stellar mass Tully-Fisher relation' at different epochs thus offers constraints for galaxy formation and evolution models seeking to explain simultaneously the properties of dark matter halos and the dynamical properties of disk galaxies \citep[e.g.][]{Dubois2014,Vogelsberger2014b,Schaye2015}.
In principle, studying whether the stellar mass Tully-Fisher relation evolves with redshift gives us information about when galaxies accumulate their stellar mass and rotation velocities.
Any offset from the local relation towards lower $M_{\star}$ values at fixed rotation velocity suggests that the dynamical mass is in place, but the stellar mass is yet to be formed \citep[e.g.][]{Puech2008,Cresci2009,Puech2010,Gnerucci2011,Swinbank2012,Straatman2017,Ubler2017}.
As has been discussed recently, information about the redshift evolution of gas fractions, $f_{g} = M_{gas}/M_{bar}$ where $M_{bar}$ is the baryonic mass, disk mass fractions, $m_{d} = M_{bar}/M_{H}$  where $M_{H}$ is the halo mass, dark matter fractions, $f_{DM}(R_{1/2}) = V^{2}_{DM}(R_{1/2})/V^{2}_{C}(R_{1/2})$ where $V_{DM}$ is the dark matter circular velocity, may also be encoded in measured offsets from the local stellar mass Tully-Fisher relation \citep{Wuyts2016b,Ubler2017,Genzel2017}.  \\

\noindent
We plot the relationship between rotation velocity and stellar mass for the isolated field sample in the left panel of Figure \ref{fig:tf_relation}.
We compare to the relationship from \cite{Harrison2017}, in which the inverse stellar mass Tully-Fisher relation relation is fit to $\simeq400$ star-forming galaxies with $9.5 < log(M_{\star}/M_{\odot}) < 10.5$ at $z\simeq0.9$, measured to be rotation-dominated with $V_{C}/\sigma_{int} > 1$.
The data are used to constrain parameters `a' and `b' in the equation $logV_{C}=b+a[logM_{\star} - 10.10]$ (see their section 4.2) finding $a_{z=0.9} = 0.33 \pm 0.11$ and $b_{z=0.9} = 2.12 \pm 0.04$.
We plot this best fit relation with the solid green line, with the green shaded region giving the typical scatter of 0.2 dex along the velocity axis.

\noindent
The rotation-dominated KDS isolated field sample galaxies are broadly consistent within the scatter with the $z\simeq0.9$ relation, but with a tendency to lie below the relation.
When considered alone, the mean velocity of the rotation-dominated isolated field sample galaxies ($V_{C} = 96.7^{+7.3}_{-7.2}km\,s^{-1}$) at their mean mass of $log(M_{\star}/M_{\odot})=9.8$ is $\simeq0.05$ dex beneath the KROSS relation at the same mass.
The dispersion-dominated KDS galaxies sit clearly separated and below the $z\simeq0.9$ relation, bringing significant scatter to lower $V_{C}$ values at fixed $M_{\star}$ as is also observed in $z\simeq0.9$ dispersion-dominated galaxies and for early-type galaxies in the local universe \citep[e.g.][]{Romanowsky2012}. 
We plot also the five galaxies from the AMAZE Clean ($z\simeq3$) sample, which are consistent within the scatter with the relation for typical star-forming galaxies at $z\simeq0.9$.
The $z\simeq0.9$ fit parameters are consistent with those measured in the local universe, $a_{z=0} = 0.278 \pm 0.010$ and $b_{z=0} = 2.142 \pm 0.004$, for a sample of $z<0.1$ disks from \cite{Reyes2011}, which are in a similar mass range to the KROSS galaxies in \cite{Harrison2017} and traced with H$\,\alpha$ emission.
In \cite{Reyes2011}, $V_{C}$ has been extracted at the same radius and we plot this $z\simeq0.1$ local relation as the dashed black line in the left panel of Figure \ref{fig:tf_relation}.
The agreement between the parameters derived at $z\simeq0$ and $z\simeq0.9$ suggests that there is no evolution of the slope/normalisation of the stellar mass Tully-Fisher relation for rotation-dominated star-forming galaxies between these redshifts. \\

\noindent
Studies of the redshift evolution of the stellar mass Tully-Fisher relation do not explain how individual, unperturbed galaxies move around the rotation velocity stellar mass plane with redshift. 
High-redshift galaxy surveys typically do not trace the progenitors of those at lower redshift, as evidenced by the larger observed stellar and baryonic masses \citep[e.g.][]{Cresci2009,Reyes2011,Wisnioski2015,Ubler2017}.
Rather, these studies trace the evolution of the relation itself, the position of which is dictated by the mean properties of typical star-forming galaxies at each redshift slice.
Some authors have reported that as redshift increases, the relation shifts towards lower masses at fixed rotation velocity \citep[e.g.][]{Puech2008,Cresci2009,Puech2010,Straatman2017,Ubler2017}.
However most find that when selecting rotation-dominated galaxies between $0.5 < z < 2$ \citep[e.g.][]{Flores2006,Miller2011,Kassin2012,Miller2012,Vergani2012,Miller2014,Contini2015a,DiTeodoro2016,Simons2016,Pelliccia2017,Molina2017,Harrison2017} the results are consistent with zero evolution of the slope and zero-point of the relation.
At $z\simeq3$ observations are limited and in \cite{Gnerucci2011} an offset of $-$1.29 dex towards lower mass at fixed rotation velocity is claimed, although there is large degree of scatter between the individual galaxies, and the AMAZE Clean sample from \cite{Gnerucci2011} defined in \cref{subsec:AMAZE} is consistent with the $z\simeq0$ relation. \\

\noindent
Conclusions on the evolution of the stellar mass Tully-Fisher relation are a strong function of sample selection; in \cite{Tiley2016} where a stricter $V_{C}/\sigma_{int} > 3$ cut is applied, evolution to lower masses at fixed rotation velocity is observed at $z\simeq0.9$, which is not the case when using the $V_{C}/\sigma_{int} > 1$ cut.
Similarly in \cite{Cresci2009} evolution of the normalisation of the stellar mass Tully-Fisher relation is observed at $z\simeq2$ using only robust rotators from the SINS sample with mean $V_{C}/\sigma_{int} = 5$.
In both cases the observed evolution is mainly the result of omitting rotation-dominated galaxies with lower $V_{C}$ values.
Conversely, fitting the stellar mass Tully-Fisher relation through a full sample including dispersion-dominated galaxies would shift the zero-point in the opposite sense.
This new, robust analyses of typical $z\simeq3.5$ star-forming galaxies is consistent with the majority of work that shows no evolution for rotation-dominated galaxies across all epochs.

\noindent
In summary, with the KDS, we have added the largest sample of galaxies observed with integral-field spectroscopy at $z>3$, and with our simple $V_{C}/\sigma_{int} > 1$ cut we conclude that the stellar mass Tully-Fisher relation defined for rotation-dominated star-forming galaxies appears to show no evolution between $z\simeq0.1$ and $z\simeq3.5$.
However, as we will show in \cref{subsec:rdf_v_over_sigma}, the $V_{C}/\sigma_{int} > 1$ subsamples become increasingly less representative of typical star-forming galaxies with increasing redshift.
We also explore this in the context of the mean rotation velocities of the comparison samples throughout the following subsection. 

\subsubsection{The evolution of maximum circular velocity}\label{subsubsec:v_evolution}

\begin{figure*}
    \centering \hspace{-1.3cm}
    \begin{subfigure}[h!]{0.5\textwidth}
        \centering
        \includegraphics[height=3.5in]{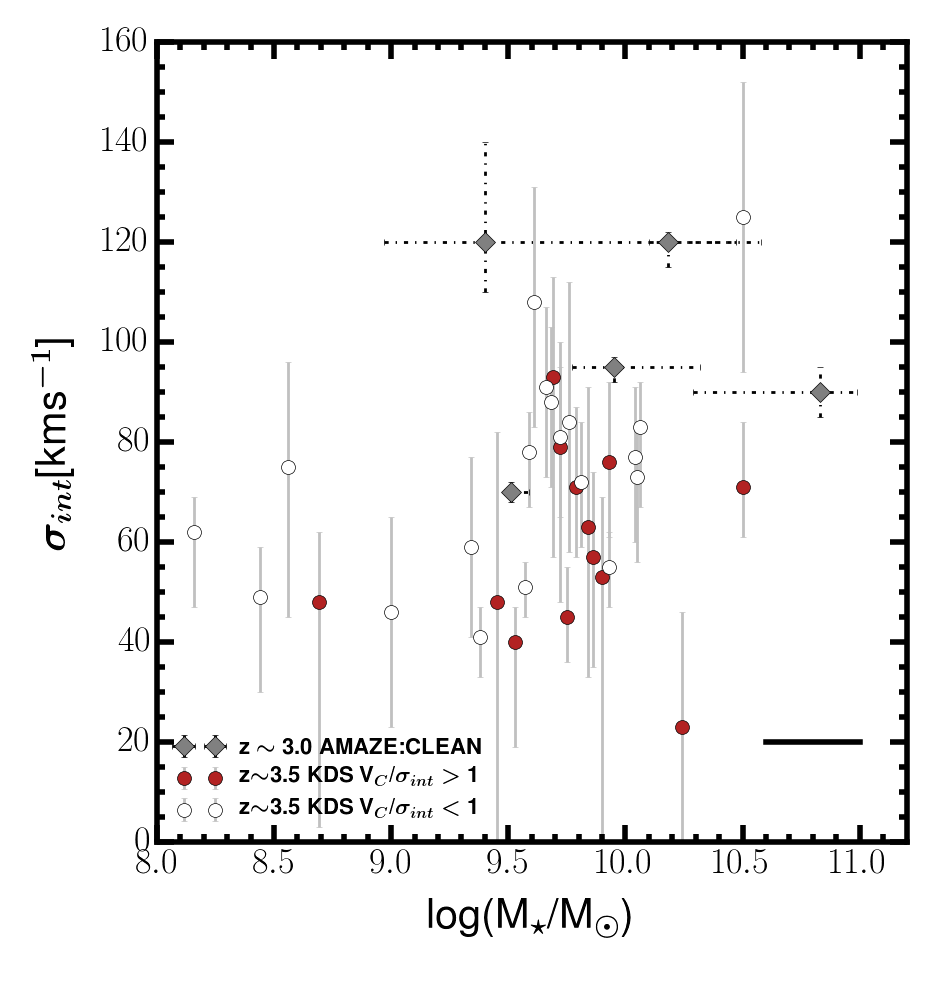}
    \end{subfigure} \hspace{0.4cm}
    \begin{subfigure}[h!]{0.5\textwidth}
        \centering
        \includegraphics[height=3.5in]{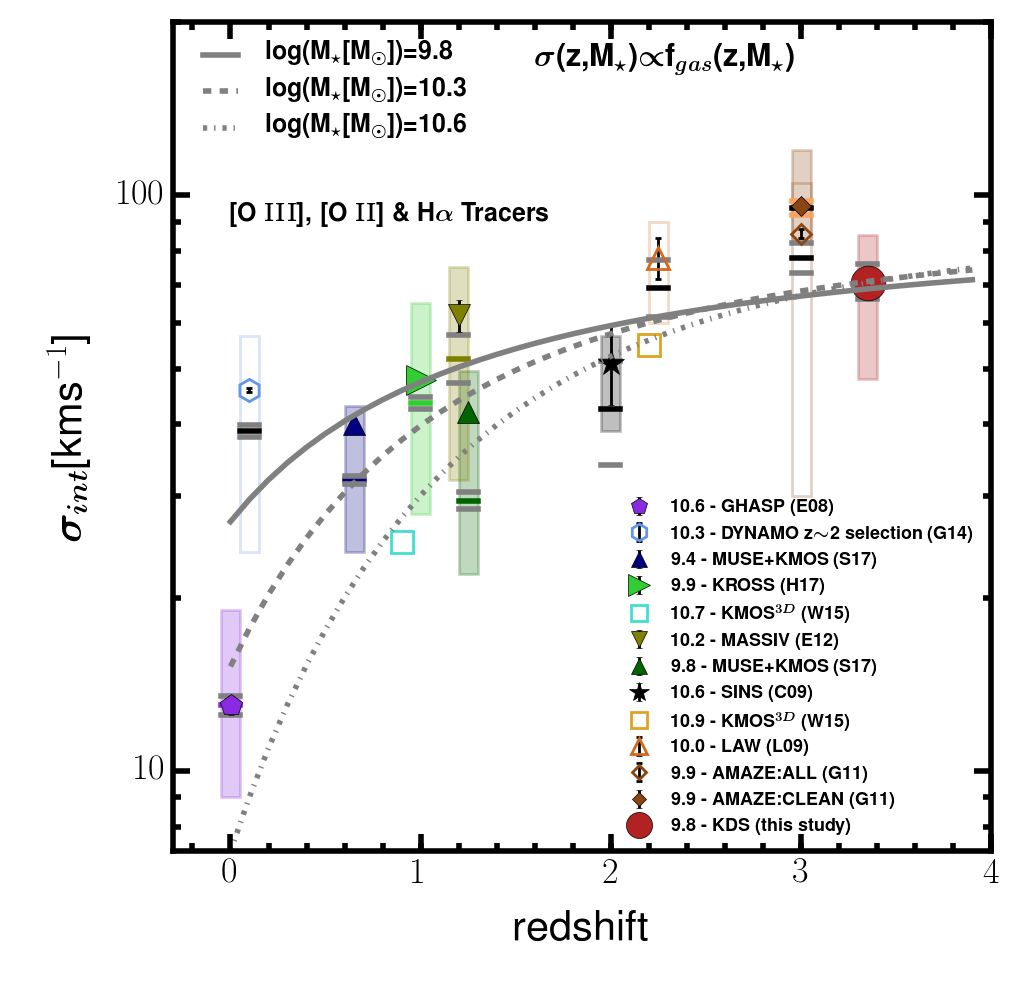}
    \end{subfigure}
    \caption{{\it Left:} We plot the intrinsic velocity dispersion against stellar mass for the rotation and dispersion-dominated isolated field sample galaxies, along with the AMAZE Clean ($z\simeq3$) sample.
    The $\sigma_{int}$ values are typically distributed between $40-90km\,s^{-1}$ and show no clear correlation with stellar mass, highlighting the complicated relationship between these two quantities. 
    {\it Right:} We present a compilation of literature $\sigma_{int}$ values plotted against redshift for surveys spanning $0 < z < 4$.
    The mean $\sigma_{int}$ values increase from $\simeq15km\,s^{-1}$ in the local universe to $> 70km\,s^{-1}$ at $z > 3$.
    The symbol convention is the same as in the right panel of Figure \protect\ref{fig:tf_relation}.
    The $\sigma_{int}(z,M_{\star})\propto f_{gas}(z,M_{\star})$ scaling relation from \protect\cite{Wisnioski2015}, the form of which is discussed in \protect\cref{discussion_increase_in_sigma}, is plotted for three different $M_{\star}$ values.
    These tracks indicate the way in which $\sigma_{int}$ evolves for samples of star-forming galaxies with different mean $M_{\star}$ values, which is a proxy for the dynamical maturity of the galaxies.
    The KDS datapoint appears to be consistent with the scenario proposed in previous work, whereby the mean $\sigma_{int}$ increases over cosmic time, with large scatter between individual galaxies, as a result of increased gas fractions \protect\citep{Wisnioski2015} and the more efficient accretion of cold gas driving disk instabilities \protect\citep[e.g.][]{Law2009,Genzel2011}.} 
    \label{fig:sigma_and_v_sigma_w_redshift}
\end{figure*}

In the right panel of Figure \ref{fig:tf_relation} we plot the fractional error weighted mean, median and distributions of $V_{C}$ values reported in the comparison (see Appendix \ref{app:comparison_samples}) and KDS isolated field samples to highlight two observations:
\begin{itemize}
    \item At each redshift slice the $V_{C}$ distributions are extremely broad, with mean values sensitive to the average stellar mass of each survey.
    \item The mean isolated field sample $V_{C}$ value is less than reported in the lower redshift comparison samples with similar mean $M_{\star}$ values.
    This is a result of averaging over a sample containing a higher fraction of dispersion-dominated galaxies.
\end{itemize}

\noindent
The right panel of Figure \ref{fig:tf_relation} indicates that the $V_{C}$ distributions for the comparison and KDS isolated field samples are wide, reflecting the mass range and the mixture of rotation/dispersion-dominated galaxies which constitute the samples.
The isolated field sample mean and median $V_{C}$ values are $76.7^{+4.9}_{-4.5}km\,s^{-1}$ and $57.0^{+6.6}_{-6.3}km\,s^{-1}$.
However, the mean and median rotation velocity of the rotation dominated galaxies are $V_{C} = 96.7^{+7.3}_{-7.2}km\,s^{-1}$ and $V_{C} = 93.0^{+8.1}_{-7.7}km\,s^{-1}$ respectively, suggesting that averaging over a sample containing a large number of dispersion dominated galaxies is bringing the averages down.

It is difficult to assess whether the $V_{C}$ values reported in the comparison samples (the modelling and extraction techniques differ as described in Appendix \ref{app:comparison_samples}) are directly comparable.
The GHASP, MASSIV, KROSS and KDS samples should in principle yield comparable $V_{C}$ values which rely on extracting from the data at a fiducial radius or using a velocity shear value, but the comparison to the dynamical models of SINS and AMAZE is not so clear.
Despite the wide distributions, the comparison samples generally have higher mean rotation velocities. \\

\noindent
In a rotationally supported system, the circular velocity is set by the requirement to support the total mass in the system, i.e. the combination of gas mass, stellar mass and dark matter mass, against gravitational collapse.
$V_{C}$ increases with increasing total mass and so we would expect the systems with larger average stellar mass, assuming that these are indicative of systems with greater total mass, to have larger $V_{C}$, which appears generally to be the case in the right panel of Figure \ref{fig:tf_relation} where the KMOS$^{3D}$, GHASP and SINS galaxies have the largest inferred $V_{C}$ values.
Using the average stellar mass values and velocity extraction radii in the isolated field sample, $log(M_{\star}/M_{\odot})=9.8$ and $2R_{1/2} = 3.2kpc$ respectively, and assuming simple, circular Keplerian orbits where the rotational motions alone support the mass of the system, we can calculate a rough lower limit of $\left<V_{C}\right>$ $>$ $95km\,s^{-1}$ for the baryonic material in the galaxies (since the inclusion of the gas mass and dark matter mass would raise this estimate).
This value is greater than the mean and median values of the isolated field sample, but comparable to the mean and median of the rotation dominated galaxies, suggesting that rotation in these systems could support, at most, the stellar mass of the galaxy.  

As mentioned above, the observation of a low mean $V_{C}$ value in the full isolated field sample is primarily explainable through averaging over a sample with a higher fraction of dispersion-dominated galaxies with lower rotation velocities.
This is seen in the left panel of Figure \ref{fig:tf_relation} where the dispersion-dominated galaxies sit clearly below and separated from the stellar mass Tully-Fisher relation.
However, it has also been discussed that pressure support, provided by turbulence and measured through the galaxy velocity dispersions, may partially compensate the gravitational force within galaxy disks and contribute to lowering the rotation velocities through an `asymmetric drift' correction \citep[e.g.][]{Burkert2010,Newman2013,Genzel2017}.
This topic is discussed further throughout \cref{sec:discussion}. \\

\noindent
We proceed to consider the cause of the high number of dispersion-dominated galaxies in our sample and the obvious candidate is the increase of velocity dispersion with redshift \citep[e.g.][]{Genzel2006,Genzel2008,ForsterSchreiber2009,Law2009,Gnerucci2011,Epinat2012,Wisnioski2015}.
There are also some galaxies in our sample where the observed velocity field has clear rotational structure and reaches flattening (e.g. see Figure \ref{fig:dispersion_dominated_galaxies} lbg\_24 with $log(M_{\star}/M_{\odot})=9.75$), with inferred $V_{C}$ smaller than $\sigma_{int}$.
This suggests again that the increase in random motions could influence the rotation velocities and that the accumulation of a massive stellar population is necessary to stabilise a galaxy \citep[e.g.][]{Law2009,Law2012b,Law2012c,Newman2013,Wisnioski2015}.
These ideas are explored in the following sections.

\subsection{Velocity dispersions}\label{subsec:results_velocity_dispersions}

In the left panel of Figure \ref{fig:sigma_and_v_sigma_w_redshift} we plot the isolated field galaxies, split into dispersion and rotation-dominated classes, along with the AMAZE Clean ($z\simeq3$) sample, in the $\sigma_{int}$ vs. $M_{\star}$ plane.
This is an interesting relationship to explore as it has been suggested that rotation velocity alone may not be a good tracer of total dynamical mass at high-redshift and that random motions traced by $\sigma_{int}$ may contribute to supporting some fraction of this total dynamical mass \citep[e.g.][]{Kassin2007,Burkert2010,Kassin2012,Law2012b,Law2012c,Ubler2017,Genzel2017}.
However, the extent to which the gaseous $\sigma_{int}$ values reported here trace stellar mass is unclear, with physical processes such as disk turbulence, gas accretion and subsequent disk instabilities contributing to the magnitude of the velocity dispersion.
All of these processes are also more prevalent at high-redshift \citep{Genzel2006,ForsterSchreiber2009,Law2009,Genzel2011,Wisnioski2015,Wuyts2016b}, leaving $\sigma_{int}$ as a challenging property to interpret. \\ 

\noindent
We observe no clear correlation between the gaseous velocity dispersion and stellar mass in any of the sub-samples, highlighting the complicated relationship between these quantities at high-redshift.
The beam-smearing corrected $\sigma_{int}$ values, with distribution median, 16th percentile and 84th percentile equal to $71.0km\,s^{-1}$, $48.0km\,s^{-1}$ and $85.4km\,s^{-1}$ respectively, are larger than in the local universe where typically $\sigma_{int} = 10-20km\,s^{-1}$ \citep{Epinat2008a} (see \cref{subsubsec:sigma_evolution} and \cref{sec:discussion} for a discussion).
Generally, the dispersion-dominated galaxies have higher $\sigma_{int}$ values in the left panel of Figure \ref{fig:sigma_and_v_sigma_w_redshift}, however the rotation-dominated galaxies also span the full distribution width.

\subsubsection{The evolution of velocity dispersion}\label{subsubsec:sigma_evolution}

\begin{figure}
    \centering \hspace{-1.13cm}
    \includegraphics[width=0.49\textwidth]{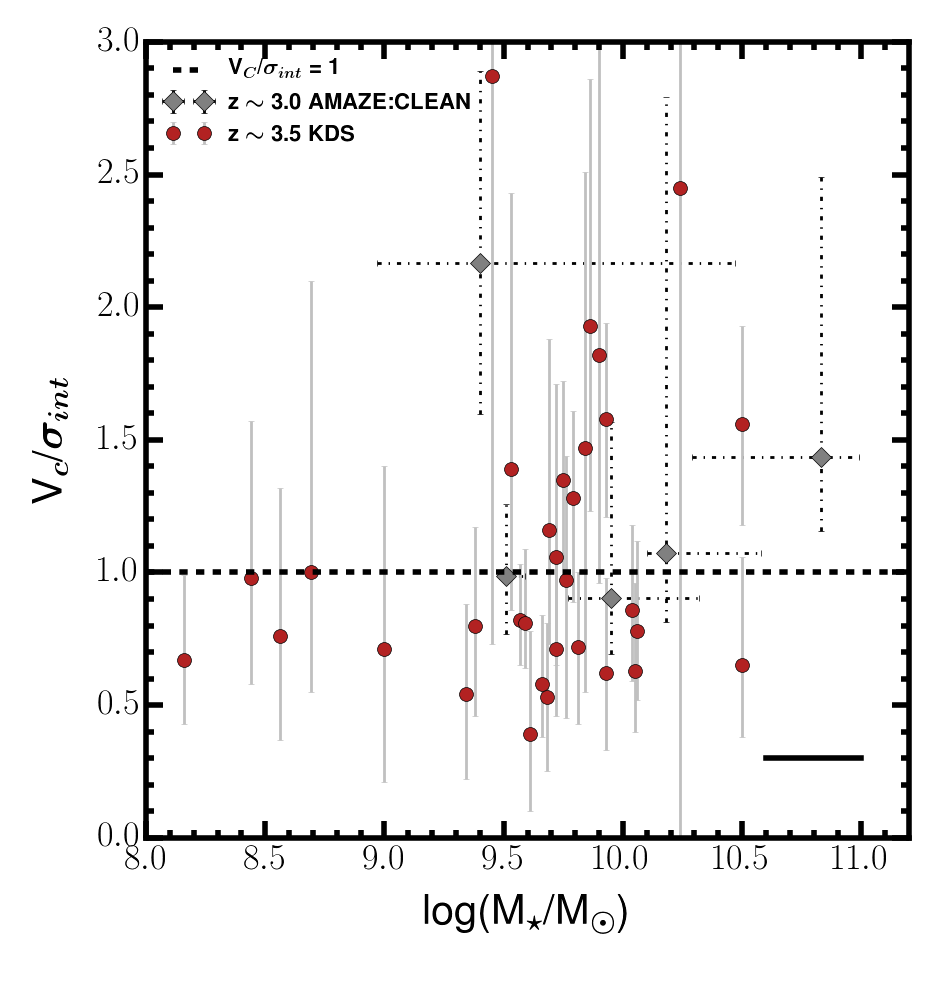}
    \caption{The distribution of galaxies in the $V_{C}$/$\sigma_{int}$ vs. $M_{\star}$ plane is shown for the isolated field sample, along with the galaxies in the AMAZE Clean ($z\simeq3$) sample.
    We mark the $V_{C}$/$\sigma_{int}=1$ line dividing rotation and dispersion-dominated galaxies in this panel, finding that more massive galaxies (e.g. $log(M_{\star}/M_{\odot}) > 9.5$) are more likely to be rotation-dominated, with dispersion-dominated galaxies spread across the stellar mass range.}
    \label{fig:v_sig_and_v}
\end{figure}

In the right panel of Figure \ref{fig:sigma_and_v_sigma_w_redshift} we plot the fractional error weighted mean, median and distributions of $\sigma_{int}$ from the comparison and KDS isolated field samples.
The values increase sharply from $\sigma_{int}\simeq10-20km\,s^{-1}$ as traced by $z\simeq0$ measurements of typical SFGs to $\sigma_{int}\simeq30-60km\,s^{-1}$ at $z\simeq1$ to $\sigma_{int}\simeq40-90km\,s^{-1}$ at $z\simeq3$, but with a wide range of values observed at each redshift.
There appears to be a trend for the surveys with lower average stellar mass (MASSIV, DYNAMO, LAW 2009, KROSS, AMAZE) to have higher velocity dispersions than the surveys with larger average stellar mass (GHASP, KMOS$^{3D}$, SINS).

In agreement with the evolution of gas fractions described in \cref{discussion_increase_in_sigma} (and see the tracks in the right panel of Figure \ref{fig:sigma_and_v_sigma_w_redshift}), we find that the mean velocity dispersion measurements from samples of typical star-forming galaxies are decreasing with redshift, with a dependence on mass such that samples with higher stellar mass have lower velocity dispersions at each epoch.
The redshift and mass dependence of these tracks is discussed in \cite{Wisnioski2015} and discussed in more detail throughout \cref{discussion_increase_in_sigma}, where we also discuss possible physical explanations for the increase in velocity dispersions with redshift. \\  

\noindent
These new data from the KDS sustain the observation that velocity dispersion is increasing with redshift and provide strong constraints on the internal dynamics of typical star-forming galaxies between $z=3-3.8$.
When considered in conjunction with data from the literature in the right panel of Figure \ref{fig:sigma_and_v_sigma_w_redshift}, it appears that the stellar mass of galaxies may play a role in mediating the velocity dispersion values (see \cref{discussion_increase_in_sigma}).

We proceed now to consider both the rotation velocities and velocity dispersions in tandem by considering $V_{C}/\sigma_{int}$ and the rotation-dominated fraction of galaxies in the isolated field sample.

\subsection{$\boldsymbol{V_{C}/\sigma_{int}}$ and the rotation-dominated fraction}\label{subsec:rdf_v_over_sigma}

\begin{figure*}
    \centering \hspace{-1.3cm}
    \begin{subfigure}[h!]{0.5\textwidth}
        \centering
        \includegraphics[height=3.5in]{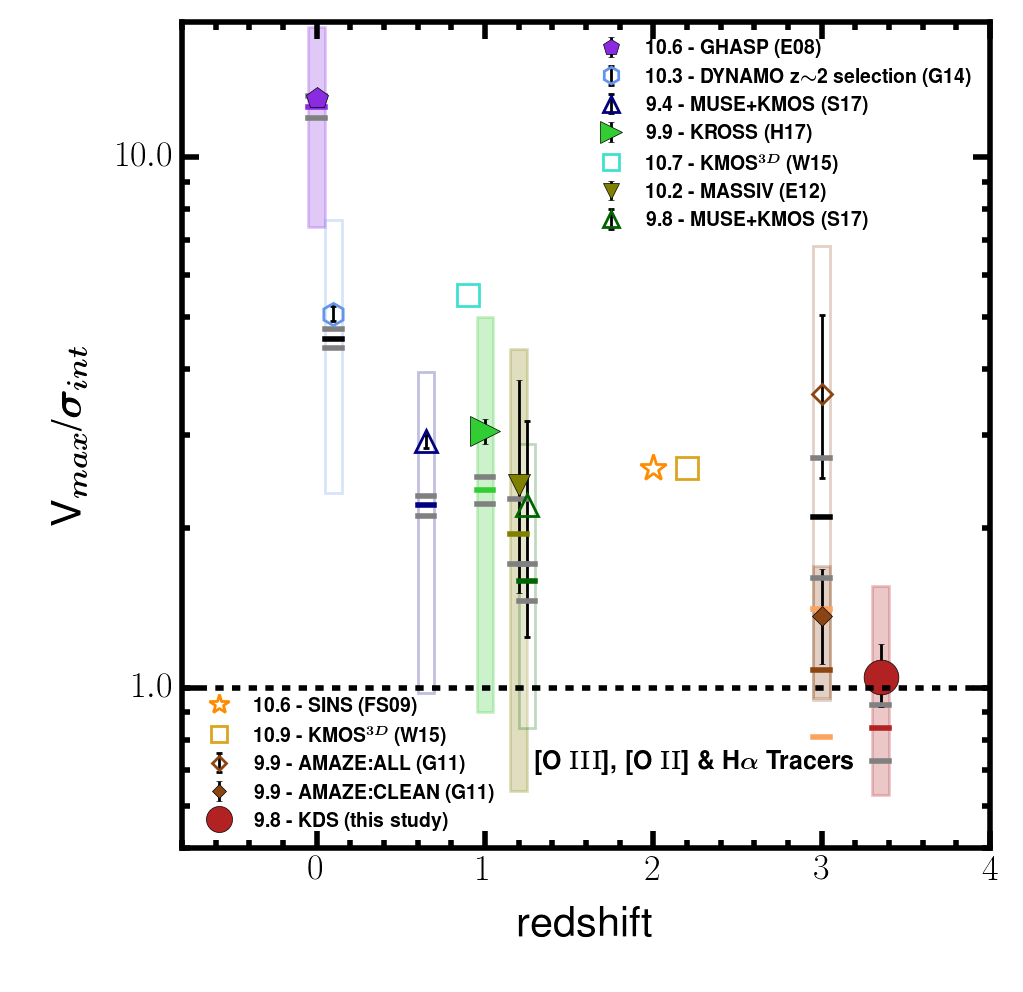}
    \end{subfigure} \hspace{+0.4cm}
    \begin{subfigure}[h!]{0.5\textwidth}
        \centering
        \includegraphics[height=3.5in]{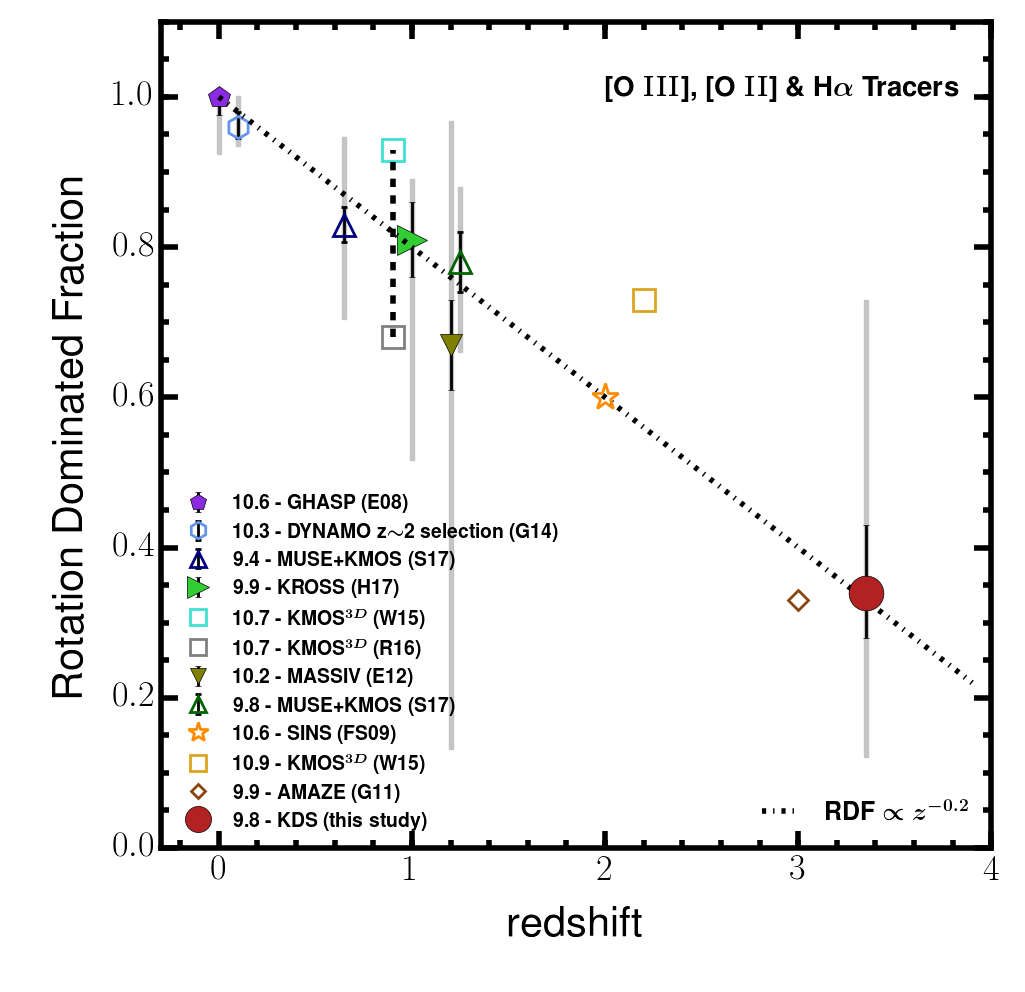}
    \end{subfigure}
    \caption{{\it Left:} $V_{C}/\sigma_{int}$ is plotted against redshift for the comparison samples spanning $0 < z < 4$.
    We see a clear decline in the mean $V_{C}/\sigma_{int}$ values with redshift, but with wide ranges in the individual measurements.
    There is a connection between $V_{C}/\sigma_{int}$ and the stellar mass of the surveys, with larger mean $M_{\star}$ surveys reporting higher mean $V_{C}/\sigma_{int}$.
    {\it Right:} We plot the associated rotation-dominated fraction, defined as the fraction of galaxies in each sample with $V_{C}/\sigma_{int} > 1$.
    The narrow shaded regions represent the maximum and miminum RDF, computed as described in Appendix \protect\ref{app:comparison_samples}.
    It appears that the rotation-dominated fraction drops from $\simeq100\%$ in the local universe to $60\%$ at intermediate redshifts and then to $\simeq35\%$ at z $\simeq3.5$, again with the trend for high $V_{C}/\sigma_{int}$ values in the larger $M_{\star}$ surveys reflected by higher rotation-dominated fractions.
    The dot-dash line is the RDF $\propto z^{-0.2}$ cosmic decline suggested in \protect\cite{Stott2016}, which appears to follow the redshift decline of the rotation dominated fraction.
    In both panels, the symbol convention is the same as in the right panel of Figure \protect\ref{fig:tf_relation}.
}
    \label{fig:rdf_and_v_sigma_w_redshift}
\end{figure*}

In Figure \ref{fig:v_sig_and_v} we plot the ratio of $V_{C}$/$\sigma_{int}$ against $M_{\star}$ for the isolated field sample, also including galaxies from the AMAZE Clean ($z\simeq3$) sample for reference.
In the local universe, disky, SFGs are observed to have $V_{C}/\sigma_{int}$ values in excess of 10 (i.e. the disks are `well settled'), due to the decline of velocity dispersion across cosmic time, as gas fractions and disk turbulence decrease \citep[e.g.][]{Epinat2008a,Epinat2008}.
Clearly the isolated field sample galaxies are in a different physical regime, with mean and median $V_{C}/\sigma_{int}$ values of $1.08^{+0.18}_{-0.15}$ and $0.97^{+0.14}_{-0.11}$ respectively.
This is despite many of the dispersion-dominated galaxies showing clear velocity gradients across the observed velocity maps (Appendix \ref{app:kinematics_plots}), with the arctangent model providing a good fit to the rotation curves.
This suggests that although rotational motions are present within the disks, they are small or comparable with the elevated $\sigma_{int}$ values observed throughout the $z\simeq3.5$ sample. \\

\noindent
The $V_{C}$/$\sigma_{int}=1$ line dividing rotation and dispersion-dominated galaxies is marked on Figure \ref{fig:v_sig_and_v}, separating the dispersion and rotation-dominated galaxies by definition.
There is a spread of rotation and dispersion-dominated galaxies across the stellar mass range, with rotation-dominated galaxies more likely to be found above $log(M_{\star}/M_{\odot}) > 9.5$.
The Rotation-Dominated Fraction (RDF), calculated with respect to the full field sample of 38 galaxies, is 0.34$\pm0.08$, which is substantially lower than in the local and intermediate redshift universe \citep[e.g.][]{Epinat2008a,Epinat2008,Green2014}.
We interpret these results in the context of the cosmic evolution of the rotation dominated fraction using the comparison samples in \cref{subsubsec:RDF_evolution}.\\

\subsubsection{The evolution of the rotation-dominated fraction}\label{subsubsec:RDF_evolution}

In the left and right panels of Figure \ref{fig:rdf_and_v_sigma_w_redshift} we plot $V_{C}/\sigma_{int}$ and the closely connected rotation-dominated fraction respectively as a function of redshift for the KDS isolated field sample and the comparison samples described in Appendix \ref{app:comparison_samples}.
In each sample, the rotation dominated fraction is computed as the fraction of galaxies with $V_{C}/\sigma_{int} > 1$, which, given the varying data quality across the comparison samples, is a simple and fair way to study the redshift evolution of this quantity. 
The filled and hollow symbols with black errorbars in the left panel of Figure \ref{fig:rdf_and_v_sigma_w_redshift} represent the mean and error on the mean, and the horizontal lines show the median and error on the median.
The shaded regions give the 16th and 84th percentiles of the distribution of individual $V_{C}/\sigma_{int}$ measurements, plotted here to indicate the range of measurements at each redshift slice.
In the right panel of Figure \ref{fig:rdf_and_v_sigma_w_redshift}, we have also computed the maximum and minimum RDF for these surveys by adding and subtracting the $V_{C}/\sigma_{int}$ error values respectively and recomputing the rotation-dominated fraction.
These are shown with the thin grey shaded regions and are intended to give an indication of the limits of the RDF at different redshift slices given the size of the errors on the individual points.\\

\noindent
The rotation-dominated fraction, traced simply by the $V_{C}/\sigma_{int}$ ratio inferred from ionised gas emission lines, drops from $\simeq100\%$ of galaxies in the local universe to $\simeq2/3$ of galaxies in the $z\simeq1-2$ universe and to $\simeq1/3$ above $z\simeq3$, albeit with individual surveys reporting wide $V_{C}/\sigma_{int}$ distributions at each redshift slice.
This is highlighted in the right panel of Figure \ref{fig:rdf_and_v_sigma_w_redshift} by overplotting the RDF $\propto z^{-0.2}$ line which appears to roughly follow the decline of the rotation-dominated with redshift as described in \cite{Stott2016}. 
The width of the $V_{C}/\sigma_{int}$ distributions indicate the galaxy diversity at each redshift slice, driven by the collection of galaxy masses, sSFRs, gas fractions, morphologies and evolutionary states which comprise each sample.
Scatter above and below the RDF $\propto z^{-0.2}$ evolution line may be attributed in part to the different mean stellar masses of the surveys, with those that have larger stellar masses tending to scatter above this line.
This draws again from the idea that higher stellar mass values suggest that the galaxies are more evolved, with more gas having been converted into stars for the bulk of the sample, which provides stability for a rotating disk and pushes the rotation-dominated fraction higher (see \cref{sec:discussion}). \\\\
\noindent
In summary, the rotation-dominated galaxies in the isolated field sample appear to conform with a redshift invariant $V_{C}$ - $M_{\star}$ relation.
Due to increased velocity dispersions, these rotation-dominated galaxies account for only $34\%$ of the field sample and so are less representative of typical star-forming galaxies at $z\simeq3.5$ than at intermediate and low redshifts. 
Explanations for the increase of $\sigma_{int}$ with redshift (and hence the decline of $V_{C}/\sigma_{int}$) with cosmic time as well as the nature of dispersion-dominated galaxies have been discussed at length throughout other studies \citep[e.g][]{ForsterSchreiber2009,Law2009,Burkert2010,Newman2013,Wisnioski2015} and we revisit these in the following discussion.

\section{DISCUSSION}\label{sec:discussion}

Throughout \cref{sec:results} we have investigated the derived kinematic parameters of the isolated field sample, showing first that, within the scatter, the rotation-dominated galaxies sit on a rotation velocity - stellar mass relationship consistent with what is observed in the local universe.
The mean rotation velocity of the isolated field sample is low as a result of averaging over a sample containing a large number of dispersion-dominated galaxies ($66\%$ of the sample) and probable pressure support from turbulent motions \citep[e.g.][]{Burkert2010}.
This increased number of dispersion-dominated galaxies in comparison to low and intermediate redshifts appears to be driven by high velocity dispersions at $z\simeq3.5$.
In this section we discuss possible physical origins for the elevated random motions traced by $\sigma_{int}$ and their impact on the galactic dynamics.

\subsection{The origin and impact of increased velocity dispersions at high-redshift}\label{discussion_increase_in_sigma}

\begin{figure*}
    \centering \hspace{-1.3cm}
    \begin{subfigure}[h!]{0.5\textwidth}
        \centering
        \includegraphics[height=3.5in]{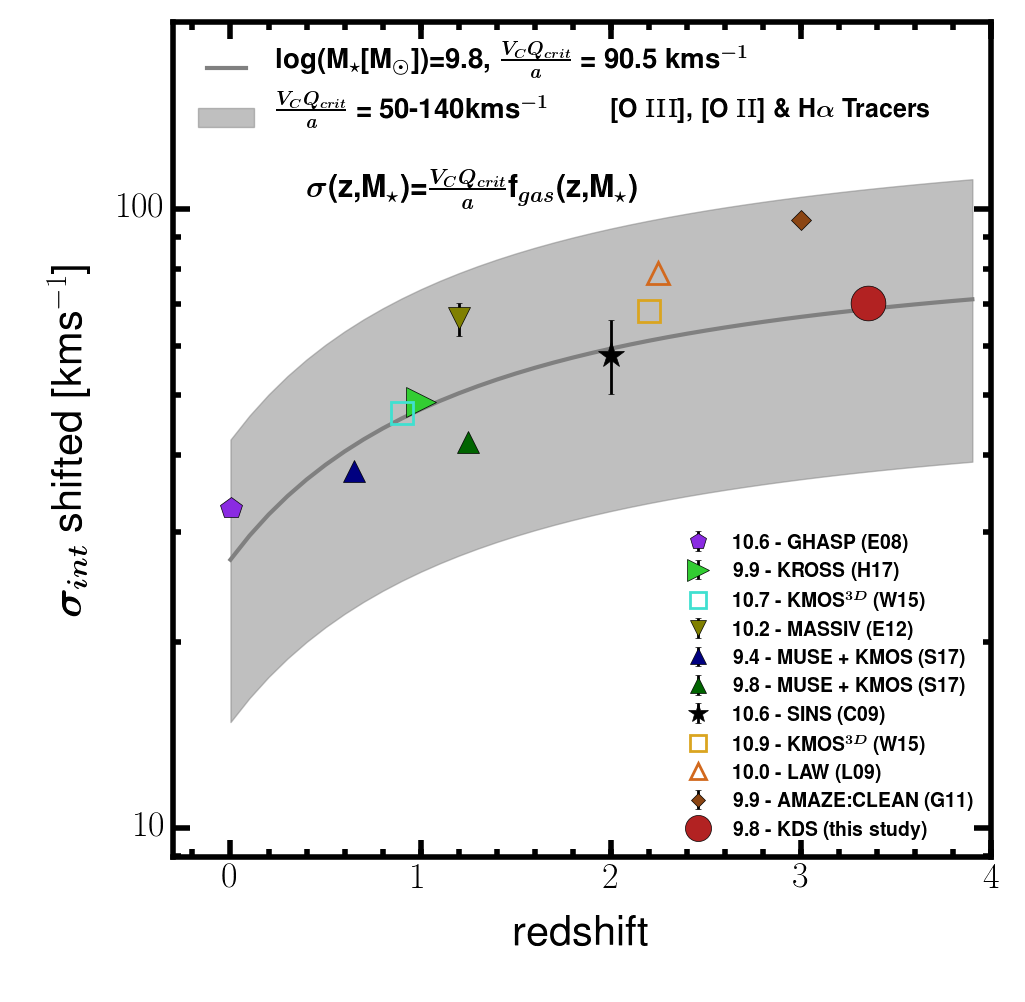}
    \end{subfigure} \hspace{+0.4cm}
    \begin{subfigure}[h!]{0.5\textwidth}
        \centering
        \includegraphics[height=3.5in]{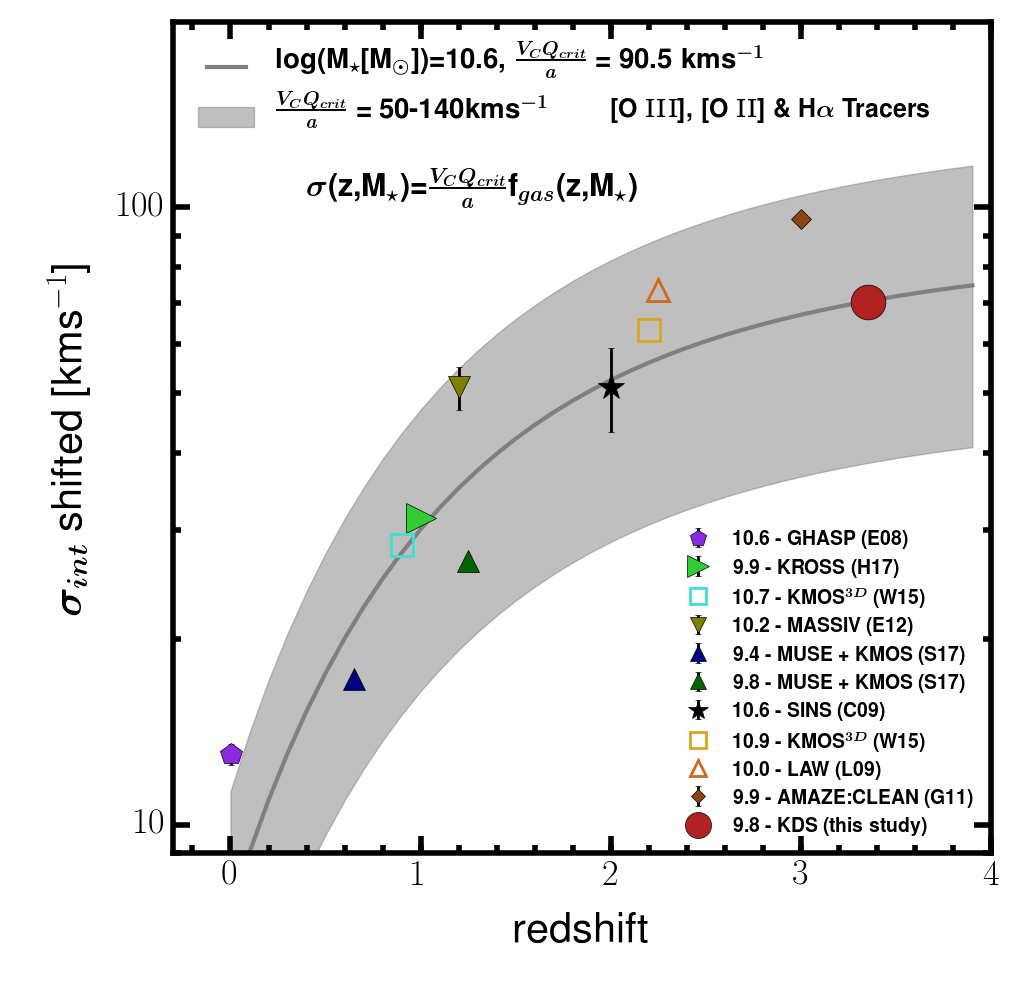}
    \end{subfigure}
    \caption{{\it Left:} We plot the $\sigma_{int}$ values for the comparison samples as well as the isolated field sample, shifted as described in the text to a reference mass of $log(M_{\star}/M_{\odot}) = 9.8$ (the median mass of the isolated field sample).
    The grey line shows the model prediction with $V_{C}Q_{crit}/a = 90.5km\,s^{-1}$, as dictated by our data.
    The grey shaded region encompasses the model predictions with lower and upper bounds using $V_{C}Q_{crit}/a = 50-140km\,s^{-1}$ respectively (corresponding roughly to the velocity range $V_{C} = 70-200km\,s^{-1}$, spanned by galaxies in the comparison samples, under the assumption that $Q_{crit} = 1.0$ and {\it a} = $\sqrt{2}$ as discussed in the text).
    The shifted points are in line with the scenario whereby the sample averaged velocity dispersions increase with redshift as a result of increasing average gas fractions. 
    {\it Right:} The same as in the left panel for a reference mass of $log(M_{\star}/M_{\odot}) = 10.6$.
    The steeper slope beyond $z\simeq2$ highlights the model decrease in gas fraction, and hence velocity dispersion, for galaxies which have accumulated a larger stellar population.}
    \label{fig:sigma_shifted}
\end{figure*}

Previous studies have claimed that the elevated velocity dispersions observed at high-redshift are a consequence of both internal and external processes.
\cite{Law2009} observe galaxies with elevated velocity dispersions at $z\simeq2.2$ (see Figure \ref{fig:sigma_and_v_sigma_w_redshift}) and attribute this to gravitational instabilities following efficient accretion, which can occur when the cold gas becomes dynamically dominant.

\cite{Genzel2011} show that for dispersion-dominated galaxies, the scales susceptible to gravitational collapse are of the order of the galactic half-light radius. 
Systems such as these form highly unstable and roughly spherical stellar distributions with large internal motions at early times, before more gradual accretion from the galactic halo forms an extended, stable, gaseous disk which instigates kinematic order \citep{Law2009,Genzel2011}.
The physical mechanisms responsible for the increase in $\sigma_{int}$ may also result in a reduction in the observed rotation velocities by partially compensating the gravitational force throughout the galactic disk \citep[e.g.][]{Burkert2010,Wuyts2016b,Ubler2017,Genzel2017,Lang2017}.
Following this, $\sigma_{int}$ decreases over time as ordered rotation begins to dominate and pressure support becomes increasingly insignificant \citep[e.g.][]{Burkert2010,Ubler2017,Genzel2017,Lang2017}.
This is in line with the evolution of gas fractions observed at intermediate and high-redshifts \citep[e.g.][]{Saintonge2013,Tacconi2013,Tacconi2017}, with the gas reservoirs of galaxies fed by the accretion of cold gas.

\noindent
\cite{Burkert2010} show, starting from the hydrostatic equation, that negative pressure gradients sourced by the disk turbulence generated through gravitational instabilities result in a decrease in observed rotation velocities. 
This has prompted others \citep[e.g.][]{Wuyts2016b,Ubler2017,Lang2017} to encorporate a velocity dispersion term in the derivation of maximum circular velocities, which contributes to (but is not entirely responsible for) the observed shift in the zero-point of the stellar mass Tully-Fisher relation between $z\simeq0-0.9$.
\cite{Ubler2017} also claim that without the inclusion of this term incorrect conclusions surrounding the evolution of the Tully-Fisher relation will be drawn.
We have not encorporated such a term when deriving the rotation velocities of the isolated field sample, but note that as a consequence of our galaxies showing some of the most extreme intrinsic dispersions observed, this would undoubtedly lead to a different distribution in the $V_{C}-M_{\star}$ plane.
The contribution of pressure to rotation velocity for the KDS sample will be explored in more detail in a future work. 
\\

\noindent
In \cite{Wisnioski2015} the authors present a scaling relation describing the increase of $\sigma_{int}$ with redshift in terms of the key observable properties: gas fraction, $f_{gas}$, gas depletion timescale, $t_{dep}$, and specific star-formation rate, sSFR.
Here we explore how well this model describes the evolution in velocity dispersion, incorporating our new measurement at $z\simeq3.5$ from the KDS.
The function describing the evolution of the velocity dispersion with redshift is shown below in Equation \ref{eq:sigma_increase}, where $Q_{crit}$ is the critical Toomre parameter for stability against gravitational collapse \citep{Toomre1964}, and {\it a} is a dimensionless parameter which depends on the assumed distribution of gas and gravitational potential \cite{ForsterSchreiber2006}. 

\begin{equation}\label{eq:sigma_increase}
    \sigma(z,M_{\star}) = \frac{V_{C}Q_{crit}}{{\it a}}f_{gas}(z,M_{\star})
\end{equation}

\noindent
The gas fraction is a function of gas depletion timescale, with an assumed redshift dependence, and specific star-formation rate, with an assumed mass and redshift dependence (see \citealt{Wisnioski2015} Equations 3-6, summarised below by Equations \ref{eq:gas_fractions} and \ref{eq:coefficients}).

\begin{equation}\label{eq:gas_fractions}
\begin{aligned}
    f_{gas} & = \frac{1}{1 + \left(t_{dep}sSFR\right)^{-1}} \\[1ex]
    & = \frac{1}{1 + \left(1.5\times10^{a(M_{\star})}(1 + z)^{b(M_{\star}) + \alpha}\right)^{-1}}
\end{aligned}
\end{equation}

\begin{equation}\label{eq:coefficients}
\begin{aligned}
   & a(M_{\star}) = -10.73 + \frac{1.26}{1 + {\it e}^{\left(10.49 - log(M_{\star}/{\rm M}_{\odot})\right)/(-0.25)}} \\[1ex]
   & b(M_{\star}) = 1.85 + \frac{1.57}{1 + {\it e}^{\left(10.35 - log(M_{\star}/{\rm M}_{\odot})\right)/(0.19)}}
\end{aligned}
\end{equation}

\noindent
This relation is an expression of the belief that high-redshift galaxies experience more intense and efficient gas inflow, with the accumulation of a more massive stellar population providing stability against perturbations.
In this equilibrium model framework \citep[e.g.][]{Dave2012,Lilly2013,Saintonge2013}, populations of disk galaxies with higher average stellar mass have lower average velocity dispersions due to this stability condition.

In the right panel of Figure \ref{fig:sigma_and_v_sigma_w_redshift} we plot Equation \ref{eq:sigma_increase} for three different input mass values ($log(M_{\star}/M_{\odot})=9.8,10.3,10.6$, roughly spanning the mass range of the comparison samples), using the updated depletion time scaling relation from \cite{Tacconi2017} (i.e. $t_{dep} \propto (1 + z)^{-0.57}$, meaning $\alpha=0.57$ in Equation \ref{eq:gas_fractions}).
The Equation $\sigma(z,M_{\star})\propto f_{gas}(z,M_{\star})$ is scaled so that the lowest mass track conincides with the KDS datapoint, finding a proportionality constant $V_{C}Q_{crit}/a = 90.5km\,s^{-1}$, which we assume when plotting all three tracks in Figure \ref{fig:sigma_and_v_sigma_w_redshift}. \\

\noindent
In this model, the velocity dispersions increase more steeply with increasing stellar mass, as shown by the solid and dashed lines in Figure \ref{fig:sigma_and_v_sigma_w_redshift}.
At higher redshift, these lines converge, suggesting that the predictions for the gas fraction become independent of stellar mass in this regime.
In general, the surveys with higher stellar mass have lower velocity dispersions, in support of the model hypothesis.

Since stellar mass appears to be a crucial factor in regulating the velocity dispersions within galaxies, and the comparison samples have disparate stellar mass ranges, we use the model to shift the $\sigma_{int}$ values according to their stellar mass to a reference mass value as per \cite{Wisnioski2015}.
This is done by computing the difference between the model predicted $\sigma_{int}$ value at the stellar mass of each comparison survey, and at the reference mass for the appropriate redshift, and shifting the reported $\sigma_{int}$ by this value.
We do this for two reference mass values; $log(M_{\star}/M_{\odot}) = 9.8$ (the median mass of the isolated field sample, left panel of Figure \ref{fig:sigma_shifted}) and $log(M_{\star}/M_{\odot}) = 10.6$ (right panel of Figure \ref{fig:sigma_shifted}).

In both panels, the solid grey line denotes the model prediction using the parameter combination determined above; $V_{C}Q_{crit}/a = 90.5km\,s^{-1}$.
The grey shaded regions encompass model predictions with lower and upper bounds defined by $V_{C}Q_{crit}/a = 50-140km\,s^{-1}$ respectively.
Assuming that $Q_{crit} = 1.0$ (as for a quasi-stable thin disk e.g. \citealt{ForsterSchreiber2006,Burkert2010}) and {\it a} = $\sqrt{2}$ (as for a disk with constant rotation velocity \citealt{Wisnioski2015}), this range corresponds to $V_{C} = 70-200km\,s^{-1}$, which is roughly the velocity range spanned by our comparison samples, and the solid grey line corresponds to $V_{C} = 128km\,s^{-1}$.
Whilst the KDS mean rotation velocity sits at the bottom of this range, possible explanations are: \textbf{(1)} The rotation velocities of the KDS galaxies are underestimated as a result of the velocity dispersion contribution discussed above; \textbf{(2)} $Q_{crit} \ne 1$ or ${\it a} \ne \sqrt{2}$ as a consequence of changing physical conditions at high-redshift. \\

\noindent
Although the details of the coefficients of Equation \ref{eq:sigma_increase} are unclear (i.e. $V_{C}$ is not fixed at one value for the comparison samples and is itself linked to stellar mass, and both $Q_{crit}$ and {\it a} could vary with redshift), the shape of the velocity dispersion evolution appears to follow that of the gas fraction, as suggested by the model.
The difference between the slopes in the left and right panels highlights the role of stellar mass in the kinematic settling of galaxies between $z\simeq3$ and the local universe. 
The isolated field sample data, in combination with comparison sample results spanning $z=0-3$, thus suggest that the time evolution of the velocity dispersion measured from ionised gas emission lines is intimately linked to the consumption of gas and the accumulation of stars in a stable disk \citep[e.g.][]{Law2009,Law2012b,Law2012c,Wisnioski2015}.
Rotation-dominated galaxies become less representative of the typical star-forming population at these redshifts as a result of the increased velocity dispersions (shown in Figure \ref{fig:rdf_and_v_sigma_w_redshift}), constituting only $\simeq1/3$ of the isolated field sample at $z\simeq3.5$.
This suggests that we are probing more chaotic and unstable systems, subject to gravitational instabilities and collapse on the disk scale length \citep{Burkert2010,Genzel2011}, residing in a much more active period of the universe in terms of both accretion and star-formation.
Recent results indicate that these systems may become strongly baryon-dominated on the galaxy disk scale at high-redshift \citep{Ubler2017,Lang2017,Genzel2017}, which could imply that the increasing contribution of dark matter to galaxy disks with decreasing redshift plays a role in the evolution of velocity dispersions.   

The precise interpretation of the velocity dispersion remains unclear, although there is mounting evidence that random motions have an increasingly significant contribution to the mass budget of high-redshift galaxies \citep[e.g.][]{Kassin2007,Law2009,Burkert2010,Kassin2012,Wuyts2016b,Lang2017,Ubler2017,Genzel2017}.
The nature of dispersion-dominated galaxies at high-redshift has been discussed in \cite{Newman2013} using Adaptive-Optics (AO) and seeing limited observations of the same galaxies.
The AO observations show that, in some cases, galaxies which are clasified as dispersion-dominated with seeing limited observations reveal much larger velocity gradients at higher spatial resolution.
This highlights that beam-smearing effects can wash-out observed rotational motions, particularly for small galaxies.
We have attempted to account for beam-smearing using half-light radius and velocity gradient dependent dynamical modelling, but we acknowledge that these observations are challenging and cannot rule out the presence of unresolved velocity gradients in our seeing-limited dataset.
Smaller galaxies do have intrinsically smaller rotation velocities, whilst there is no observed dependence of velocity dispersion on galaxy size \citep[e.g.][]{Newman2013}.
Dispersion-dominated galaxies are therefore likely to be at an earlier evolutionary stage and unstable to gravitational instabilities which generate turbulence which is then maintained by the potential energy of the gas in the disk \citep{Burkert2010,Newman2013}.

Bringing the above points together, turbulent motions consume an increasingly significant fraction of the energy budget at high-redshift when typical star-forming galaxies are at earlier evolutionary stages.
The KDS isolated field sample presents an extension of these trends to higher redshift, where galaxies are even more likely to be dispersion-dominated, and pressure support may be more significant than explored at $z\simeq2.3$ \citep{Wuyts2016b,Ubler2017,Genzel2017,Lang2017}.    
However, the precise interpretation of $\sigma_{int}$ measured from ionised gas emission lines remains unclear, and future observations of both stellar and gaseous velocity dispersions in tandem with {\it JWST} will allow for unambigous investigations into the nature of gaseous velocity dispersions at high-redshift. \\

\noindent
In summary, the isolated field sample galaxies are diverse and kinematically immature, with $\simeq2/3$ of the sample being dominated by random motions which may contribute to supporting a portion of the dynamical mass \citep[e.g.][]{Kassin2007,Burkert2010,Kassin2012,Newman2013,Straatman2017,Ubler2017,Lang2017}.
In the context of the equilibrium model explored here, these galaxies will evolve gradually with redshift along the velocity-mass relationship into stable disk galaxies which have converted their gas reservoirs into stellar mass \citep[e.g.][]{Lilly2013,Tacconi2013,Wisnioski2015,Tacconi2017}. 
Future work at higher resolution and redshift will trace the extent to which gas-derived velocity dispersions trace the stellar dispersions and will investigate even earlier and more chaotic stages in the lifetimes of late-type galaxies.

\section{CONCLUSIONS}\label{sec:conclusion}
We have presented new dynamical measurements of 32 typical star-forming galaxies spanning both cluster and field environments at $z\simeq3.5$ as part of the KMOS Deep Survey, based on IFU data observed with KMOS.
These measurements push back the frontier of IFU observations in the early universe and provide more robust constraints on the internal and rotational dynamics of $9.0 < log(M_{\star}/M_{\odot})< 10.5$ typical star-forming galaxies at these redshifts.
By using a combined morpho-kinematic classification based on broad-band {\em HST} imaging and our IFU data, we have separated interacting galaxies from the sample, finding merger rates consistent between the field pointings ($\simeq16\%$) and a very high merger rate in the cluster pointing ($\simeq89\%$).
We have made beam-smearing corrected measurements of $V_{C}$ and $\sigma_{int}$ for the remaining isolated field galaxies, and we interpret these in the context of previous dynamical studies using IFU data (Appendix \ref{app:comparison_samples} and Table \ref{tab:evolution_numbers}).
The main conclusions of this work are summarised as follows:

\begin{itemize}
    \item We use a fractional error weighted mean to derive the mean kinematic parameters of the KDS isolated field sample of $V_{C} = 76.7^{+4.9}_{-4.5}km\,s^{-1}$ and $\sigma_{int} = 70.8^{+3.3}_{-3.1} km\,s^{-1}$ (right panels of Figures \ref{fig:tf_relation} and \ref{fig:sigma_and_v_sigma_w_redshift} respectively).
    Rotation (dispersion) dominated galaxies are defined as those with $V_{C}/\sigma_{int} > (<) 1$.  
    \item We plot the mean $V_{C}$ values of the comparison samples and the KDS isolated field sample, finding significant diversity in $V_{C}$ measurements in each survey (right panel of Figure \ref{fig:tf_relation}).
    The spread in measurements is a result of the samples containing a mixture of rotation and dispersion-dominated galaxies with varying $M_{\star}$ values, and the low mean KDS value is primarily the result of averaging over a sample with a high fraction of dispersion-dominated galaxies and indicating the impact pressure support may have on reducing rotation velocities.
    At higher $M_{\star}$, the accumulation of a stellar population provides stability, leading to an interplay between $\sigma_{int}$ and $V_{C}$ mediated across redshift by the cosmic decline of gas fractions, sSFRs and accretion \citep[e.g.][]{Law2012b,Wisnioski2015}.
    \item When the $V_{C}$ values are viewed as a function of mass in the inverse stellar mass Tully-Fisher Relation, the rotation-dominated galaxies are correlated with mass and lie within the errors on the same relation as derived for rotation-dominated galaxies with $9.0 < log(M_{\star}/M_{\odot}) < 11.0$ at $z\simeq0.9$ in the KROSS survey (Figure \ref{fig:tf_relation}). 
    This relation is consistent with star-forming galaxies at $z=0$ and so we report no significant evolution in the slope or zero-point of the smTFR between $z=0-3.5$.
    Consistency of sample selection and measurement techniques are crucial factors when determining evolution in the $V_{C}$ vs. $M_{\star}$ plane by comparing samples at different redshifts.
    The dispersion-dominated galaxies in the isolated field sample scatter below the trend.
    \item When considering the rotation-dominated galaxies alone, the mean and median rotation velocities are $V_{C} = 96.7^{+7.3}_{-7.2}km\,s^{-1}$ and $V_{C} = 93.0^{+12.0}_{-12.5}km\,s^{-1}$ respectively, which are roughly equivalent to a simple lower limit virial theorem calculation using the mean mass and half-light radius of the sample.
    This suggests that pressure forces, which are more significant at high-redshift, may play a role in supporting the total mass in all of the galaxies in the isolated field sample \citep[e.g.][]{Burkert2010}.  
    \item 13/32 isolated field sample galaxies are rotation-dominated and 19/32 are dispersion-dominated, with a tendency for galaxies with higher $M_{\star}$ to have larger values of $V_{C}/\sigma_{int}$ (Figure \ref{fig:v_sig_and_v}).
    This gives a rotation-dominated fraction of $34 \pm 8 \%$ in the isolated field sample, substantially less than surveys at lower redshift, although there is significant diversity amongst the individual $V_{C}/\sigma_{int}$ measurements at each redshift slice (left and right panels of Figure \ref{fig:rdf_and_v_sigma_w_redshift}).
    When viewed as a function of redshift, the rotation dominated fraction appears to follow the scaling relation RDF$\propto z^{-0.2}$.
    \item We plot mean $\sigma_{int}$ values computed in a consistent way for SFG samples spanning a wide redshift baseline (the comparison samples described in Appendix \ref{app:comparison_samples}), finding a sharp increase in $\sigma_{int}$ values between $z=0-1$ and a fairly shallow increase thereafter, mediated by the mean $M_{\star}$ of the galaxy samples (right panel of Figure \ref{fig:sigma_and_v_sigma_w_redshift}).
    This is in line with a simple equilibrium model prescription in which the gas fractions and the impact of several physical mechanisms such as accretion of gas from the IGM, stellar feedback and turbulence increase with redshift and combine to increase random motions within high-redshift galaxies.

\end{itemize}

\section*{Acknowledgements}

OJT acknowledges the financial support of the Science and Technology Facilities Council through a studentship award. 
MC and OJT acknowledge the KMOS team and all the personnel of the European Southern Observatory Very Large Telescope for outstanding support during the KMOS GTO observations.
CMH, AMS and RMS acknowledge the Science and Technology Facilities Council through grant code ST/L00075X/1.
RJM acknowledges the support of the European Research Council via the award of a Consolidator Grant (PI: McLure).
JSD acknowledges the support of the European Research Council via the award of an Advanced Grant (PI J. Dunlop), and the contribution of the EC FP7 SPACE project ASTRODEEP (Ref.No: 312725).
AMS acknowledges the Leverhulme Foundation.
JM acknowledges the support of a Huygens PhD fellowship from Leiden University. DS acknowledges financial support from the Netherlands Organization for Scientific research (NWO) through a Veni fellowship and from FCT through an FCT Investigator Starting Grant and Start-up Grant (IF/01154/2012/CP0189/CT0010).
This work is based on observations taken by the CANDELS Multi-Cycle Treasury Program with the NASA/ESA {\em HST}, which is operated by the Association of Universities for Research in Astronomy, Inc., under NASA contract NAS5-26555.
This work is based on observations taken by the 3D-{\em HST} Treasury Program (GO 12177 and 12328) with the NASA/ESA {\em HST}, which is operated by the Association of Universities for Research in Astronomy, Inc., under NASA contract NAS5-26555.
Based on data obtained with the European Southern Observatory Very Large Telescope, Paranal, Chile, under Large Program 185.A-0791, and made available by the VUDS team at the CESAM data center, Laboratoire d'Astrophysique de Marseille, France.





\clearpage 
\bibliographystyle{mnras}
\bibliography{/Users/owenturner/Documents/PhD/KMOS/Latex/Bibtex/library.bib}
\clearpage



\appendix

\section{Comparison samples}\label{app:comparison_samples}

\begin{table*}
\centering
\begin{threeparttable}
\caption{The mean and median kinematic properties used throughout Figures \protect\ref{fig:tf_relation}, \protect\ref{fig:sigma_and_v_sigma_w_redshift} and \protect\ref{fig:rdf_and_v_sigma_w_redshift} for the different surveys.
The errors on the mean and median represent the statistical errors from bootstrap resampling.
For each of the comparison samples, the cross or tick after each of the mean kinematic properties indicates whether we have defined this as a fair value to compare against, with further details provided throughout Appendices \protect\ref{subsec:GHASP} - \protect\ref{subsec:AMAZE}.}
\label{tab:evolution_numbers}
\begin{tabular}{llllllllll}

 \hline
Survey & $\left< z \right> $ & $\left< log\left(\frac{M_{\star}}{M_{\odot}}\right)\right>$ & $\left<V_{C}\right>$ & med$\left(V_{C}\right)$ & $ \left< \sigma_{int} \right>$ & med$\left(\sigma_{int}\right)$ & $\left< \frac{V_{C}}{\sigma_{int}} \right>$ & med$\left(\frac{V_{C}}{\sigma_{int}}\right)$ & RDF  \\[1ex]
 \hline
 GHASP & 0.001 & 10.6   &       $189.0^{+3.5}_{-3.0}$ \cmark  & $159.4^{+12.0}_{-15.0}$ &  $13.0^{+0.5}_{-0.5}$ \cmark & $13.0^{+0.5}_{-0.5}$ & $12.9^{+0.5}_{-0.4}$ \cmark &           $12.5^{+0.6}_{-0.6}$&               $1.00^{+0.00}_{-0.02}$  \cmark   \\[1ex]
 DYNAMO & 0.1 & 10.3    &       $183.0^{+1.0}_{-1.0}$ \xmark  & $164.0^{+3.0}_{-5.0}$   &  $45.9^{+0.3}_{-0.3}$ \xmark & $39.0^{+0.9}_{-1.0}$ & $5.1^{+0.2}_{-0.2}$  \xmark &           $4.6^{+0.2}_{-0.2}$ &               $0.96^{+0.02}_{-0.02}$  \xmark   \\[1ex]
 MUSE+KMOS & 0.65 & 9.4 &       $103.8^{+1.5}_{-1.6}$ \xmark  & $73.0^{+2.5}_{-2.5}$    &  $40.0^{+0.3}_{-0.3}$ \cmark & $32.0^{+0.6}_{-0.5}$ & $2.9^{+0.1}_{-0.1}$  \xmark &           $2.21^{+0.1}_{-0.1}$&               $0.83^{+0.02}_{-0.02}$  \xmark   \\[1ex]
 $^{*}$KROSS & 0.9 & 9.9 &      $117.0^{+4.0}_{-4.0}$ \cmark  & $109.0^{+5.0}_{-5.0}$   &  -                    \cmark & -                    & $3.1^{+0.2}_{-0.2}$  \cmark &           $2.4^{+0.1}_{-0.1}$ &               $0.81^{+0.05}_{-0.05}$  \cmark   \\[1ex]
 KMOS$^{3D}$ & 1.0 & 10.7  &     170.0                        &  -                      &  25.0                 \xmark &  -                   & 5.5                  \xmark &            -                  &               0.93                    \xmark   \\[1ex]
 MASSIV & 1.2 & 10.2 &          $132.1^{+10.4}_{-8.2}$ \cmark & $103.0^{+13.4}_{-11.0}$ &  $61.8^{+3.8}_{-4.2}$ \cmark & $52.0^{+5.2}_{-4.7}$ & $2.4^{+1.4}_{-0.9}$  \cmark &           $2.0^{+0.3}_{-0.2}$ &               $0.67^{+0.06}_{-0.06}$  \cmark   \\[1ex]
 MUSE+KMOS & 1.25 & 9.8 &       $75.5^{+3.0}_{-3.3}$   \xmark & $54.0^{+3.0}_{-2.8}$    &  $42.0^{+0.5}_{-0.5}$ \cmark & $29.5^{+1.0}_{-1.0}$ & $2.2^{+1.0}_{-1.0}$  \xmark &           $1.59^{+0.1}_{-0.1}$&               $0.78^{+0.04}_{-0.04}$  \xmark   \\[1ex]
 SINS(C09) & 2.0 & 10.6 &       $232.0^{+12.8}_{-12.7}$\cmark & $240.0^{+18.0}_{-17.2}$ &  $51.2^{+8.0}_{-7.9}$ \cmark & $42.5^{+9.1}_{-8.5}$ & $5.0^{+0.9}_{-1.0}$  \xmark &           $4.7^{+1.0}_{-1.0}$ &               1.0                     \cmark   \\[1ex]
 SINS(F09) & 2.0 & 10.6 &       $201.3^{+4.3}_{-4.0}$  \cmark & $174.0^{+12.1}_{-10.3}$ &  -                    \xmark & -                    & 2.6                  \xmark &           -                   &               0.60                    \xmark   \\[1ex]
 KMOS$^{3D}$ & 2.2 & 10.9 &     170.0                  \xmark & -                       &  55.0                 \xmark & -                    & 2.6                  \xmark &           -                   &               0.73                    \xmark   \\[1ex]
 LAW 09 & 2.3 & 10.0 &          -                      \xmark & -                       &  $78.0^{+6.5}_{-6.5}$ \xmark & $69.0^{+8.3}_{-7.5}$ & -                    \xmark &           -                   &               0.73                    \xmark   \\[1ex]
 AMAZE (Full) & 3.0 & 10.0 &    $217.0^{+59.1}_{-40.2}$\xmark & $129.0^{+47.5}_{-42.8}$ &  $85.9^{+1.5}_{-1.4}$ \xmark & $78.0^{+4.5}_{-4.8}$ & $3.6^{+1.5}_{-1.1}$  \xmark &           $2.1^{+0.6}_{-0.5}$ &               0.33                    \xmark   \\[1ex]
 AMAZE (Clean) & 3.0 & 10.0 &   $140.8^{+40.1}_{-30.2}$\cmark & $129.0^{+45.0}_{-33.4}$ &  $95.7^{+2.0}_{-2.0}$ \cmark & $95.0^{+2.6}_{-2.6}$ & $1.4^{+0.3}_{-0.3}$  \cmark &           $1.1^{+0.3}_{-0.3}$ &               -                       \xmark   \\[1ex]
 \textbf{KDS} & 3.5 & 9.8 &     $75.7^{+4.4}_{-4.4}$   \cmark & $57.5^{+5.5}_{-5.5}$    &  $70.3^{+3.3}_{-3.1}$ \cmark & $71.0^{+5.0}_{-4.8}$ & $1.1^{+0.2}_{-0.1}$  \cmark &            $0.84^{+0.1}_{-0.1}$&               $0.37^{+0.08}_{-0.08}$ \cmark    \\[1ex]
 \hline
\end{tabular}
\begin{tablenotes}
      \small
      \item $^{*}$ The intrinsic velocity dispersion values for the KROSS sample will be presented in Johnson et al. {\it in prep.} 
    \end{tablenotes}
  \end{threeparttable}
  \end{table*}

In \cref{sec:results} we compare the isolated field sample results with the results of surveys tracing dynamics with ionised gas emission across a wide redshift baseline, to determine the evolving physical state of SFGs as the age of the universe increases.
The galaxy selection criteria in these surveys, with the exception of DYNAMO \citep{Green2014} as discussed below, consistently picks out representative star-forming galaxies at each epoch and the dynamical properties are traced by observing either the [O~{\sc III}]$\lambda$5007, [O~{\sc II}]$\lambda$3727,3729 or $H\,\alpha$ ionised gas emission lines.

Given the range of modelling and kinematic parameter extraction methods it is important to verify the extent to which the results from these surveys can be directly compared and treated as forming an evolutionary sequence, which we test in the following subsections by considering each survey in turn.

We make use of tabulated data from the surveys, where available, to compute sample averages using the following method. 
The fractional error weighted mean of $V_{C}/\sigma_{int}$, $\sigma_{int}$ and $V_{C}$ in each of the samples is computed (i.e. we do not want the derived values with extremely large errors to dominate the averages).
The errors on these mean values are computed in a statistical sense, by generating bootstrapped samples, with replacement, with size equivalent to the original survey sample size and with values perturbed by a random number drawn from a gaussian distribution with width given by the error on the original point.
The same process is applied to compute the errors on the sample medians and we report the 16th and 84th percentiles of the distributions of each of the above quantities as an indicator of the distribution width, and hence galaxy diversity, at each redshift slice.
These results are listed in Table \ref{tab:evolution_numbers} and discussed in detail throughout \cref{sec:results}, where we will make statements about dynamical evolution by connecting the dots of these different surveys, assuming that on average they are tracing a population of SFGs which evolve across cosmic time.
Throughout the plots in \cref{sec:results} we highlight which values can be directly compared, due to consistency of sample selection and measurement methods, using filled and hollow symbols.
This is also highlighted with the ticks and crosses beside the mean values in Table \ref{tab:evolution_numbers}. 

\subsection{GHASP ($\bmath{z\simeq0}$)}\label{subsec:GHASP}
The GHASP survey \citep[E08]{Epinat2008a,Epinat2008} makes use of Fabry-Perot observations of 203 spiral and irregular galaxies with median $log(M_{\star}/M_{\odot})=10.6$ in the local universe ($z\simeq0$) to produce 3D H$\,\alpha$ datacubes.
The data extend out to several half-light radii and in each case tilted ring models are fit to determine $V_{C}$ at large radii where the rotation curves have flattened, with $\sigma_{int}$ given as the dispersion in the rotation velocity across the whole field.
In this survey, the mean and median kinematic properties computed from the tabulated literature values are directly comparable to the KDS and we used filled symbols in the plots throughout \cref{sec:results} to highlight this.

\subsection{DYNAMO ($\bmath{z\simeq0.1}$, but selected with physical properties resembling $\bmath{z\simeq2}$)}\label{subsec:DYNAMO}
In the DYNAMO survey \citep[G14]{Green2014}, H$\,\alpha$ IFU data of 67 galaxies with median $log(M_{\star}/M_{\odot})=10.3$ at $z\simeq0.1$ are presented.
The selection criteria is such that half of the sample have high sSFRs representative of SFGs at $z\simeq2$, and are also found to have higher $f_{gas}$ values than locally, mimicking the physical conditions of $z\simeq2$ SFGs.
$V_{C}$ values are extracted at a fixed radius from an arctangent model fit to the velocity fields ($1.6R_{1/2}$ as measured in the r-band) and $\sigma_{int}$ is taken as the luminosity weighted average of the beam-smearing corrected dispersion field.
We make use of tabulated data from this survey, but use hollow symbols throughout the plots in \cref{sec:results} to highlight the difference in selection critera.

\subsection{MUSE and KMOS ($\bmath{z\simeq0.65}$ and $\bmath{z\simeq1.25}$)}\label{subsec:MUSE_and_KMOS}
In \cite{Swinbank2017} a collection of $\simeq400$ star-forming galaxies observed in [O~{\sc II}]$\lambda$3727,3729 with MUSE and $H\,\alpha$ emission with KMOS is described.
Of this parent sample, inclination corrected velocities and intrinsic velocity dispersions are derived for 179 galaxies, to which we make comparisons.
The observations span the redshift range $0.2 < z < 1.6$ and we split the data into two subsets with $z < 1.0$ (129 galaxies with mean redshift of $z = 0.65$, mean stellar mass of $log(M_{\star}/M_{\odot})=9.80$) and $z > 1.0$ (50 galaxies with mean redshift of $z = 1.25$, mean stellar mass $log(M_{\star}/M_{\odot})=9.80$).
The reported velocities are extracted at $3R_{D}$, comparable to the extraction radius for the KDS sample, but these have not been beam-smearing corrected.
The intrinsic velocity dispersions have been corrected for beam-smearing effects.
For this reason, the velocity dispersions are deemed directly comparable to the KDS, reflected by the filled symbols throughout the plots in \cref{sec:results}, but the points involving velocity measurements are represented by hollow symbols due to the lack of a beam-smearing correction.

\subsection{KROSS ($\bmath{z\simeq0.9}$)}\label{subsec:KROSS}
We make comparisons to the most recent KROSS results \citep[H17]{Harrison2017}, which presents KMOS $H\,\alpha$ observations of $\simeq600$ SFGs with median $log(M_{\star}/M_{\odot})=9.9$ and $z\simeq0.9$.
$V_{C}$ values are extracted from exponential disk fits to the data at $2R_{1/2}$ (differing from arctangent fits by a median of 0.5\%) with the $\sigma_{int}$ values also extracted at $2R_{1/2}$ when the data extend to this radius and when they do not $\sigma_{int}$ is taken to be the median of the $\sigma_{obs}$ map.
These quantities are then beam-smearing corrected as described in H17 section 3.3.3. 
Due to the consistency of modelling and measurement techniques, the $z\simeq0.9$ galaxies in this sample are directly comparable to the KDS and are plotted throughout \cref{sec:results} with filled symbols.

\subsection{KMOS$^{3D}$ ($\bmath{z\simeq1.0}$ and $\bmath{z\simeq2.3}$)}\label{subsec:kmos_3d}
The KMOS$^{3D}$ results described in \cite[W15]{Wisnioski2015} describe KMOS H$\,\alpha$ observations of $\simeq191$ massive SFGs across two redshift slices. 
At $z\simeq1.0$ the galaxies have $log(M_{\star}/M_{\odot})=10.7$ and at $z\simeq2.3$ the median is $log(M_{\star}/M_{\odot})=10.9$, over an order-of-magnitude above the KDS median of $log(M_{\star}/M_{\odot})=9.8$. The rotation velocity is taken as $V_{C} = \frac{1}{2sin(i)}(v_{obs_{max}} - v_{obs_{min}})$ and $\sigma_{int}$ is extracted from the $\sigma_{obs}$ map far from the kinematic centre, where the effects of beam-smearing are negligible.
No tabulated values are provided for this survey, and so we cannot re-compute the mean and median values.
We make use of values quoted in W15 at both redshift intervals throughout \cref{sec:results}, plotting these with hollow symbols due to the inconsistency of measurement method.

\subsection{MASSIV ($\bmath{z\simeq1.2}$)}\label{subsec:MASSIV}
The MASSIV sample \citep[E12]{Epinat2012} uses SINFONI H$\,\alpha$ observations of 50 SFGs with median $log(M_{\star}/M_{\odot})=10.2$ at $z\simeq1.2$.
An arctangent function is fit to the data and $V_{C}$ is extracted at $\simeq1.7R_{1/2}$ and the $\sigma_{int}$ value is derived from the $\sigma_{obs}$ map by subtracting in quadrature the beam-smearing correction value as per \cref{subsubsec:beam_smearing_corrected_dispersions}.
We make use of tabulated values presented in this survey to compute the mean and median kinematic properties, plotted with filled symbols throughout \cref{sec:results}.

\subsection{SINS ($\bmath{z\simeq2.0}$)}\label{subsec:SINS}
The SINS survey \citep[FS09]{ForsterSchreiber2009} presents SINFONI $H\,\alpha$ observations of 80 massive galaxies with median $log(M_{\star}/M_{\odot})=10.6$ at $z\simeq2.0$.
Galaxies in this sample are classified as rotation or dispersion-dominated following FS09 section 9.5.1 which classifies rotation-dominated galaxies using the observed velocity and integrated velocity dispersion criteria $V_{obs}/(2\sigma_{tot}) > 0.4$. \\

\noindent
$V_{C}$ values for rotation-dominated galaxies are computed using a combined velocity gradient + width approach \citep{ForsterSchreiber2006} and for dispersion-dominated galaxies using the velocity width.
Only integrated velocity dispersions are presented for the full sample.
In addition, \citep[C09]{Cresci2009} model robust rotators in the SINS sample using the IDL code DYSMAL, which derives rotation curves given an input radial mass distribution.
In this approach, the $V_{C}$ value comes from the best-fitting model parameter and $\sigma_{int}$ is calculated using Equation \ref{eq:sins_sigma}, with the best-fitting $\sigma_{01}$ used (reflecting thin and thick disks; see their text for more detail) and with $\sigma_{02}$, an additional component of isotropic velocity dispersion throughout the disk, left as a free parameter in the fitting.

\begin{equation}\label{eq:sins_sigma}
\begin{split}
\sigma_{01} = \sqrt{\frac{v^{2}(R)h_{z}}{R}} \quad OR \quad \sigma_{01} = \frac{v(R)h_{z}}{R} \\
& \sigma_{int} = \sqrt{\sigma_{01}^{2} + \sigma_{02}^{2}}
\end{split}
\end{equation}

\noindent
It is unclear whether the $V_{C}$ values for the full sample from FS09 are directly comparable, but we make use of the tabulated values for completeness and plot these with filled symbols throughout \cref{sec:results}.
The velocity dispersion values, $V_{C}/\sigma_{int}$ and the rotation dominated fraction from FS09 are plotted with hollow symbols throughout \cref{sec:results} (using the results listed in W15 for the point locations) due to the difference in measurement method for the velocity dispersion values.

The $V_{C}$ and $\sigma_{int}$ values in C09 are likely biased towards rotation-dominated galaxies with a well-settled disk.
Nonetheless, we plot these values with filled symbols in \cref{sec:results} as a directly comparable dataset, omitting the $V_{C}/\sigma_{int}$ values and the rotation dominated fraction, which are listed for the full FS09 sample.

\subsection{LAW 09 ($\bmath{z\simeq2.3}$)}\label{subsec:law_09}
In \cite[L09]{Law2009} OSIRIS [O~{\sc III}]$\lambda$5007 and H$\,\alpha$ observations are collected for 13 galaxies with median $log(M_{\star}/M_{\odot})=10.0$ at $z\simeq2.3$.
The `velocity shear' is computed as $v_{shear} = \frac{1}{2}(v_{max} - v_{min})$ without inclination correction and $\sigma_{int}$ is the flux-weighted mean of $\sigma_{obs}$.
We note that the $\sigma_{mean}$ values and errors tabulated in the \cite{Law2009} paper are the flux-weighted mean and standard deviation of the measurements in individual spaxels in each galaxy, not corrected for beam-smearing in any way.
These measurements are expected to be larger than the $\sigma_{int}$ measurements at similar redshift which have been beam-smearing corrected, or extracted at the galaxy outskirts where the effects of beam-smearing are much smaller.
We use tabulated data $\sigma_{int}$ data in L09 to compare this distribution with other results, but plot with a hollow symbol due to the lack of a beam-smearing correction.

\subsection{AMAZE ($\bmath{z\simeq3.0}$)}\label{subsec:AMAZE}
The AMAZE sample \cite[G11]{Gnerucci2011} presents SINFONI [O~{\sc III}]$\lambda$5007 measurements for 33 galaxies with median $log(M_{\star}/M_{\odot})=9.9$ at $z\simeq3$, closest in redshift to the KDS galaxies.
In this study, rotation curves and intrinsic velocity dispersions are derived from a modelled exponential mass distribution, with the extracted $V_{C}$ value taken as the large radius limit of the rotation curve and the $\sigma_{int}$ as the maximum of the difference in quadrature between the $\sigma_{obs}$ map and the $\sigma_{model}$ map (which also takes into account instrumental resolution and beam-smearing; see their Equation 8).
A caveat when comparing with the G11 values is that dynamical properties are computed for 11 rotation-dominated galaxies in their sample, with the remaining 22 galaxies not analysed. \\

\noindent
Also, for 5 galaxies in G11, $V_{C}$ is not well constrained and no errorbar is given; for these objects we take the fractional error on $V_{C}$ equal to 1 when computing the sample averages.
The $\sigma_{int}$ errors for individual galaxies are generally very small, and for 3 galaxies the $\sigma_{int}$ value is consistent with 0 km\,s$^{-1}$.
For the mean value computation for the full sample we set these equal to the SINFONI resolution limit of 30 km\,s$^{-1}$.
A total of 6/11 galaxies either have no errors on $V_{C}$ or have $\sigma_{int}$ values consistent with 0 km\,s$^{-1}$, which we refer to as the AMAZE `Unconstrained' sample and represent these with hollow symbols throughout the plots in \cref{sec:results}.

\noindent
This Unconstrained sample also contains the galaxy s\_sa22a-M38, common between the KDS and G11, classified here as a merger due to double {\em HST} components and twin peaks in the object spectrum, but classified in G11 as a rotating galaxy with $V_{C} = 346 km\,s^{-1}$ (although the authors discuss the possibility that the galaxy could be either a close pair or two clumps embedded within a rotating disk).
We use tabulated values from the full analysed sample of 11 galaxies, plotting these with hollow symbols throughout \cref{sec:results} as a consequence of the caveats listed above.
The 5/11 galaxies which are not in the Unconstrained sample are referred to as the AMAZE `Clean sample' and are directly comparable to the KDS.
These are plotted with filled symbols throughout \cref{sec:results}.
We plot the quoted rotation-dominated fraction of 33$\%$ as a hollow symbol, noting that this could be higher if any of the 22/33 galaxies not analysed are rotation-dominated. \\

\noindent
This comparison sample summary shows that there are many different approaches for computing the same intrinsic kinematic parameters, dictated by data quality and model preference, however, despite this diversity, over the past decade the studies appear to be converging on `modelling out' the effects of beam-smearing using similar approaches (e.g. as described in \citealt{Davies2011}, where removing the effects of beam-smearing with spectrally and spatially convolved models is also the least biased approach).
This trend towards consistency is encouraging, and, as described in \cref{subsec:3d_modelling}, we have also shaped our approach towards extracting intrinsic kinematic parameters after correcting for beam-smearing effects quantified in the modelling.
We stress that, keeping in mind sample selection, mass ranges and kinematic parameter extraction methods are crucial when comparing results between different surveys.

\section{Intrinsic parameter distributions}

\begin{figure}
\centering \hspace{-1.13cm}
\includegraphics[width=0.49\textwidth]{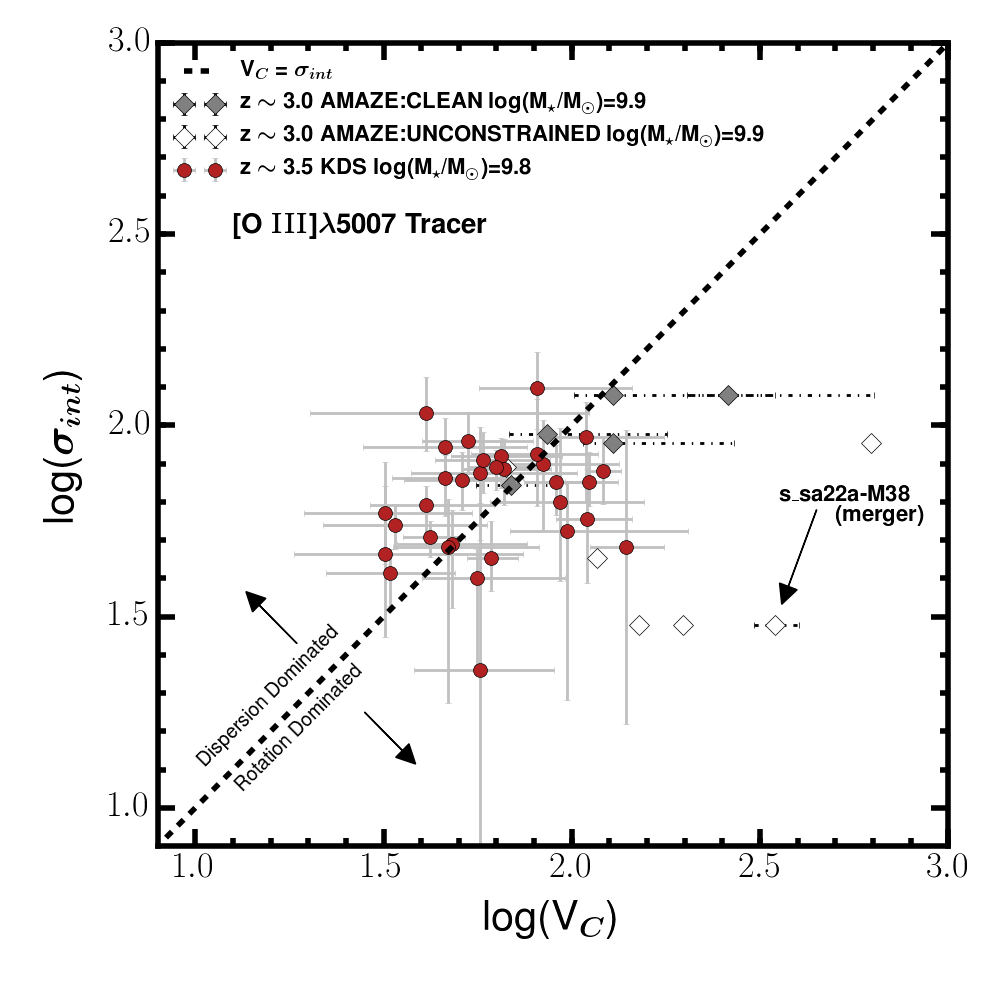}
\caption{The distribution of the isolated field sample in the $\sigma_{int}$ vs. $V_{C}$ plane is plotted with the red symbols.
The AMAZE Clean ($z\simeq3$) sample galaxies (see \protect\cref{subsec:AMAZE}) are plotted with grey symbols and the AMAZE Unconstrained sample with hollow symbols.    There is no strong correlation between the two parameters in either of these surveys and we note that the KDS and AMAZE Clean galaxies are generally in agreement.
The $V_{C}/\sigma_{int} = 1$ line divides the galaxies classified as rotation-dominated and dispersion-dominated.
The galaxy s\_sa22a-M38, classified here as a merger but classified in \protect\cite{Gnerucci2011} as a rotating galaxy with $V_{C} = 346 km\,s^{-1}$ is marked on the plot.}
\label{fig:intrinsic_parameters}
\end{figure}

In Figure \ref{fig:intrinsic_parameters} we plot the distribution of the isolated field sample in the $\sigma_{int}$ vs. log($V_{C}$) plane. 
The $V_{C}$ = $\sigma_{int}$ line roughly bisects the KDS sample, and separates the dispersion-dominated and rotation-dominated galaxies by definition.
The isolated field sample galaxies are clustered around a relatively tight region in both $V_{C}$ and $\sigma_{int}$ and there is good agreement between the KDS and AMAZE Clean ($z\simeq3$) samples.
The galaxies in the AMAZE Unconstrained sample generally have much larger $V_{C}$ values despite occupying a similar stellar mass range to our isolated field sample, with this being a consequence of large model extrapolations beyond the observed data.
We follow the bootstrapping procedure described in Appendix \ref{app:comparison_samples} to compute error weighted averages and statistical errors.

\section{Kinematic parameters error estimates}\label{app:kin_error_estimates}

\subsection{3D modelling kinematic parameter error estimates}\label{appsubsec:model_errors}
As mentioned in \cref{subsec:3d_modelling}, the MCMC sampling provides distributions for each of the model parameters, from which we can extract the 84th and 16th percentile values as the +/- 1-$\sigma$ errors.
During the following procedure, $PA_{kin}$ is fixed to the maximum-likelihood value for all model evaluations.
We proceed to reconstruct the beam-smeared and intrinsic dynamical models for each of the isolated field galaxies using the 16th and 84th percentile parameters, with the lower velocity 16th percentile model constructed using the upper limit on the inclination, and the 84th percentile model using the lower limit.
To clearly distinguish the 1-$\sigma$ error regions for both the beam-smeared and intrinsic models, the region between the 16th and 84th percentile evaluations is shaded blue and red respectively in the velocity extaction panels throughout Figures \ref{fig:rotation_dominated_galaxies} and \ref{fig:dispersion_dominated_galaxies}.
The +/- 1-$\sigma$ error values for both $V_{C}$ are then calculated using the equations below, which also take into account the measurement errors, with $\bar{\sigma}_{v_{obs}}$ equal to the average observational uncertainty extracted along $PA_{kin}$.
The subscripts `16th' and `84th' denote the parameters used to construct the model from which the velocity has been extracted.

\begin{equation}\label{eq:VC_plus}
   \delta V_{C}+ = \sqrt{\left(V_{C-84} - V_{C}\right)^{2} + \left(\bar{\delta}V_{obs}\right)^{2}}
\end{equation}

\begin{equation}\label{eq:VC_minus}
   \delta V_{C}- = \sqrt{\left(V_{C} - V_{C-16}\right)^{2}  + \left(\bar{\delta}V_{obs}\right)^{2}}
\end{equation}

\noindent
The upper and lower errors on the $\sigma_{int}$ are calculated using a similar approach.
To encorporate the uncertainty introduced in the modelling by assuming a fixed value of $\sigma_{int}=50 km\,s^{-1}$, we make two further model evaluations using both the 16th and 84th percentile parameters with $\sigma_{int} = 40 km\,s^{-1}$ and $\sigma_{int} = 80 km\,s^{-1}$, with these representing reasonable minimum and maximum mean instrinsic sigma values (chosen by considering the range in $\sigma_{int}$ at similar redshift, see e.g. Figure \ref{fig:sigma_and_v_sigma_w_redshift}).
When assuming the broader $\sigma_{int} = 80 km\,s^{-1}$ value in each spaxel the beam-smearing correction decreases, as the impact of convolution with spectral lines in adjacent spaxels with shifted velocity centres is less severe.
Conversely, assuming $\sigma_{int} = 40 km\,s^{-1}$ increases the beam-smearing.
We calculate the minimum and maxmimum intrinsic velocity dispersion, $\sigma_{int-84-40}$ and $\sigma_{int-16-80}$, using equations \ref{eq:sig_84_40} and \ref{eq:sig_16_80} respectively:

\begin{equation}\label{eq:sig_84_40}
   \sigma_{int-84-40} = \sqrt{(\sigma_{obs} - \sigma_{bs-84-40})^{2} - \sigma_{sky}^{2}}
\end{equation}

\begin{equation}\label{eq:sig_16_80}
   \sigma_{int-16-80} = \sqrt{\left(\sigma_{obs} - \sigma_{bs-16-80}\right)^{2} - \sigma_{sky}^{2}}
\end{equation}

\noindent
where $\sigma_{bs-84-40} = \sigma_{model-84-40} - 40$ and $\sigma_{bs-16-80} = \sigma_{model16-80} - 80$ are the beam-smeared maps.
The lower and upper errors on $\sigma_{int}$ are then given by the following:

\begin{equation}\label{eq:sig_plus_error}
   \delta\sigma_{int}+ = \sqrt{\left(\sigma_{int} - \sigma_{int-84-40}\right)^{2} + \left(\bar{\sigma}_{\sigma_{obs}}\right)^{2}}
\end{equation}

\begin{equation}\label{eq:sig_minus_error}
   \delta\sigma_{int}- = \sqrt{\left(\sigma_{int-16-80} - \sigma_{int}\right)^{2} + \left(\bar{\sigma}_{\sigma_{obs}}\right)^{2}}
\end{equation}

\noindent
with $\bar{\sigma}_{\sigma_{obs}}$ equal to the average of the velocity dispersion measurement errors.
Once these quantities have been measured, the upper and lower errors on the ratio $V_{C}/\sigma_{int}$ can be computed using equations \ref{eq:v_over_sig_plus} and \ref{eq:v_over_sig_minus}.

\begin{equation}\label{eq:v_over_sig_plus}
   \delta\frac{V_{C}}{\sigma_{int}}+ = \frac{V_{C}}{\sigma_{int}}\sqrt{\left(\frac{\delta V_{C}+}{V_{C}}\right)^{2} + \left(\frac{\delta\sigma_{int}+}{\sigma_{int}}\right)^{2}}
\end{equation}

\begin{equation}\label{eq:v_over_sig_minus}
   \delta\frac{V_{C}}{\sigma_{int}}- = \frac{V_{C}}{\sigma_{int}}\sqrt{\left(\frac{\delta V_{C}-}{V_{C}}\right)^{2} + \left(\frac{\delta\sigma_{int}+-}{\sigma_{int}}\right)^{2}}
\end{equation}

\section{Kinematics Plots}\label{app:kinematics_plots}
In Figures \ref{fig:rotation_dominated_galaxies}, \ref{fig:dispersion_dominated_galaxies} and \ref{fig:merger_galaxies} we plot the kinematic grids for the KDS galaxies classified as rotation-dominated, dispersion-dominated and merger candidates respectively.
The figure captions provide more information on each of the panels of these grids.

\begin{figure*}
    \centering

    \includegraphics[width=0.95\textwidth]{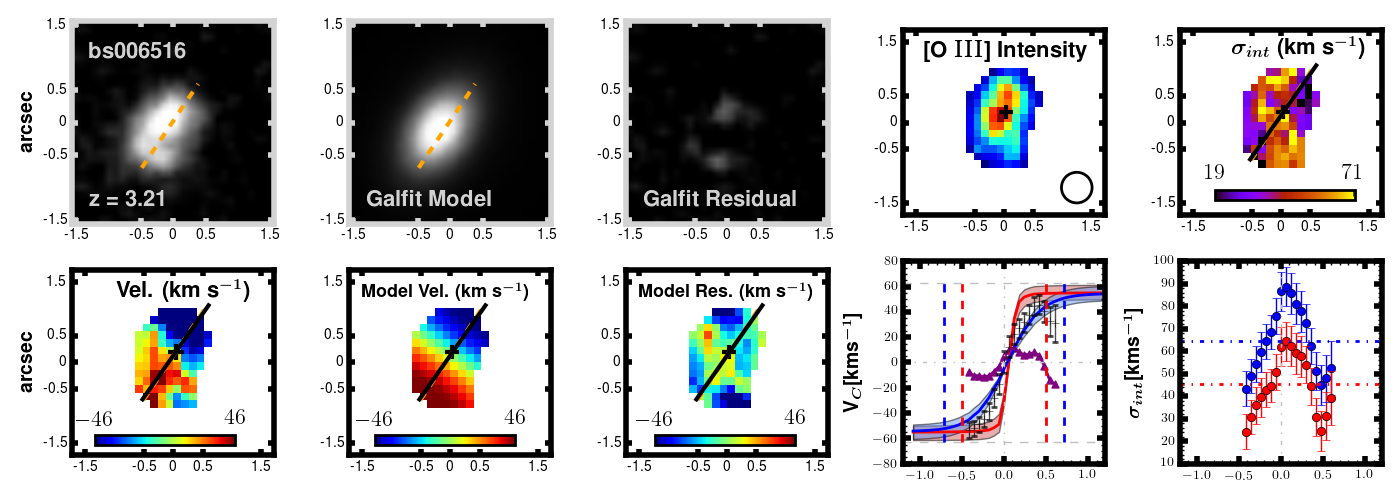}
    \includegraphics[width=0.95\textwidth]{GOODS_ROTATION/combine_sci_reconstructed_bs008543_grid_paper.png}
    \includegraphics[width=0.95\textwidth]{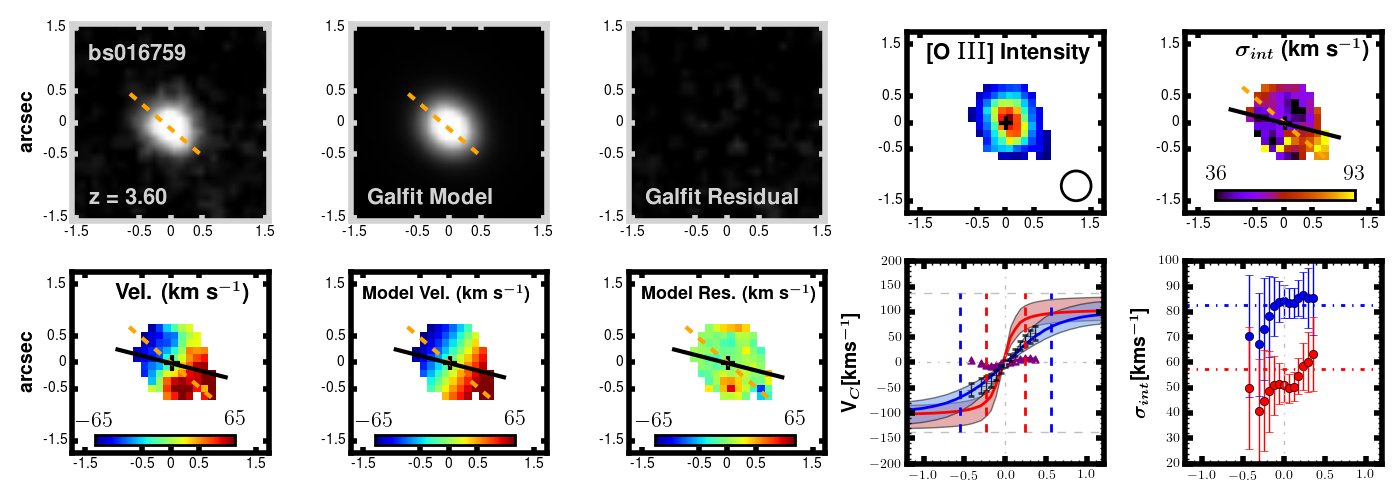}

    \caption{We plot the kinematic grids for the 13/32 isolated field sample galaxies classified as rotation-dominated.
    For each galaxy we plot the {\em HST} image, {\scriptsize GALFIT} model and {\scriptsize GALFIT} residuals, along with the observed [O~{\sc III}]$\lambda$5007 intensity map, dispersion map and velocity map extracted from the datacubes with gaussian-fits to the individual spaxels as described in \protect\cref{subsubsection:spaxel_fitting}.
    The solid black line and dashed orange line plotted in various panels show $PA_{kin}$ and $PA_{morph}$ respectively.
    On the bottom row we plot the beam-smeared model velocity and data-model velocity residuals with colour bar limits tuned to the velocity model.
    Also on the bottom row, we plot extractions along $PA_{kin}$ for both the velocity and velocity dispersion.
    The grey points with errorbars on the velocity extraction plot are from the observed velocity map, the blue line and blue shaded regions represent the beam-smeared model fit to the data and errors respectively, whereas the red line and shaded regions are the intrinsic model from which $V_{C}$ is extracted.
    The purple symbols represent the extraction along $PA_{kin}$ from the residual map, and the two vertical dashed lines denote the intrinsic (inner) and convolved (outer) $2R_{1/2}$ values, with $V_{C}$ extracted at the intrinsic value and $V_{BS}$ extracted at the convolved value of the intrinsic and observed profiles respectively.
    The intrinsic models flatten at small radii, and so small changes in $R_{1/2}$ have negligible impact on the final, extracted $V_{C}$ values.
    The blue points on the velocity dispersion extraction plot show the values extracted from the observed dispersion map and the red are from the beam-smearing corrected map as per Equation \protect\ref{eq:dispersion_comp}, and the vertical dashed lines have the same meaning as for the velocity extraction plot.
    Also plotted with the blue and red dot-dash horizontal lines are the median values of the observed and intrinsic (i.e. $\sigma_{int}$) dispersion maps respectively.}
    \label{fig:rotation_dominated_galaxies}

\end{figure*}

\begin{figure*}\ContinuedFloat
    \centering

    \includegraphics[width=0.95\textwidth]{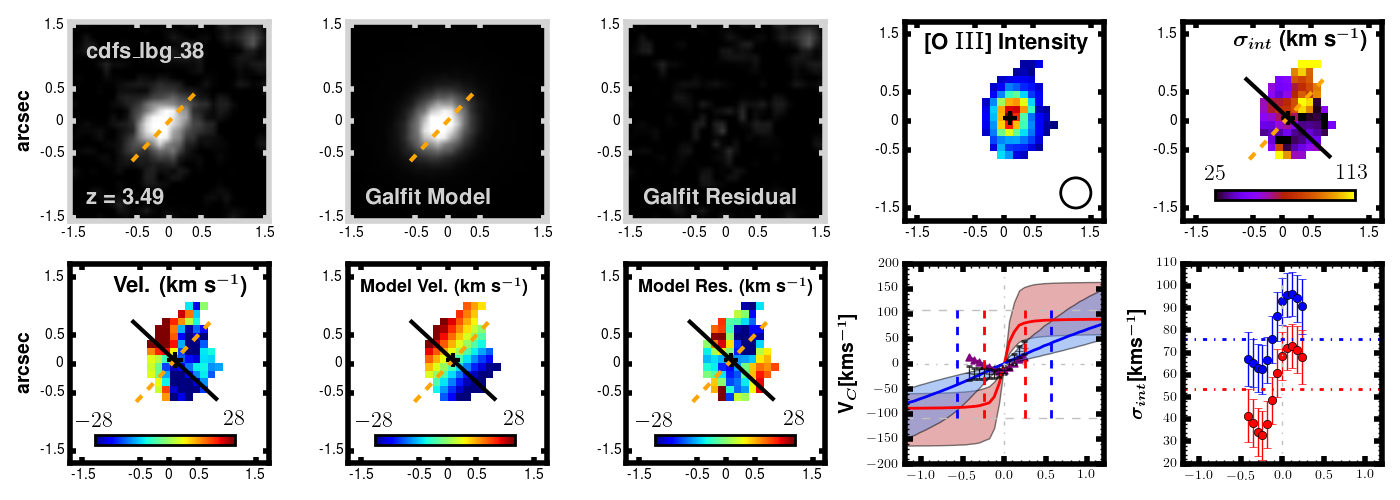}
    \includegraphics[width=0.95\textwidth]{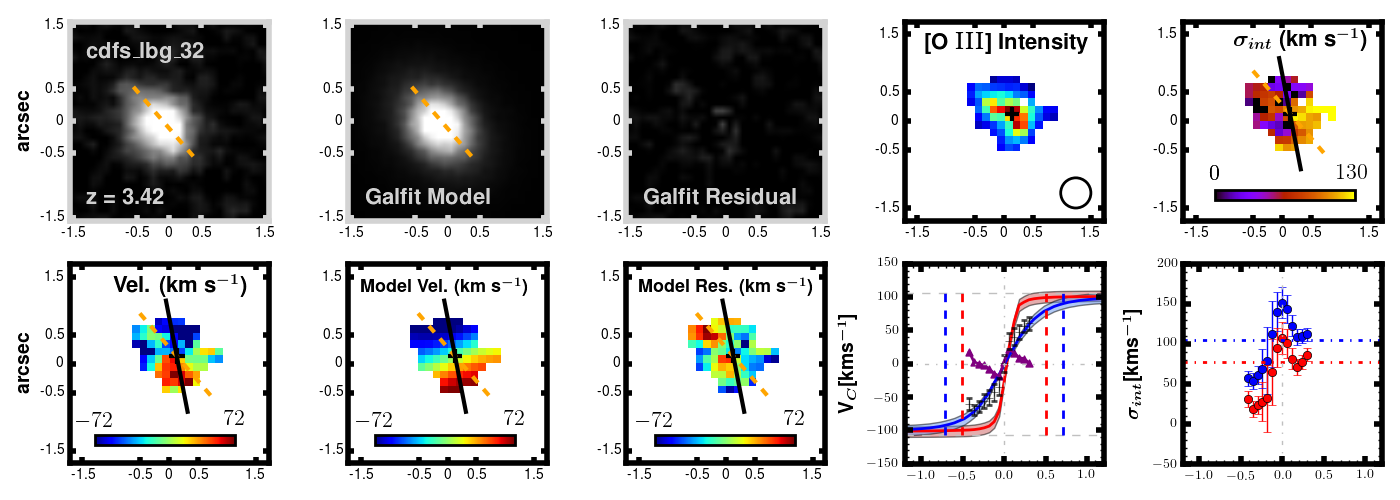}
    \includegraphics[width=0.95\textwidth]{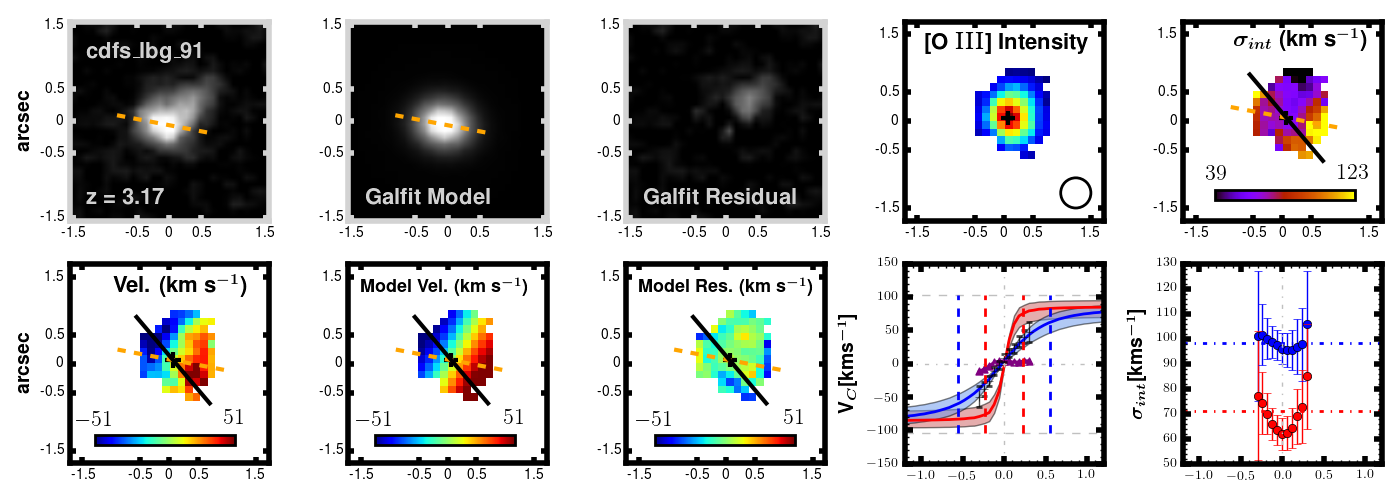}
    \includegraphics[width=0.95\textwidth]{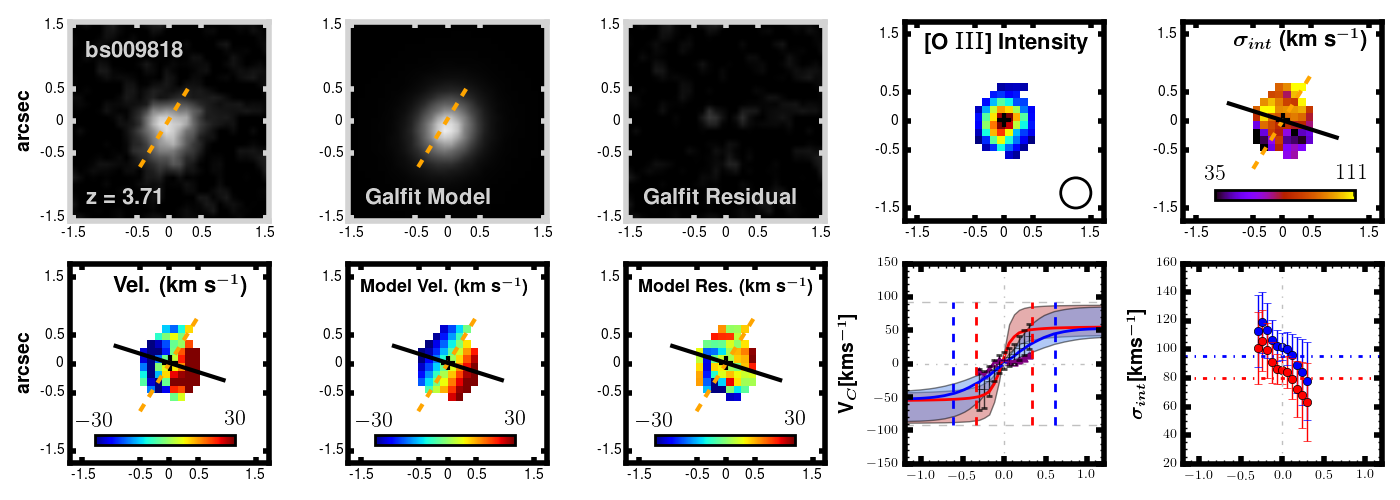}

    \caption{\textbf{Continued.}}

\end{figure*}

\begin{figure*}\ContinuedFloat
    \centering

    \includegraphics[width=0.95\textwidth]{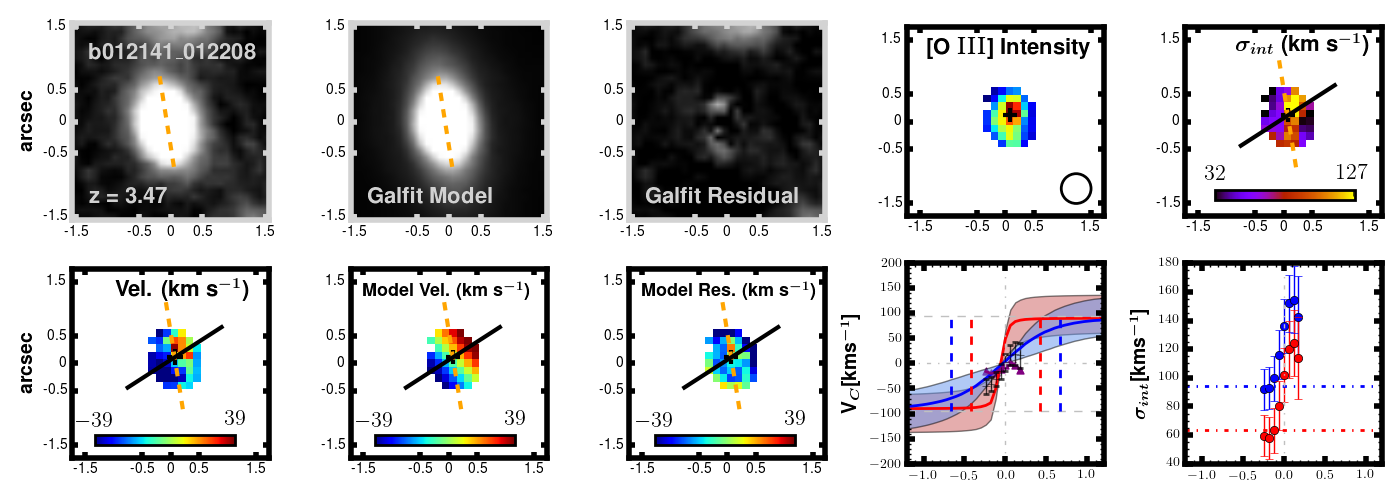}
    \includegraphics[width=0.95\textwidth]{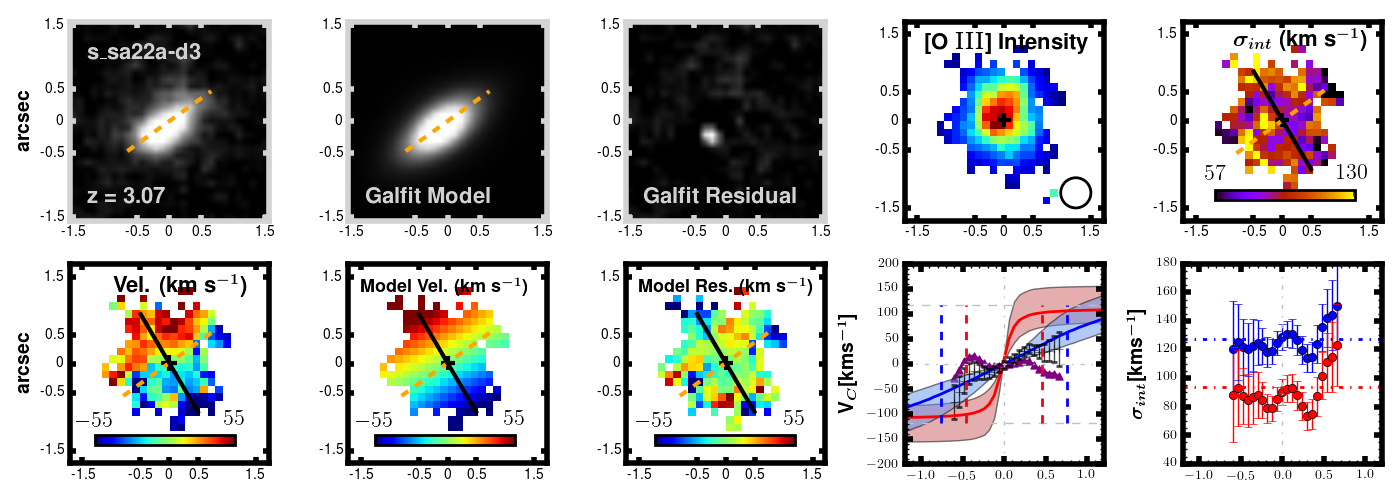}
    \includegraphics[width=0.95\textwidth]{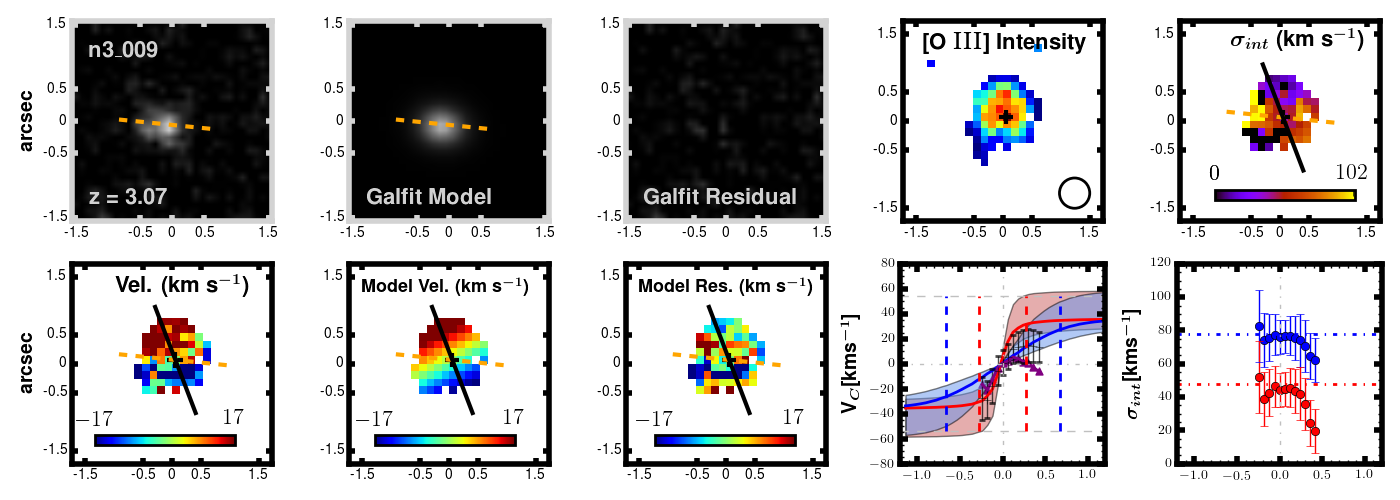}
    \includegraphics[width=0.95\textwidth]{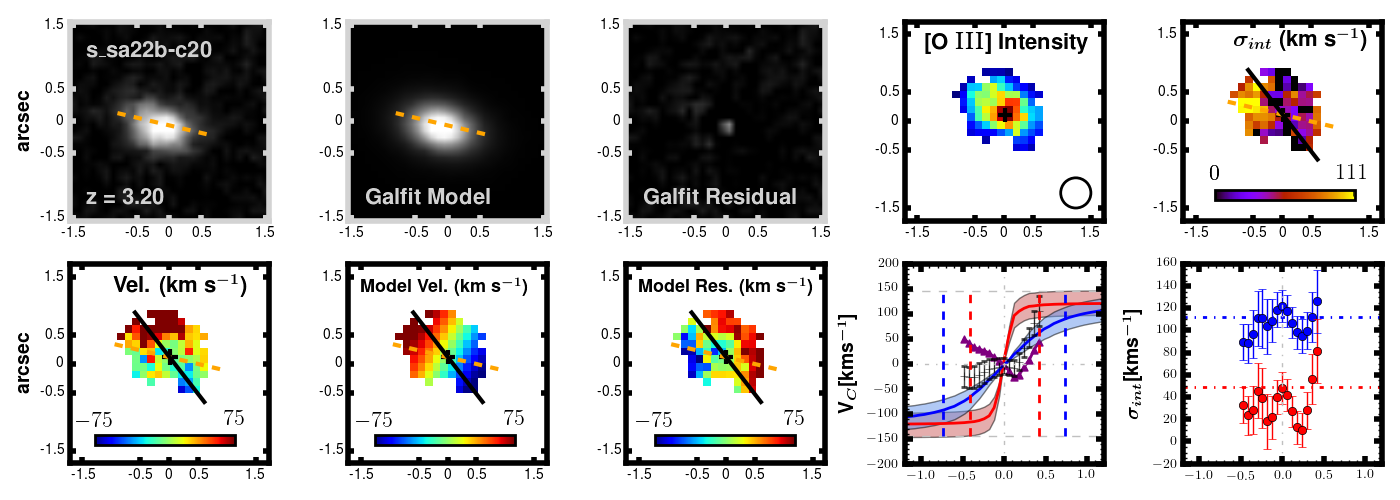}

    \caption{\textbf{Continued.}}

\end{figure*}

\begin{figure*}\ContinuedFloat
\centering

    \includegraphics[width=0.95\textwidth]{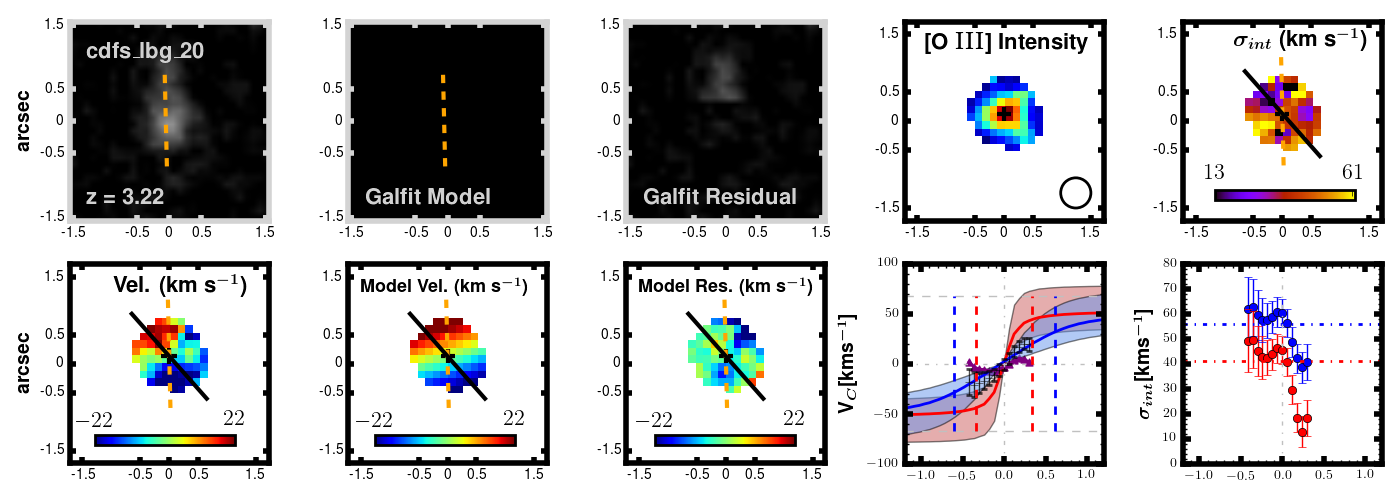}
    \includegraphics[width=0.95\textwidth]{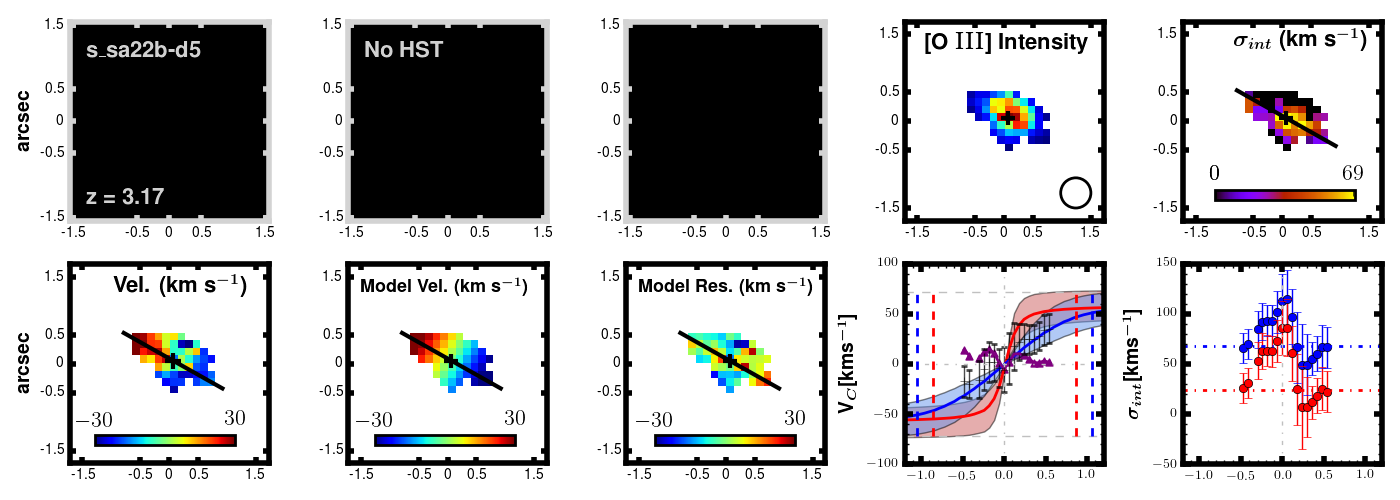}

    \caption{\textbf{Continued.}}

\end{figure*}

\begin{figure*}
    \centering

    \includegraphics[width=0.95\textwidth]{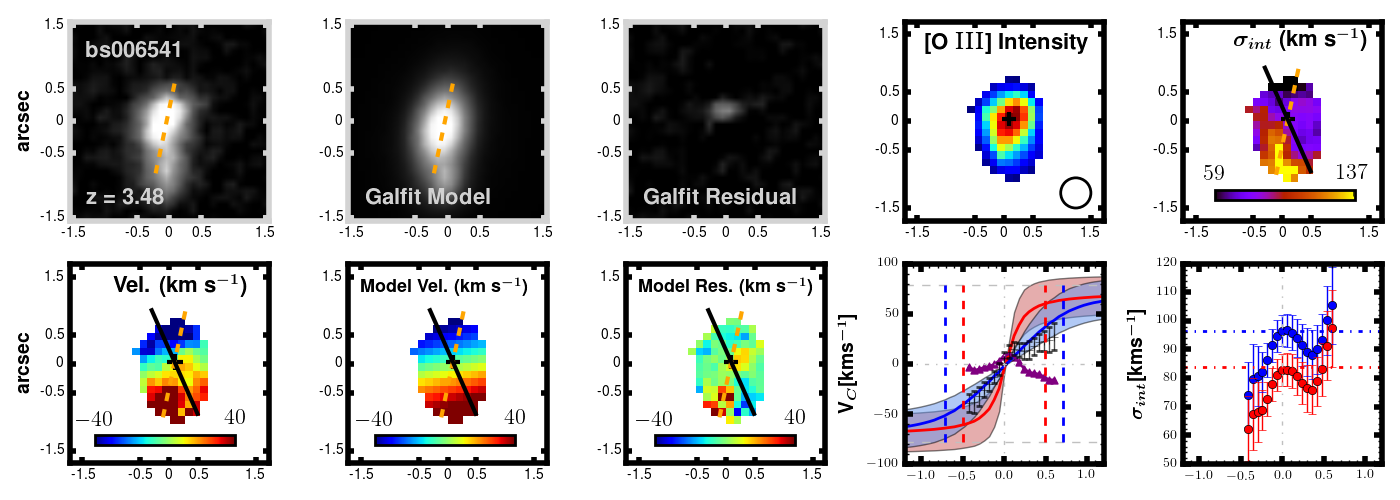}
    \includegraphics[width=0.95\textwidth]{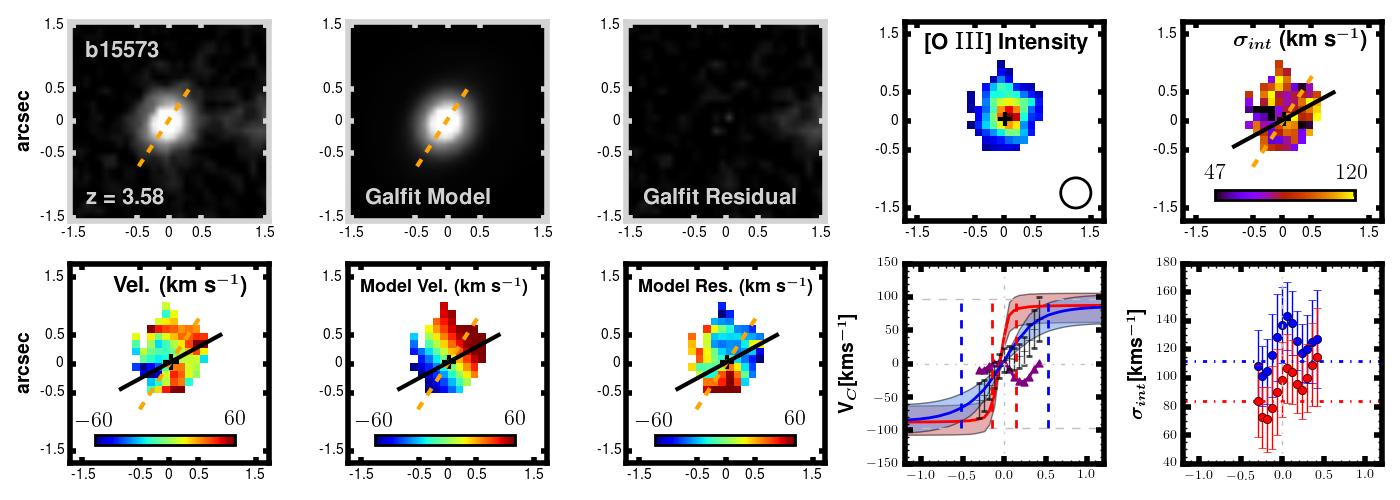}
    \includegraphics[width=0.95\textwidth]{GOODS_DISPERSION/combine_sci_reconstructed_cdfs_lbg_24_grid_paper.png}
    \includegraphics[width=0.95\textwidth]{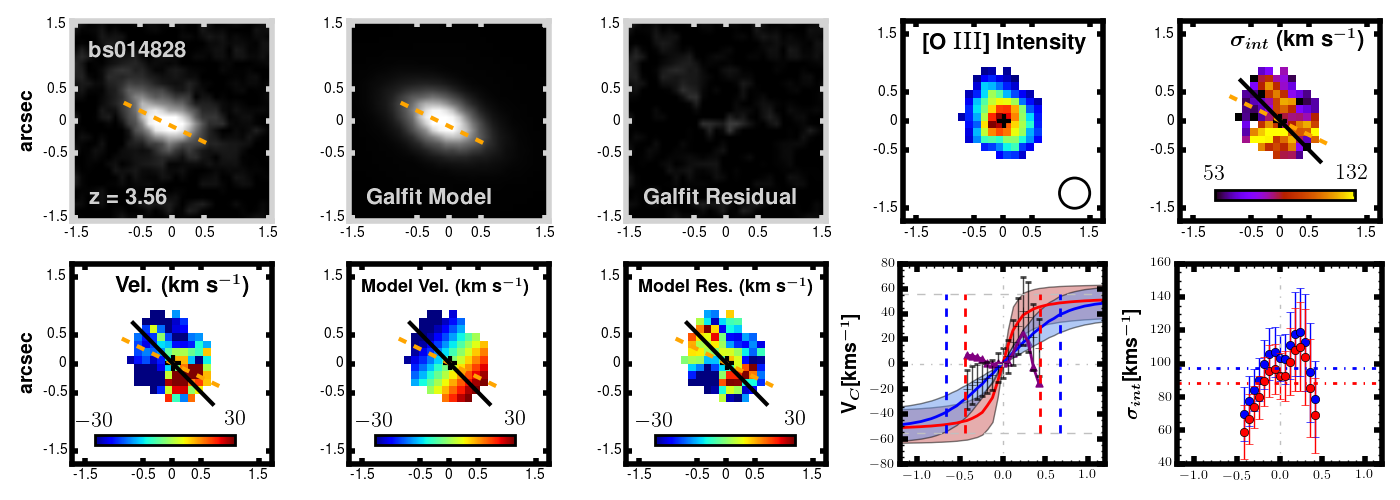}

    \caption{The same as in figure \protect\ref{fig:rotation_dominated_galaxies} but for the dispersion-dominated galaxies.}
    \label{fig:dispersion_dominated_galaxies}

\end{figure*}

\begin{figure*}\ContinuedFloat
\centering

    \includegraphics[width=0.95\textwidth]{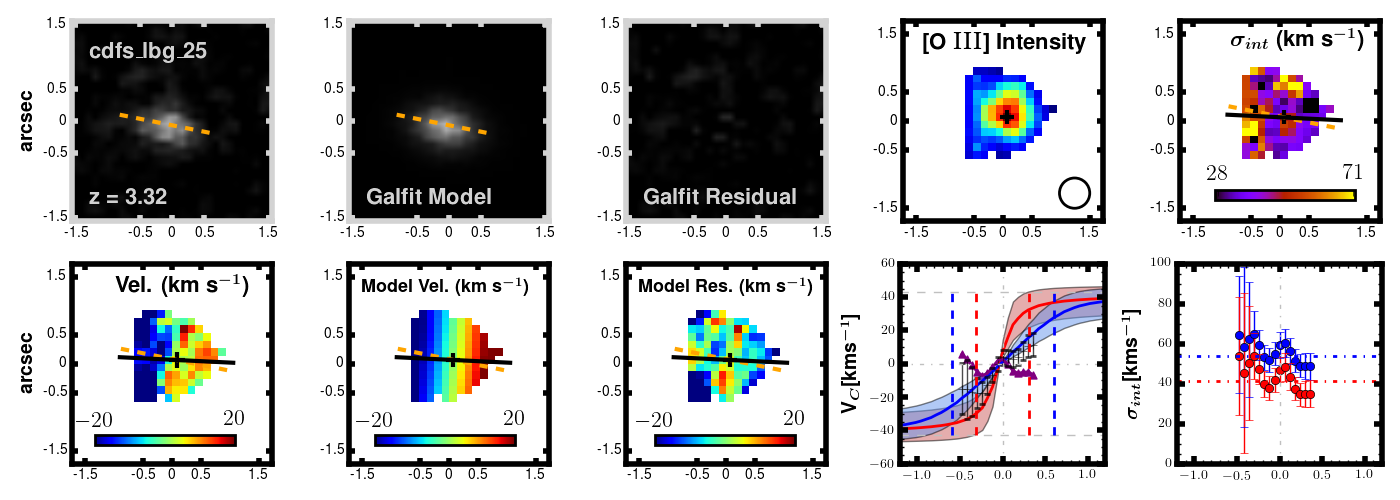}
    \includegraphics[width=0.95\textwidth]{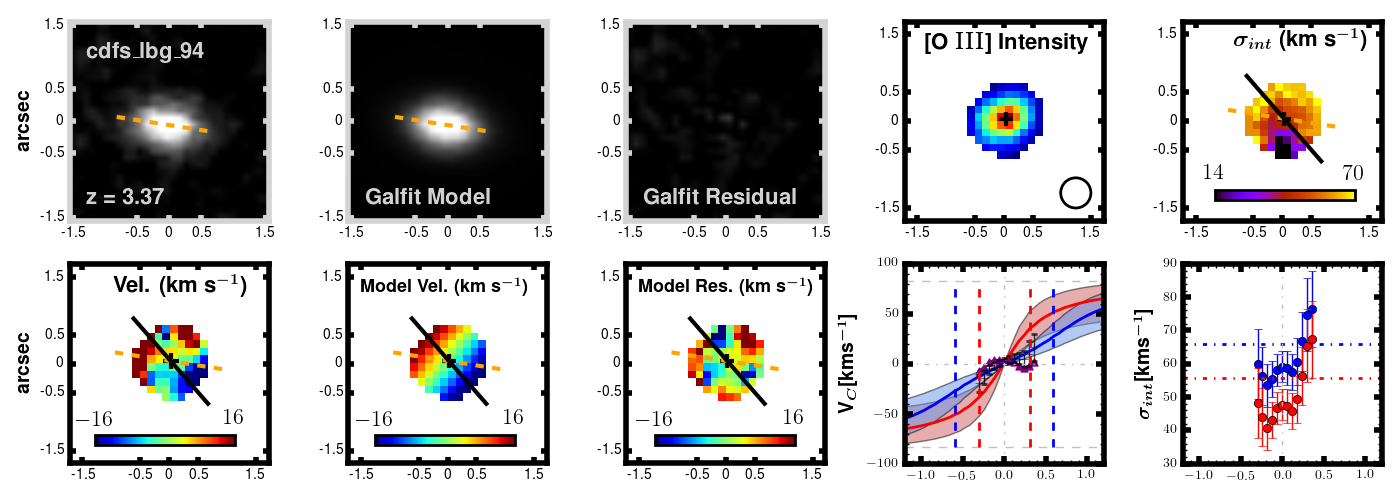}
    \includegraphics[width=0.95\textwidth]{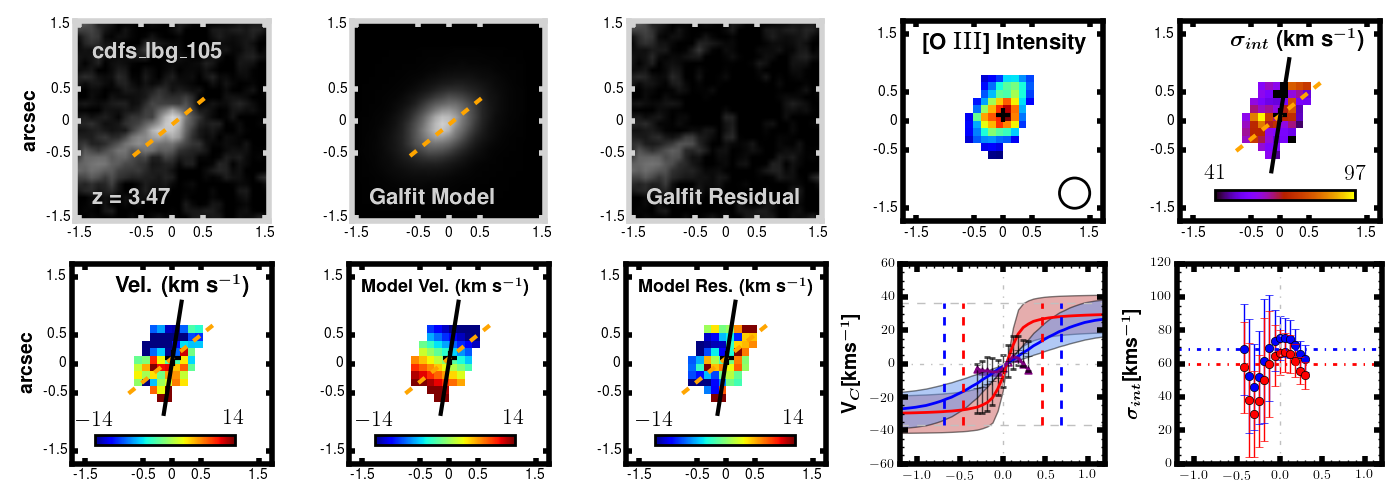}
    \includegraphics[width=0.95\textwidth]{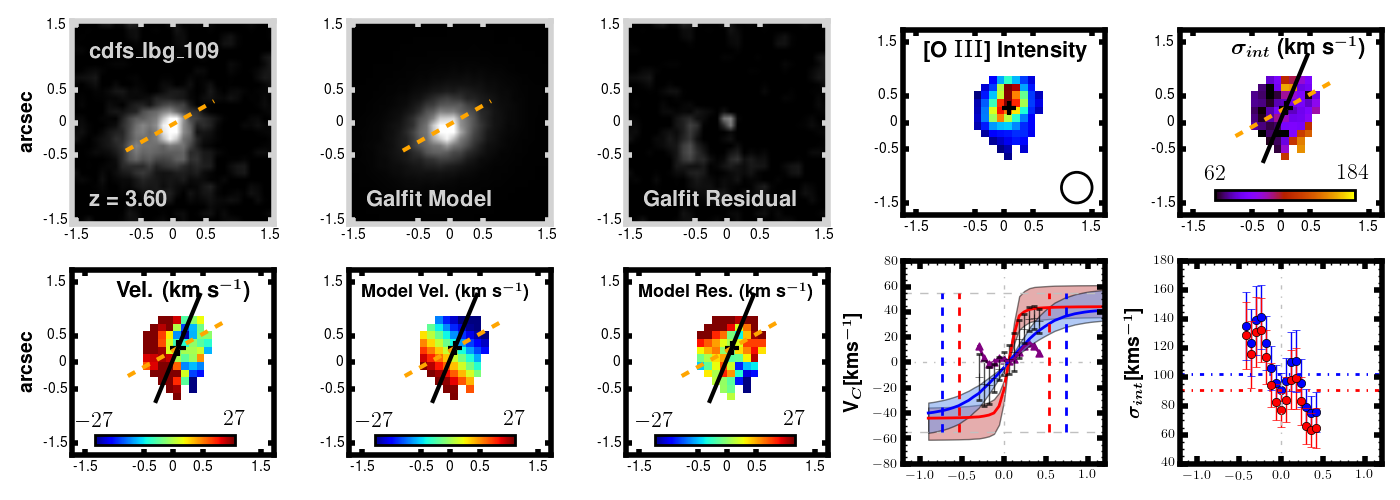}

    \caption{\textbf{Continued.}}

\end{figure*}

\begin{figure*}\ContinuedFloat
    \centering

    \includegraphics[width=0.95\textwidth]{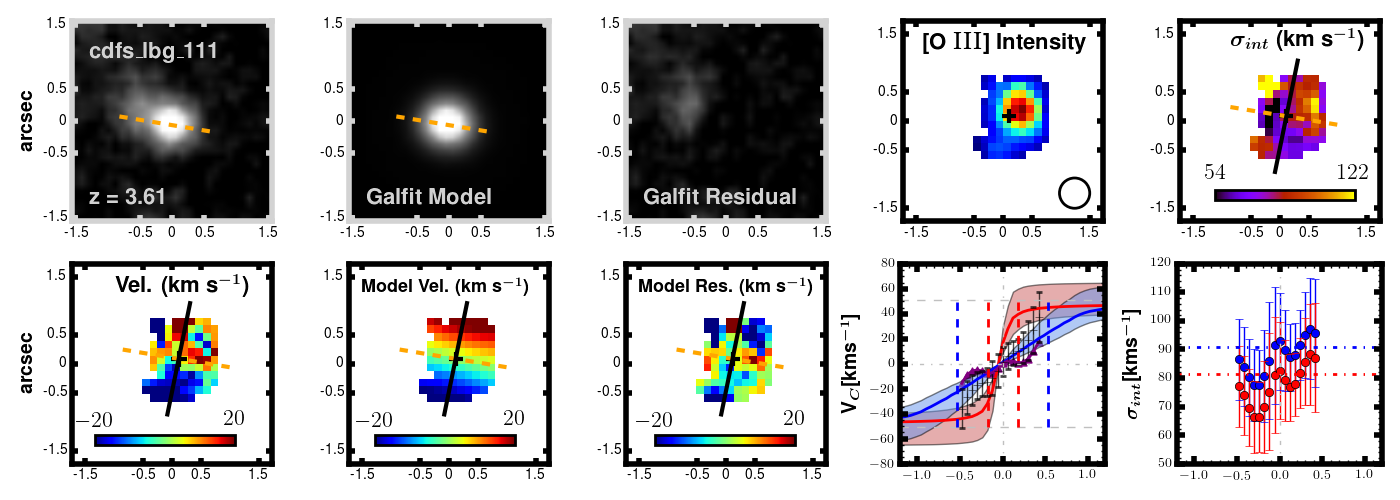}
    \includegraphics[width=0.95\textwidth]{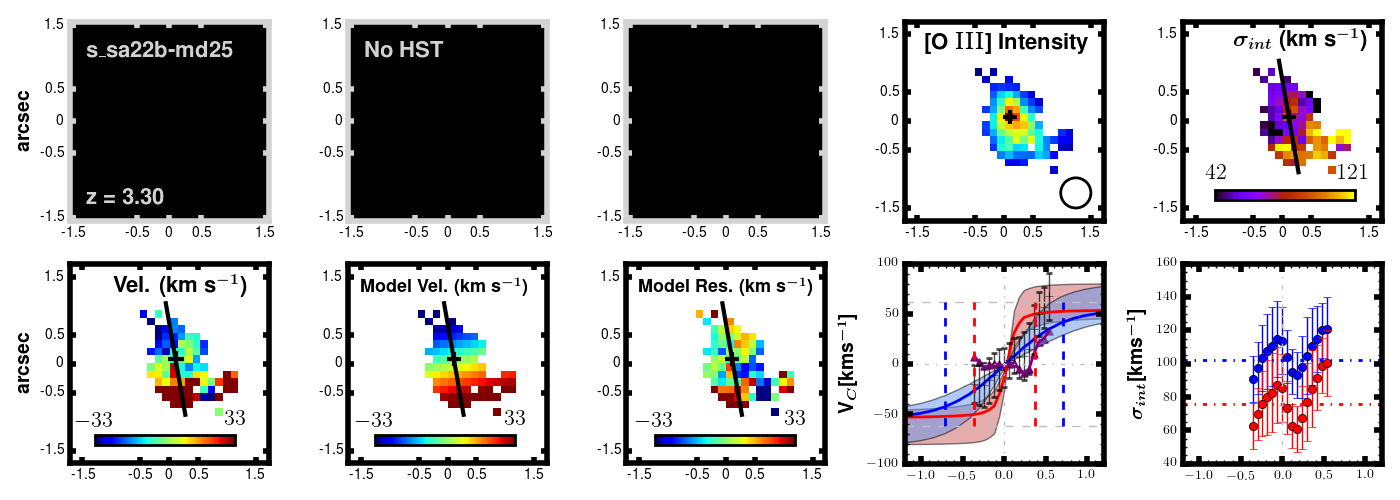}
    \includegraphics[width=0.95\textwidth]{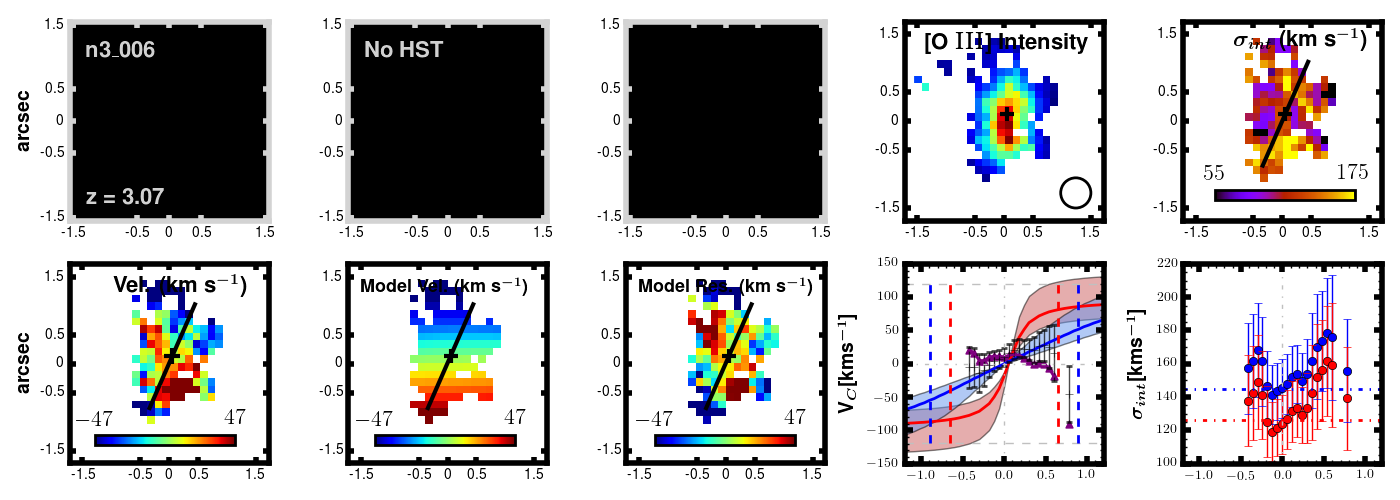}
    \includegraphics[width=0.95\textwidth]{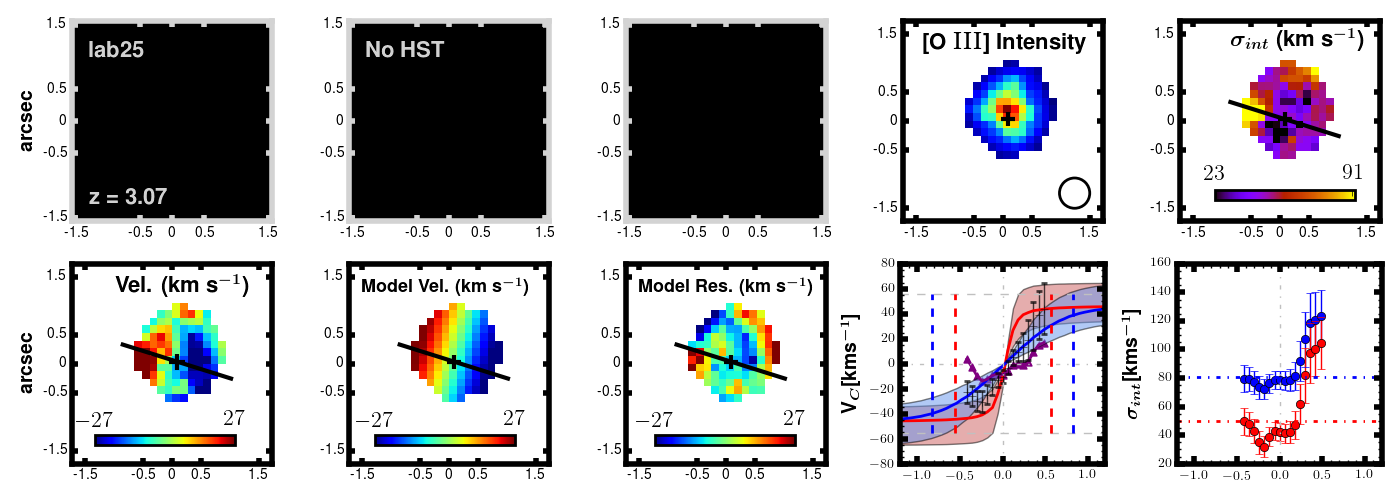}

    \caption{\textbf{Continued.}}

\end{figure*}

\begin{figure*}\ContinuedFloat
    \centering

    \includegraphics[width=0.95\textwidth]{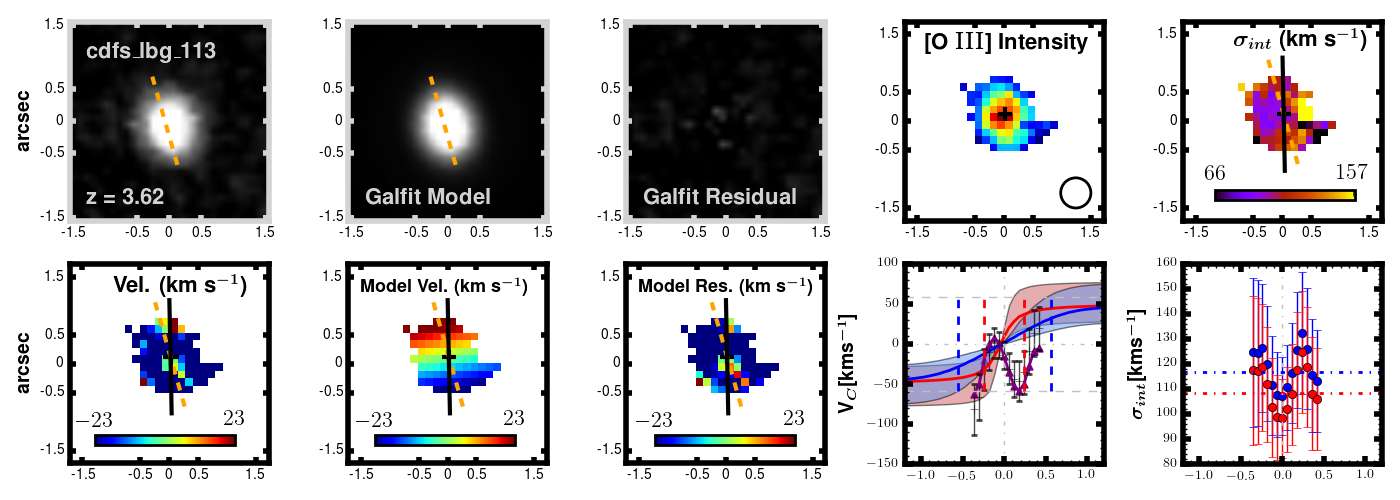}
    \includegraphics[width=0.95\textwidth]{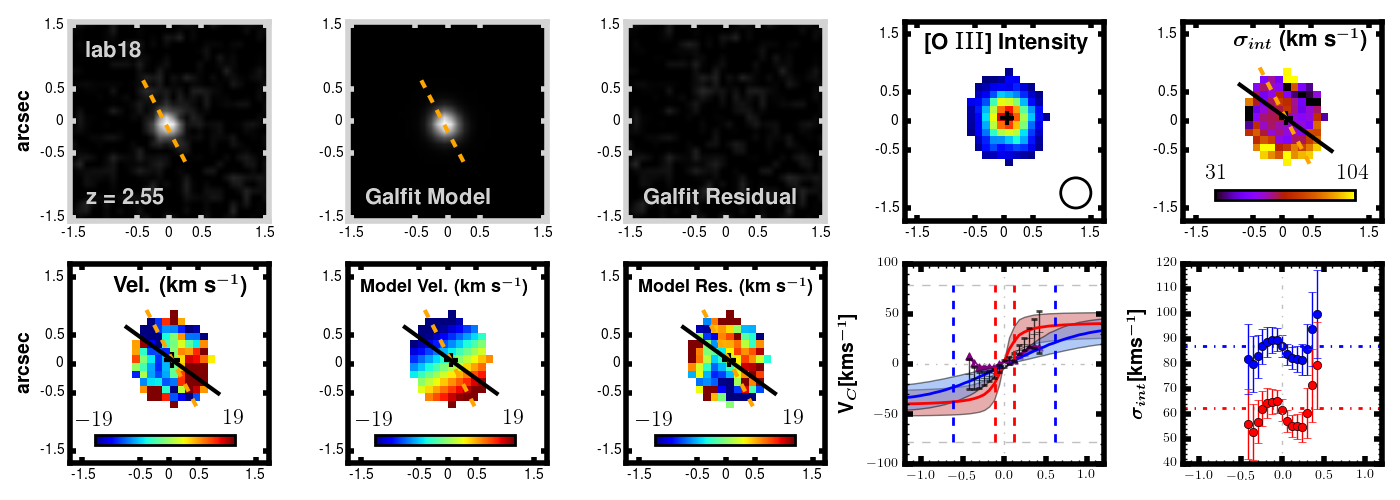}
    \includegraphics[width=0.95\textwidth]{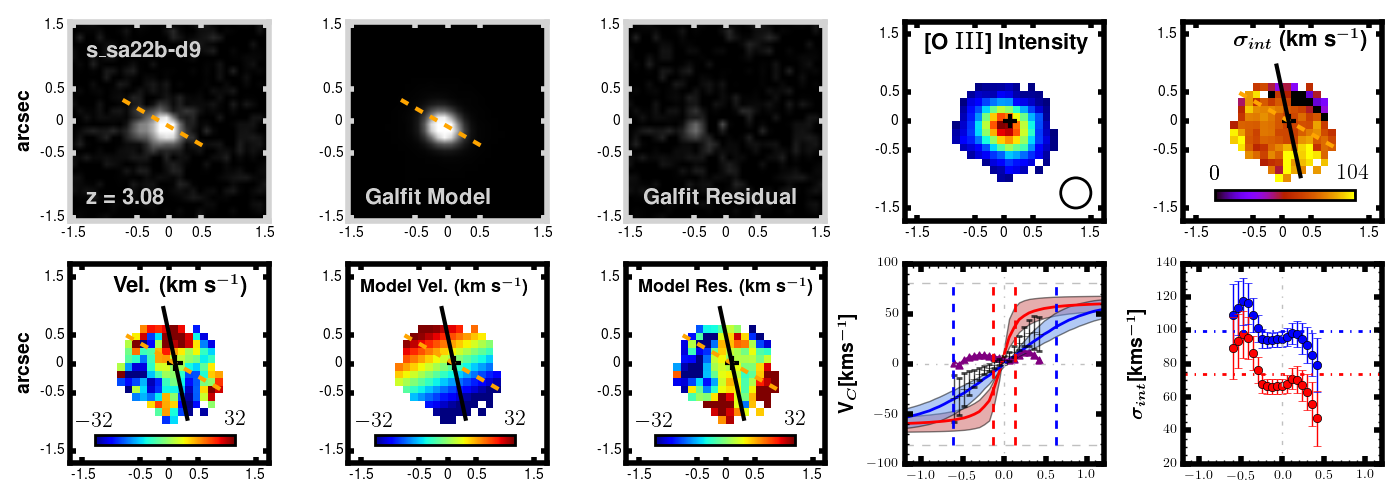}
    \includegraphics[width=0.95\textwidth]{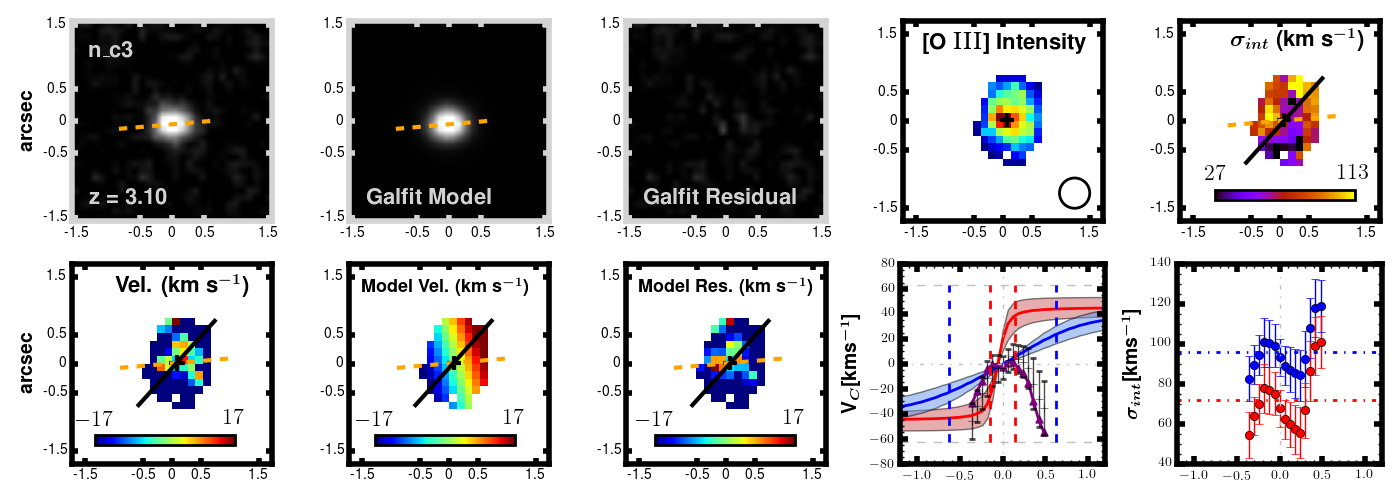}

    \caption{\textbf{Continued.}}

\end{figure*}

\begin{figure*}\ContinuedFloat
    \centering

    \includegraphics[width=0.95\textwidth]{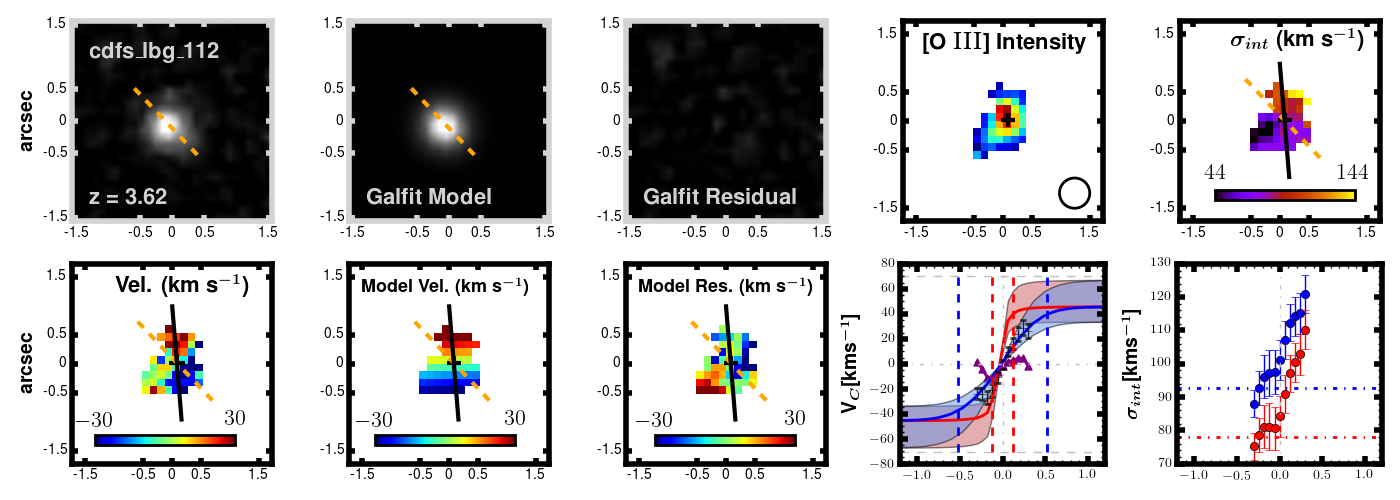}
    \includegraphics[width=0.95\textwidth]{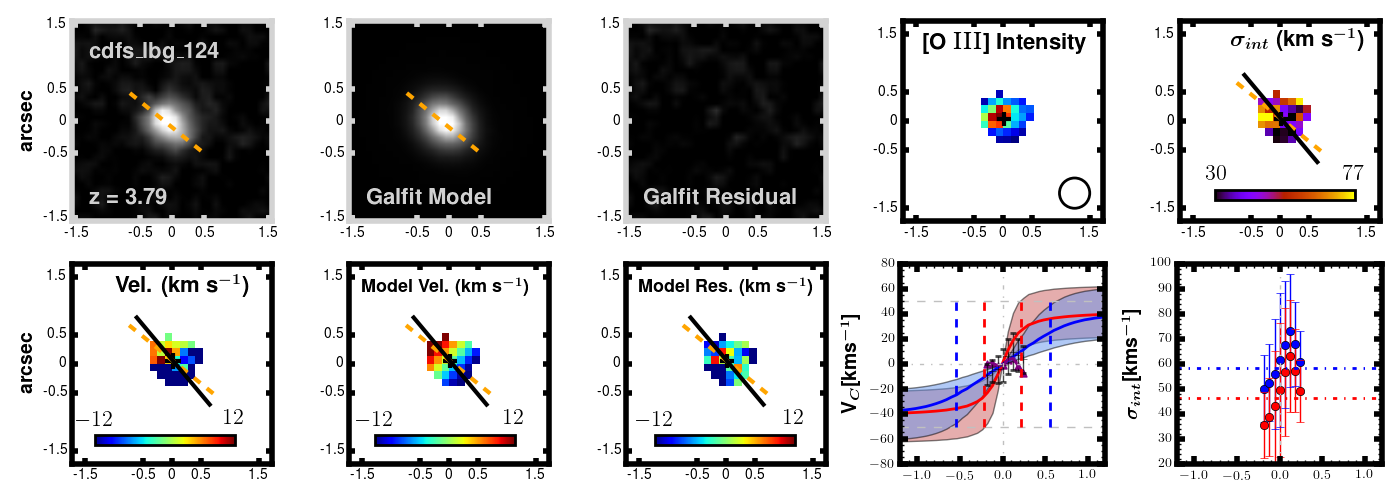}
    \includegraphics[width=0.95\textwidth]{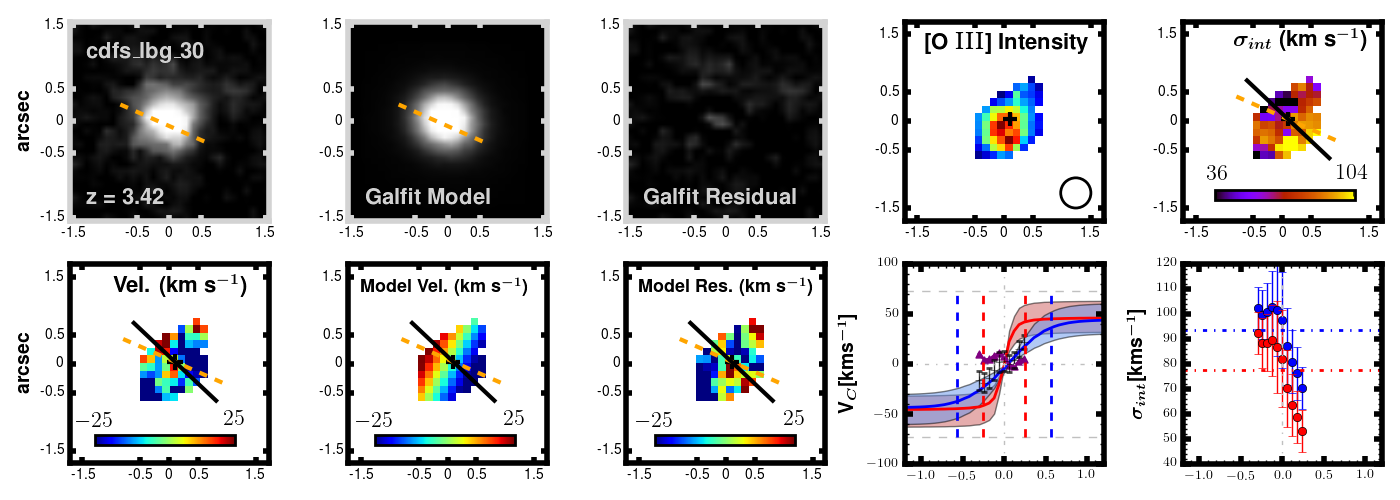}

    \caption{\textbf{Continued.}}

\end{figure*}

\begin{figure*}
    \centering

    \includegraphics[width=0.95\textwidth]{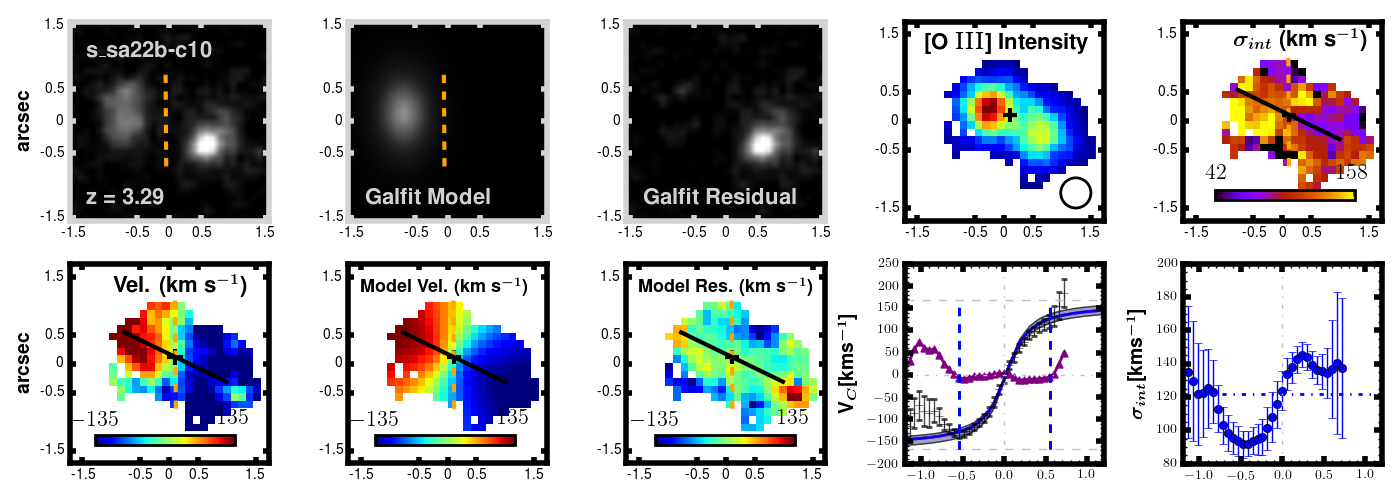}
    \includegraphics[width=0.95\textwidth]{GOODS_MERGERS/combine_sci_reconstructed_cdfs_lbg_14_grid_paper.png}
    \includegraphics[width=0.95\textwidth]{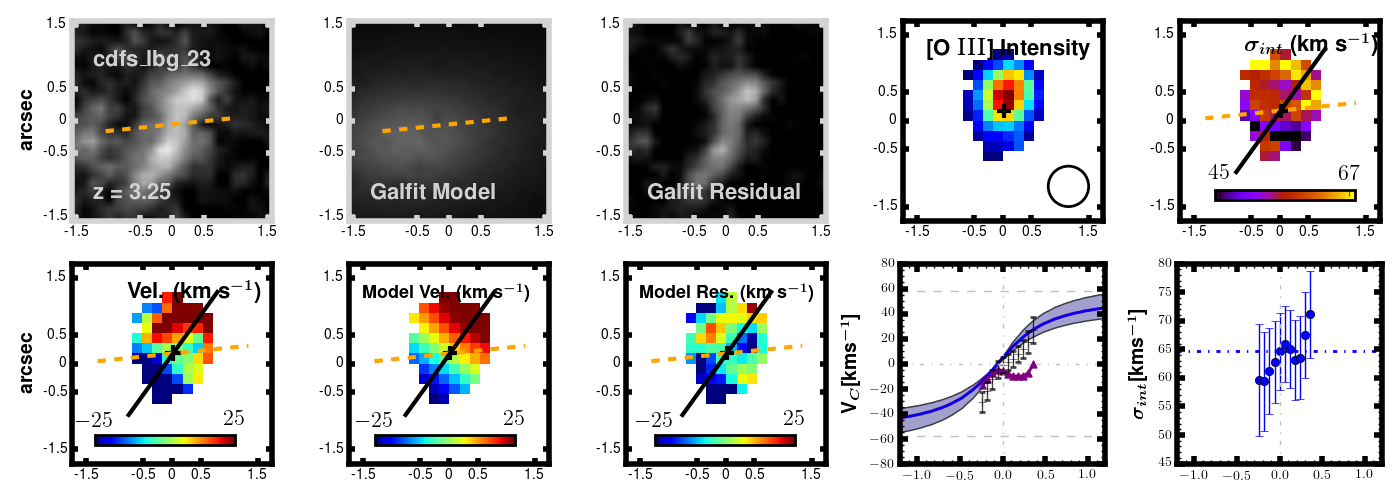}

    \caption{The same as for figure \protect\ref{fig:rotation_dominated_galaxies} but for the merger candidates.
    In this case we plot only fits to the data in the velocity extraction plot with the blue line, rather than attempting the full beam-smearing analysis.
    Several of the galaxies here mimic rotation from a purely kinematic perspective, but have two or more {\em HST} components and an accompanying double peak in the object spectrum at the object centre, leading to large velocity dispersions when single gaussian-fits are attempted.}
    \label{fig:merger_galaxies}
\end{figure*}

\begin{figure*}\ContinuedFloat
    \centering

    \includegraphics[width=0.95\textwidth]{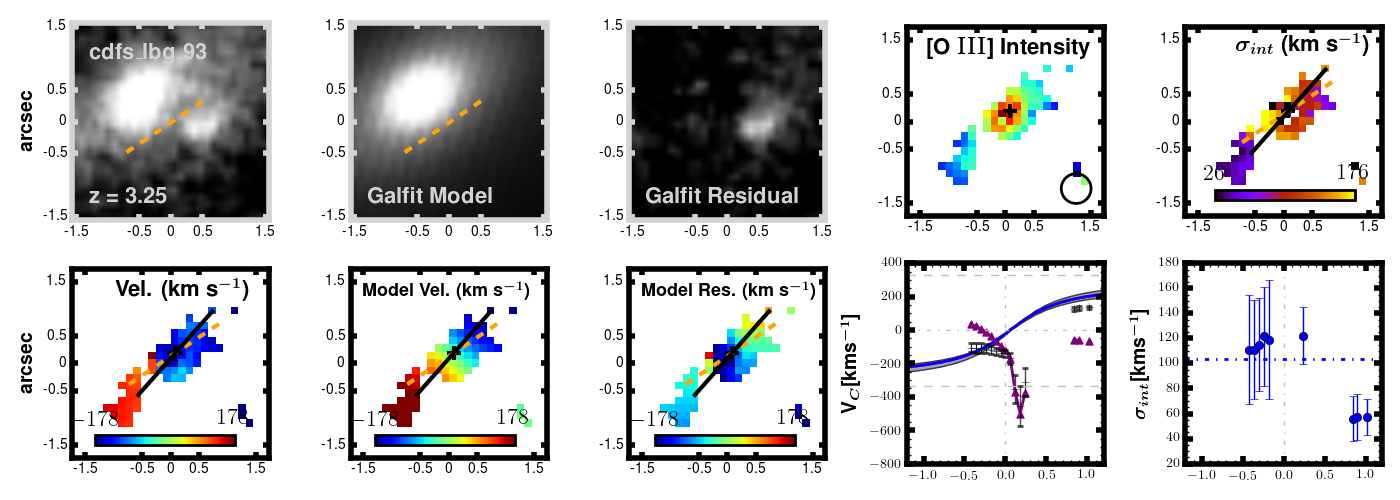}
    \includegraphics[width=0.95\textwidth]{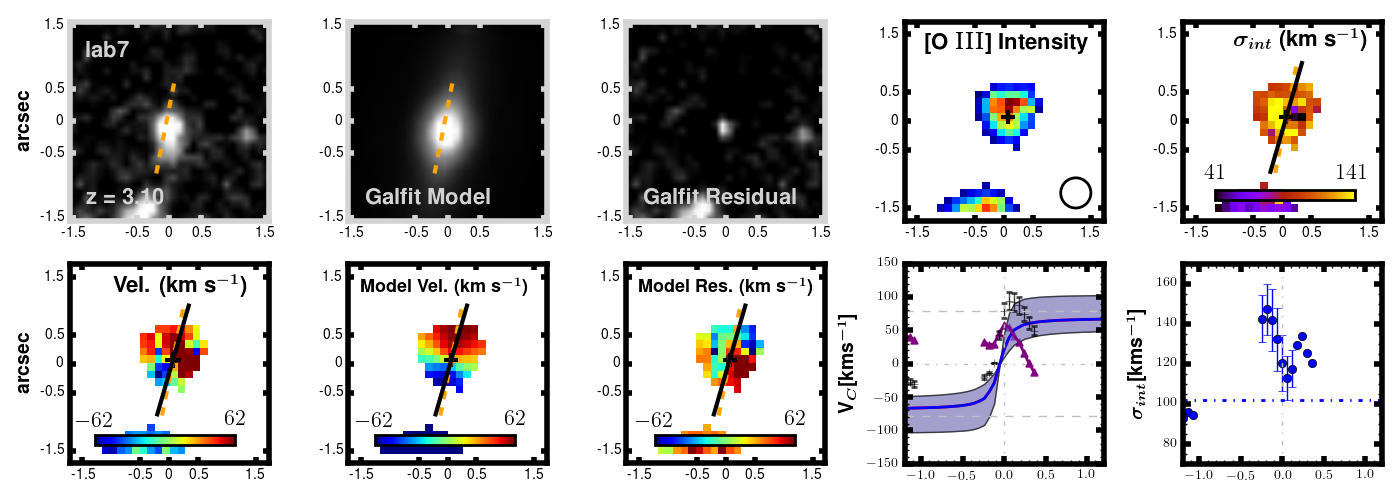}
    \includegraphics[width=0.95\textwidth]{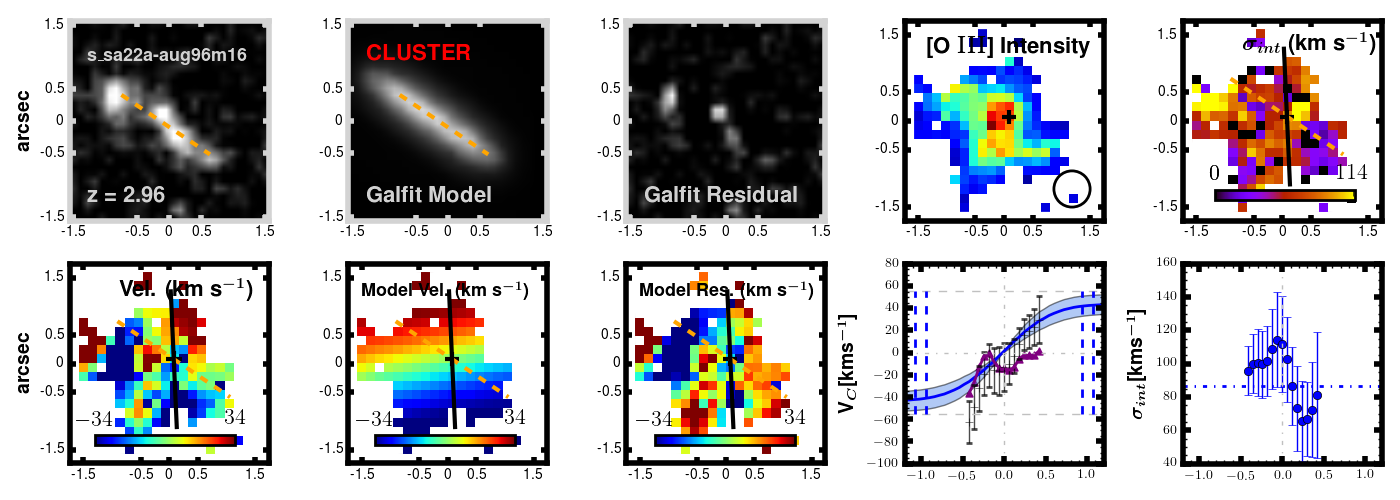}
    \includegraphics[width=0.95\textwidth]{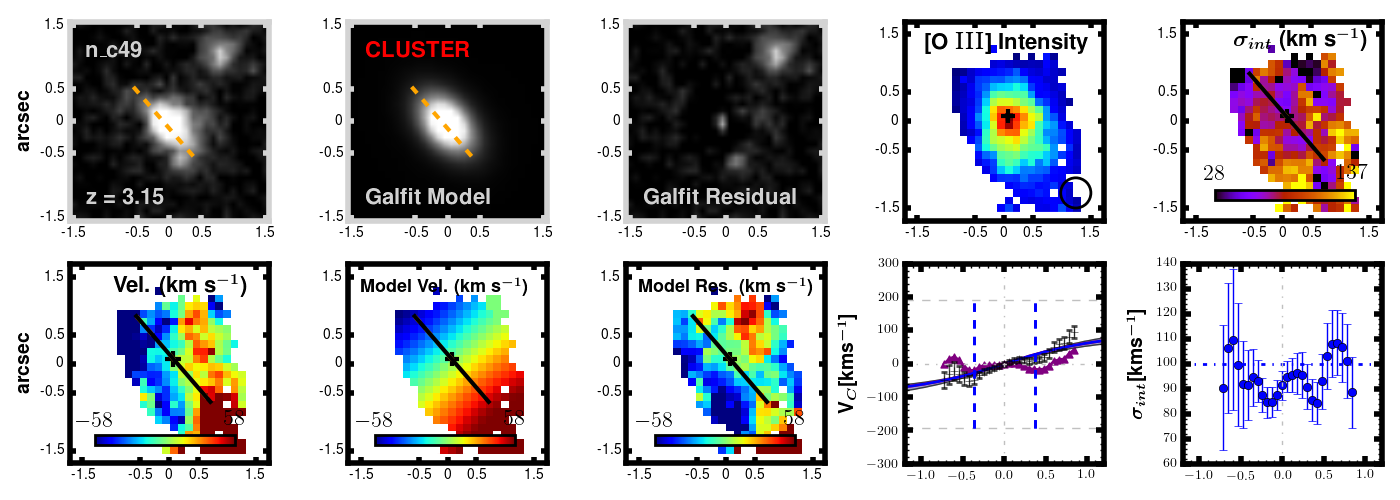}    

    \caption{\textbf{Continued.}}

\end{figure*}

\begin{figure*}\ContinuedFloat
    \centering

    \includegraphics[width=0.95\textwidth]{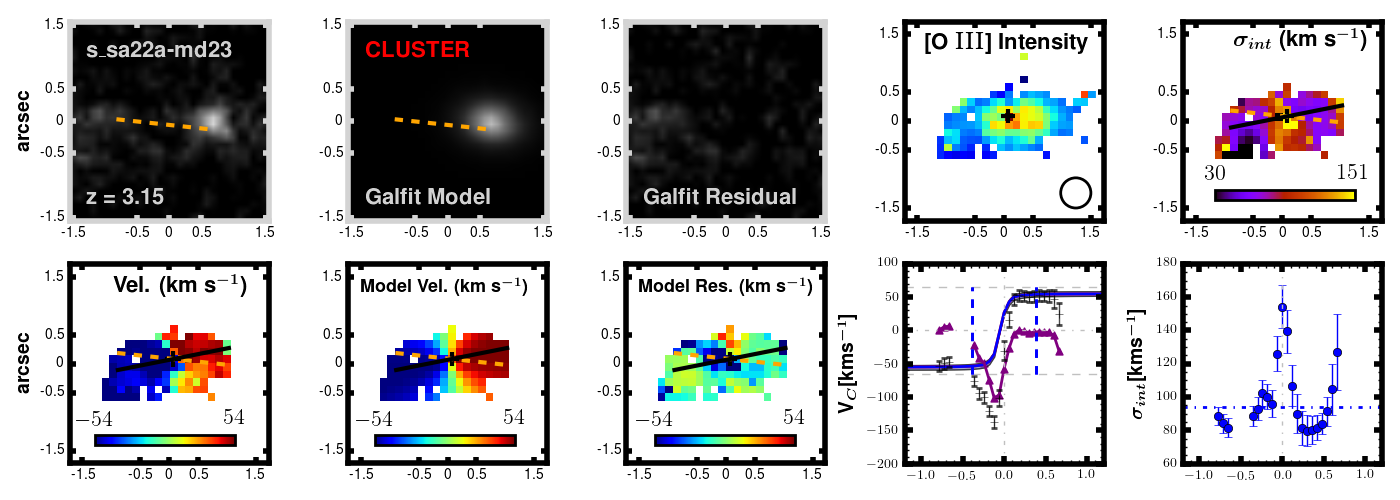}
    \includegraphics[width=0.95\textwidth]{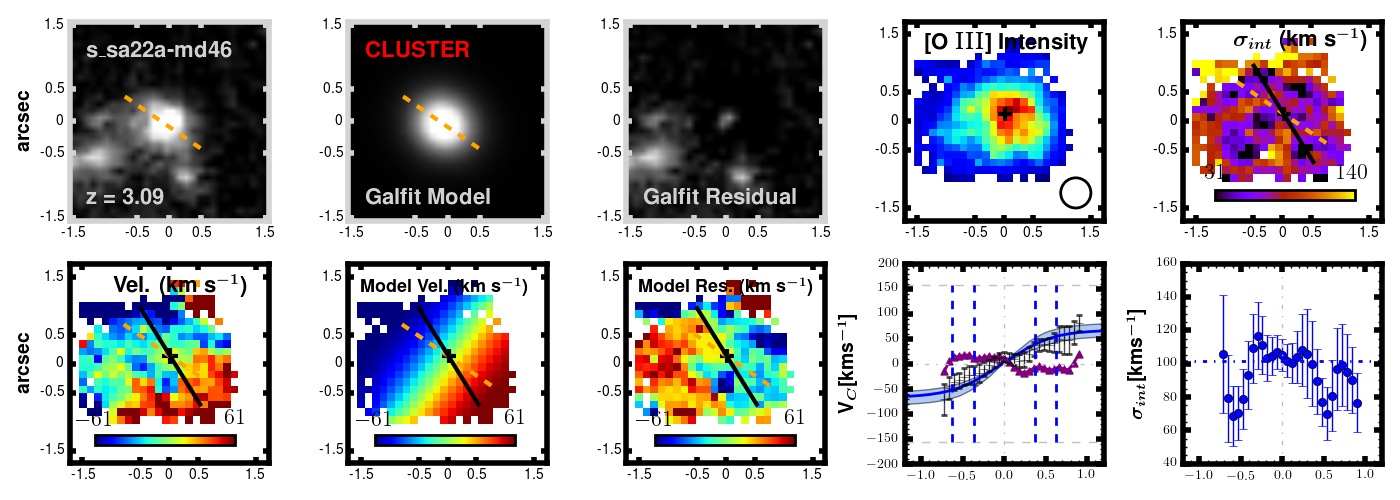}
    \includegraphics[width=0.95\textwidth]{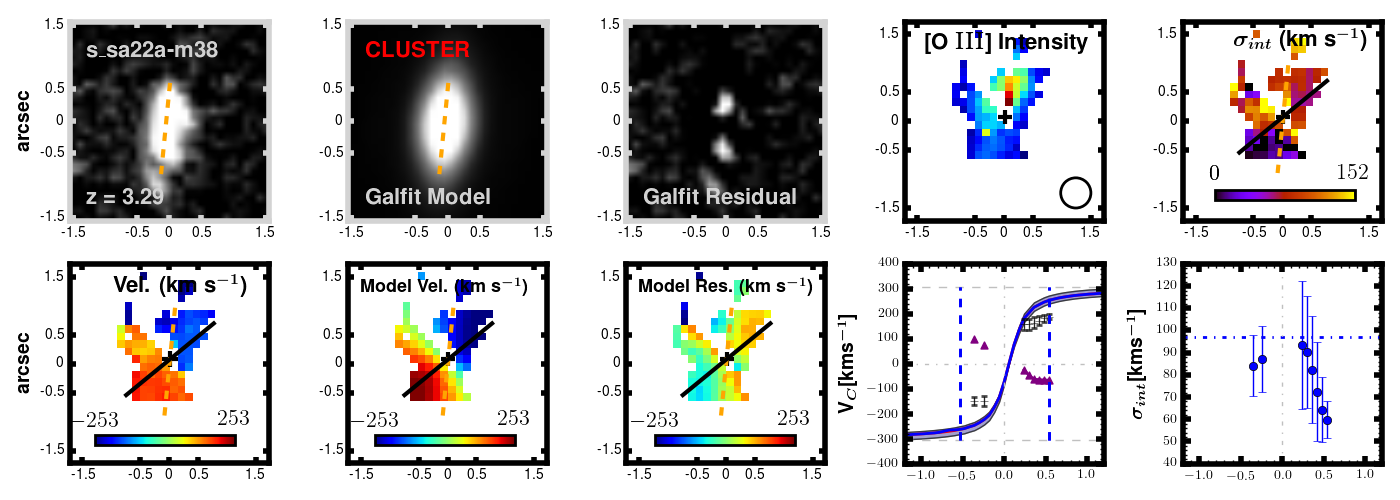}    
    \includegraphics[width=0.95\textwidth]{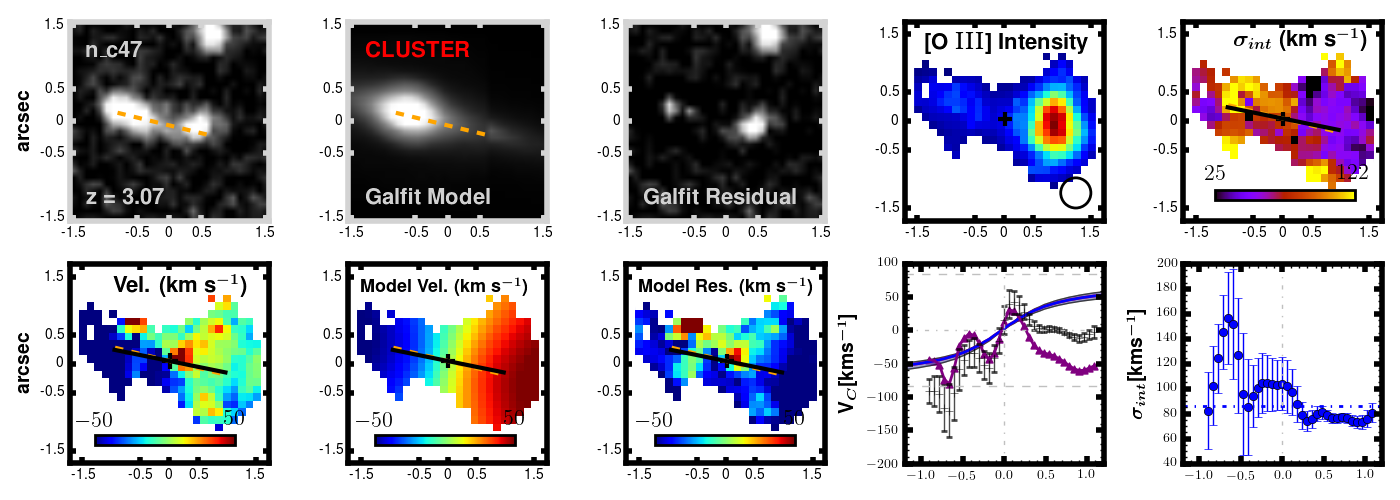}

    \caption{\textbf{Continued.}}

\end{figure*}

\begin{figure*}\ContinuedFloat
    \centering

    \includegraphics[width=0.95\textwidth]{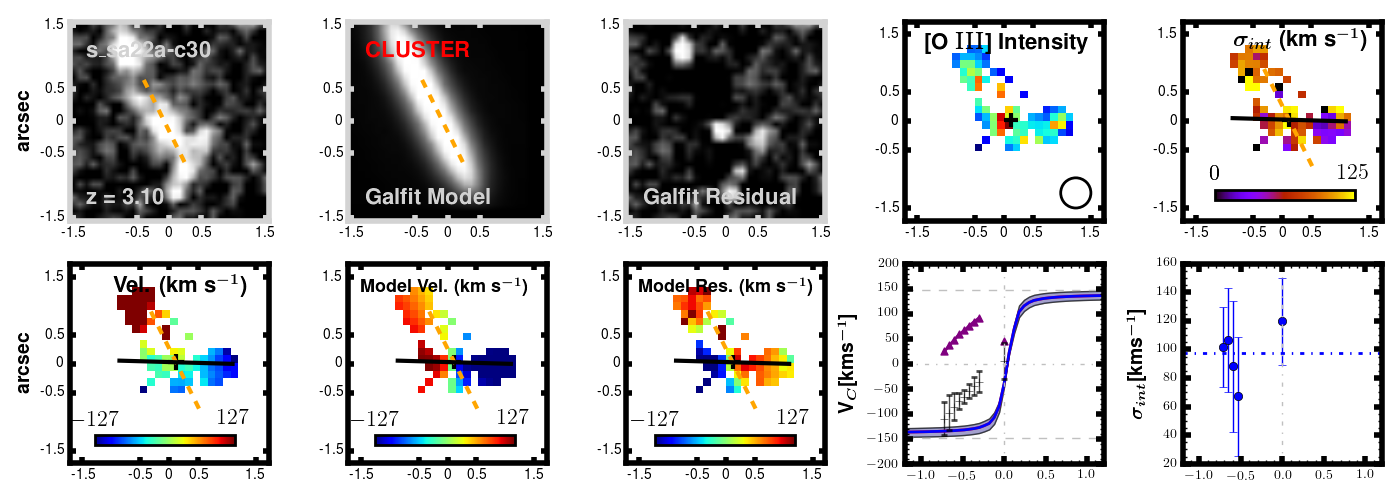}
    \includegraphics[width=0.95\textwidth]{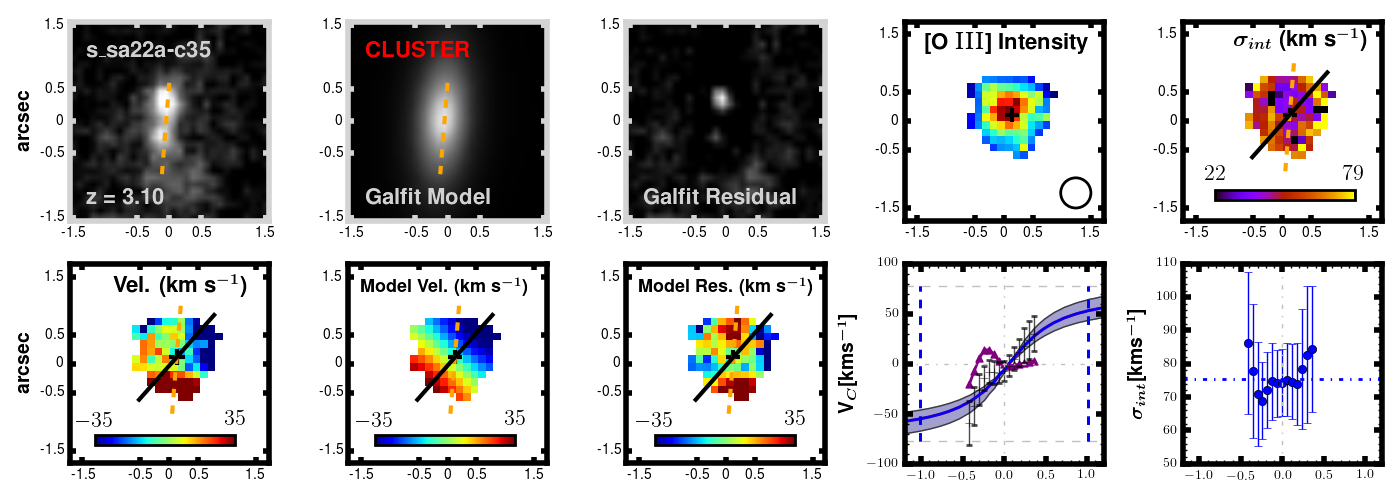}
    \includegraphics[width=0.95\textwidth]{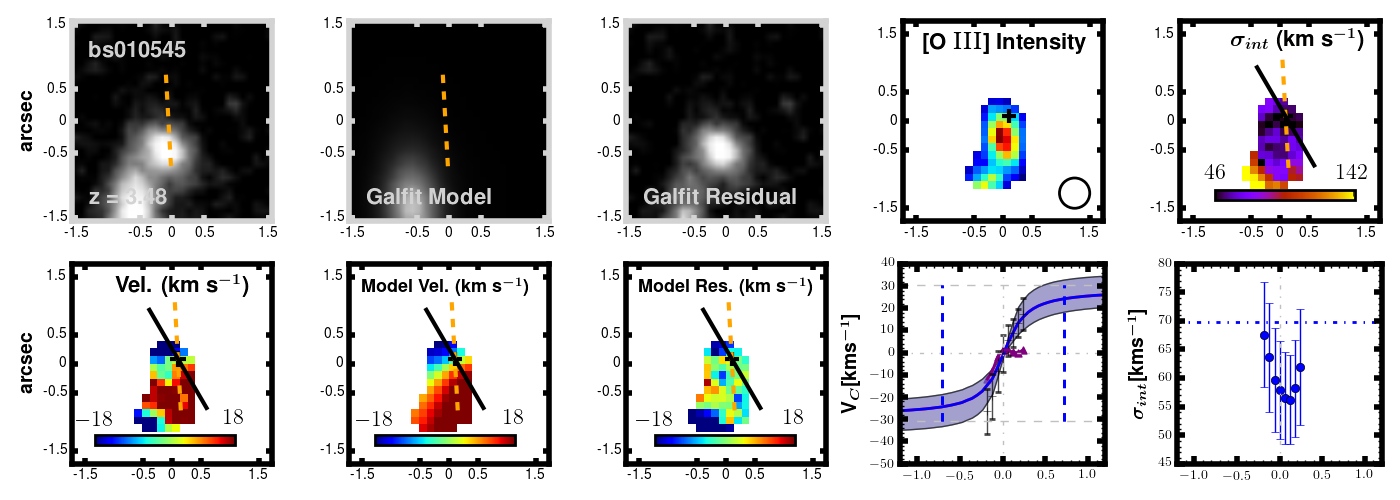}

    \caption{\textbf{Continued.}}

\end{figure*}


\bsp    
\label{lastpage}
\end{document}